\documentclass[a4paper]{IEEEtran}
\usepackage{acronym}

\usepackage[percent]{overpic}
\usepackage{array}
\usepackage{epstopdf}
\usepackage{fancyhdr}

\usepackage{lmodern}
\usepackage{textcomp}
\usepackage{algorithmic}
\usepackage{bigstrut}
\usepackage{rotating}
\usepackage{tabularx}
\usepackage{pifont}
\usepackage{multirow}

\usepackage{graphics,graphicx,amsopn,amsmath,amssymb,amsfonts}
\usepackage{cite}
\usepackage{citesort}

\usepackage{colortbl}

\newcommand{\cmark}{\ding{51}}%
\newcommand{\xmark}{\ding{55}}%

\newcolumntype{L}[1]{>{\raggedright\let\newline\\\arraybackslash\hspace{0pt}}m{#1}}
\newcolumntype{C}[1]{>{\centering\let\newline\\\arraybackslash\hspace{0pt}}m{#1}}
\newcolumntype{R}[1]{>{\raggedleft\let\newline\\\arraybackslash\hspace{0pt}}m{#1}}

\def\BibTeX{{\rm B\kern-.05em{\sc i\kern-.025em b}\kern-.08em
    T\kern-.1667em\lower.7ex\hbox{E}\kern-.125emX}}

\begin{document}


\title{\LARGE \bf Non-Coherent and Backscatter Communications: Enabling Ultra-Massive Connectivity in 6G Wireless Networks}



\author{
Syed Junaid Nawaz,
Shree Krishna Sharma,
Babar Mansoor,
Mohmammad N. Patwary, and
Noor M. Khan
\thanks{S. J. Nawaz and B. Mansoor are with the Department of Electrical and Computer Engineering, COMSATS University Islamabad (CUI), Islamabad 45550, Pakistan. (e-mail:{\it junaidnawaz@ieee.org} and {\it babar\_mansoor@comsats.edu.pk})}
\thanks{S. K. Sharma is with the SnT - securityandtrust.lu, University of Luxembourg, Kirchberg, Luxembourg 1855, Luxembourg. (e-mail: {\it shree.sharma@uni.lu})}
\thanks{M. N. Patwary is with the
Faculty of Science \& Engineering, University of Wolverhampton, Wolverhampton WV1 1LY, UK. e-mail: {\it patwary@wlv.ac.uk}}
\thanks{N. M. Khan is with the Department of Electrical Engineering, Capital University of Science and Technology, Islamabad, Pakistan. (e-mail:{\it noor@ieee.org})
}
\thanks{Corresponding author: Syed Junaid Nawaz (e-mail: {\it junaidnawaz@ieee.org}).}
\thanks{Accepted for Publication in IEEE Access -- DOI: 10.1109/ACCESS.2021.3061499}
}


\markboth
{Preprint -- Accepted in IEEE Access -- S. J. Nawaz {\it et al.}: Non-Coherent \& Backscatter Communications}
{}
\maketitle

\begin{abstract}
With the commencement of the 5\textsuperscript{th} generation (5G) of wireless networks, researchers around the globe have started paying their attention to the imminent challenges that may emerge in the beyond 5G (B5G) era. Various revolutionary technologies and innovative services are offered in 5G networks, which, along with many principal advantages, are anticipated to bring a boom in the number of connected wireless devices and the types of use-cases that may cause the scarcity of network resources. These challenges partly emerged with the advent of massive machine-type communications (mMTC) services, require extensive research innovations to sustain the evolution towards enhanced-mMTC (e-mMTC) with the scalable network cost in 6\textsuperscript{th} generation (6G) wireless networks.
Towards delivering the anticipated massive connectivity requirements with optimal energy and spectral efficiency besides low hardware cost,
this paper presents an enabling framework for 6G networks, which utilizes two emerging technologies, namely, non-coherent communications and backscatter communications (BsC).
Recognizing the coherence between these technologies for their joint potential of delivering e-mMTC services in the B5G era, a comprehensive review of their state-of-the-art is conducted. The joint scope of non-coherent and BsC with other emerging 6G technologies is also identified, where the reviewed technologies include unmanned aerial vehicles (UAVs)-assisted communications, visible light communications (VLC), quantum-assisted communications, reconfigurable large intelligent surfaces (RLIS), non-orthogonal multiple access (NOMA), and machine learning (ML)-aided intelligent networks. Subsequently, the scope of these enabling technologies for different device types (e.g., UAVs, body implants, etc), service types (e.g., e-mMTC), and optimization parameters (e.g., spectrum, energy, cost) is analyzed. Finally, in the context of the proposed non-coherent and BsCs based framework for e-mMTCs, some promising future research directions and open research challenges are highlighted.
\end{abstract}
\textbf{keywords:} 5G, 6G, backscatter, mMTC, non-coherent

\section*{List of Abbreviations}
\footnotesize
\begin{acronym}[CA-COOK] 

\acro{3GPP}{3\textsuperscript{rd} Generation partnership project}
\acro{3-D}{3-Dimensional}
\acro{5G}{5\textsuperscript{th} Generation}
\acro{6G}{6\textsuperscript{th} Generation}

\acro{AcBsC}{Acoustic backscatter communications}
\acro{ADC}{Analog-to-Digital Converter}
\acro{AI}{Artificial intelligence}
\acro{AMPS}{Advanced mobile phone service}
\acro{AoA}{Angle-of-arrival}
\acro{AP}{Access point}
\acro{APM}{Amplitude-phase modulation}
\acro{APSK}{Amplitude-phase shift keying}
\acro{ASK}{Amplitude shift keying}
\acro{AWGN}{Additive white Gaussian noise}

\acro{B2B}{Backscatter-device-to-backscatter-device}
\acro{B5G}{Beyond 5G}
\acro{BAN}{Body area networks}

\acro{BsC}{Backscatter Communications}

\acro{BS}{Base station}

\acro{BER}{Bit error rate}

\acro{CA-COOK}{Chip average chaotic on-off keying}
\acro{CDMA}{Code division multiple access}
\acro{CoMP}{Coordinated multipoint}
\acro{CSI}{Channel State Information}

\acro{D2D}{Device-to-device}
\acro{DCC}{Direct chaotic communication}
\acro{DF}{Decode-and-forward}
\acro{DFDD}{Decision-Feedback differential detector}
\acro{DSP}{Digital signal processor}
\acro{DPSK}{Differential phase shift keying}

\acro{e-MBB}{Enhanced mobile broadband}

\acro{e-mMTC}{Enhanced massive machine-type communications}

\acro{ESPRIT}{Estimation of signal parameters via rotational invariance techniques}

\acro{FDD}{Frequency division duplex}
\acro{FDMA}{Frequency division multiple access}
\acro{FSK}{Frequency shift keying}

\acro{GaN}{Gallium Nitride}
\acro{GDoF}{Generalized Degrees of Freedom}
\acro{GSM}{Global system for mobile}

\acro{HAP}{High-altitude platform}

\acro{IM}{Index modulation}
\acro{IoT}{Internet-of-things}
\acro{ICT}{Information and communication technologies}

\acro{ITU}{International telecommunication union}

\acro{LAP}{Low-altitude platform}
\acro{LIS}{Large intelligent surfaces}
\acro{LoS}{Line-of-sight}
\acro{LRS}{Large reflective surfaces}

\acro{LTE}{Long-term evolution}

\acro{MAC}{Medium access control}
\acro{MAP}{Maximum a posteriori probability}

\acro{MEC}{Mobile edge computing}
\acro{MFSK}{Multiple FSK}
\acro{MIMO}{Multiple-input multiple-output}
\acro{MIS}{Multiple-input single-output}
\acro{ML}{Machine Learning}
\acro{MLh}{Maximum-likelihood}
\acro{mMTC}{Massive machine-type communications}
\acro{mMIMO}{Massive MIMO}
\acro{mmWave}{Millimeter Wave}

\acro{MAP}{Maximum aposteriori probability}
\acro{MRC}{Maximum ratio combining}
\acro{MS}{Multiple symbol}
\acro{MSMA}{Multiple subcarrier multiple access}
\acro{MSDD}{MS differential detector}
\acro{MSDSD}{MS differential sphere detector}
\acro{MUD}{Multi-user detection}
\acro{MUSIC}{Multiple signal classification}

\acro{NFV}{Network function virtualization}
\acro{nLoS}{non-LoS}
\acro{NOMA}{Non-orthogonal multiple access}

\acro{OAM}{Orbital angular momentum}
\acro{OMA}{Orthogonal multiple access}

\acro{OFDM}{Orthogonal frequency division multiplexing}
\acro{OFDMA}{Orthogonal frequency division multiple access}
\acro{OOK}{On-off keying}

\acro{PAM}{Pulse amplitude modulation}
\acro{PDU}{Protocol data unit}
\acro{PLL}{Phase lock loop}
\acro{PLS}{Physical layer security}

\acro{PeM}{Permutation modulation}
\acro{PSK}{Phase shift keying}

\acro{QAM}{Quadratural amplitude modulation}
\acro{QBsC}{Quantum BsC}
\acro{QC}{Quantum computing}
\acro{QI}{Quantum illumination}
\acro{QoS}{Quality of service}
\acro{QPSK}{Quadrature phase shift keying}
\acro{QSAs}{Quantum search algorithms}

\acro{RF}{Radio frequency}
\acro{RFID}{RF identification}
\acro{RLIS}{Reconfigurable LIS} 
\acro{RMS}{Root mean square}

\acro{SC}{Selection combining}
\acro{SD}{Software-defined}
\acro{SDMA}{Space division multiple access}
\acro{SDN}{SD networks}
\acro{SDP}{Semidefinite programming}
\acro{SER}{Symbol error rate}
\acro{SIC}{Successive interference cancellation}
\acro{SIMO}{Single-input multiple-output}
\acro{SISO}{Single-input single-output}
\acro{SS}{Symbol-by-symbol}
\acro{SWIPT}{Simultaneous wireless information and power transfer}

\acro{SM}{Spatial modulation}

\acro{SNR}{Signal-to-noise ratio}

\acro{SR}{Symbiotic Radio}

\acro{SSK}{Space-shift keying}

\acro{TACD}{Total access communication system}
\acro{T2T}{Tag-to-tag}
\acro{TDMA}{Time division multiple access}
\acro{THz}{Terahertz}
\acro{THID}{THz indentification}
\acro{ToA}{Time-of-arrival}

\acro{UAV}{Unmanned aerial vehicles}
\acro{UDN}{Ultra-dense network}
\acro{UHEE}{Ultra-high energy efficiency}
\acro{UHF}{Ultra high frequency}
\acro{URLLC}{Ultra-reliable low-latency communications}

\acro{VLC}{Visible light communications}
\acro{VLBsC}{Visible light BsC}

\acro{W-CDMA}{Wideband-CDMA}
\acro{WiFi}{Wireless fidelity}
\acro{WSN}{Wireless sensor networks}

\end{acronym}

\section{Introduction} 

The future generations of wireless networks are believed to become an essential part of everyone's daily life subserved through the revolution in network services derived from the notion of ``everything-as-a-service". The plethora of new services emerged from the idea will require highly efficient information and communication technologies (ICT) as the core elements of the echo system. Living in future civil societies will require network connectivity as an essential element, just as air is to breathe. The provision of network connectivity to everything at all-time with ultra-high reliability, ultra-high energy efficiency, ultra-low latency and ultra-high-speed may require the development of new revolutionary technologies along with the evolution in current technologies.

 The rollout of 5\textsuperscript{th} Generation (5G) wireless networks started with the Release-15 of 3\textsuperscript{rd} Generation Partnership Project (3GPP) \cite{3GPPrel15} in 2019. The standardization of 5G further matured with Release-16 of 3GPP, which is regarded as 5G phase 2. The 5G networks have taken an immense leap forward with the introduction of various new technology innovations, e.g., massive multiple-input multiple-output (mMIMO), NOMA, full-duplex, mmWave, ultra-dense small-cells, intelligent network operations, mobile edge computing (MEC), software-defined networks (SDN), to name a few \cite{8951153}. The notable target features of 5G networks as defined by the International Telecommunication Union (ITU) include enhanced mobile broadband (eMBB), massive machine-type communication (mMTC), and ultra-reliable low-latency communications (URLLC).

\subsection{Beyond 5G (B5G) Wireless Networks}

The advent of new technologies and services of 5G networks is predicted to massively increase the number of devices connected to the network and the volume of data generated in the network. The global mobile-users and machine type subscriptions are predicted to reach 17.1bn and 97bn by the year 2030 \cite{union2015imt}, respectively. Moreover, including the contribution from mMTC, an annual increase of 55\% in mobile data traffic is estimated for the years between 2020 and 2030, which is expected to result in the generation of $5.016$ ZetaBytes data traffic per month by the year 2030. It can be foreseen that the capacity of 5G wireless networks may reach its limit in a decade or so \cite{QML_6G_Junaid}.

The demands of ultra-high energy efficiency (UHEE) to concede massively populated and densely composed communication networks will emerge as a complex challenge than ever before in the beyond 5G (B5G) (or 5G+) wireless networks. B5G refers to the network specifications and requirements represented by the era beyond the launch of 5G and before the anticipated launch of 6G. 5G is in its very fist stage of development and several modifications are expected for the standards to get matured during B5G period. On the other hand, the term 6G has been recently introduced in the research community and it may take a decade or so before it gets implemented in practice. Moreover, the realization of information detection accuracy in a densely connected network operating in a non-synchronous fashion may emerge as another stern challenge. In \cite{chianiopen}, various open research challenges for B5G wireless networks are reviewed. The telecommunication engineers and researchers from around the globe have now started postulating the hypothesis of new technologies and exploration of new spectrum to meet the envisioned drastically increasing demands of 6G and beyond wireless networks.

Among the recent literature on 6G vision,  \cite{QML_6G_Junaid} has proposed Quantum Computing (QC), Machine Learning (ML), and Quantum ML based framework for 6G communication networks. Also, the scope of QC-assisted ML solutions to meet the reconfigurability demands of 6G communication networks at network-side, network-edge, air-interface, and user-side have been comprehensively discussed. In \cite{8782879}, the need for B5G wireless network has been motivated, and critical technical challenges have been discussed. In addition, the potential solutions associated with spectrum, energy, core network design, security, and testbed development of 6G networks have been discussed. A study on needs, technical requirements, and solutions for 6G networks has been presented in \cite{Tariq6G_2019}, in which various interesting potential technologies for 6G wireless networks have been suggested. Furthermore, in \cite{8412482}, the evolution trend of mobile network generations, from 1G to 5G, has been evaluated to forecast the need and requirements for 6G networks.

In Table \ref{tab_6G_Technologies_Comparison}, we provide a summary of the state-of-the-art of the potential candidate technologies for 6G wireless networks. Among the many indicated challenges and potential technologies, this paper focuses on the potential technologies associated with energy efficiency, network capacity, and reliability demands of 6G wireless networks. To this end, non-coherent and backscatter communications (BsC) are mainly focused.


\subsection{B5G Network Capacity and Energy Efficiency}

Towards the realization of green-, smart-, and automated-World, 5G wireless networks are expected to contribute a manifold increase in network capacity and energy efficiency compared to 4G networks through different revolutionary technologies. However, the envisioned massive increase in the operational scale and dimensions of the future networks, partly emerged from the advent of mMTCs in 5G, necessitates the exploration of methods for further enhancement in network capacity and energy efficiency. One promising future technology to improve the converge, capacity, and energy efficiency of the networks is the introduction of mobility at the base stations (BS) served through wireless backhaul, e.g., unmanned aerial vehicles (UAVs) based serving stations. The efficiency quantification of such massively connected 3-dimensional (3-D) future wireless networks, in terms of bits/sec/Joule/meter$^3$ (b/s/J/m$^3$), is a highly compelling research problem. The paradigm of battery-less mMTC through BsC links can help in achieving the UHEE in 6G wireless networks. Quantum technologies assisted BsC has also very recently emerged as a promising way forward. The BsC and ambient BsC are believed to have a strong potential in not only enhancing the volumetric network capacity and energy efficiency but also the spectral efficiency through their spectrum sharing paradigm \cite{memon2019ambient}.

Subsequently, the emerged paradigm of Symbiotic Radio (SR) shares not only the radio-spectrum but also the radio-source, which has a strong potential in further enhancing the network operational cost and energy efficiency \cite{long2018symbiotic}. The accurate knowledge of radio channel characteristics in such radio systems is very critical in precise suppression of interference and detection of information. The creation of a favorable radio propagation environment through the designing and deployment of large intelligent surfaces (LISs) and reconfigurable LISs (RLISs) in a particular communication scenario is another emerging paradigm in the quest of achieving ultrahigh energy and spectral efficiency in future wireless networks \cite{huang2018large}. The tweaking of transmit power and scattering/reflection coefficient of the dominant scattering objects in the environment can help in achieving the UHEE. To this end, this article thoroughly reviews and analyzes the state-of-the-art of the BsC, ambient BsC, Quantum BsC, SR, and LISs technologies for addressing the network capacity and energy efficiency demands of the 6G wireless networks.

\subsection{B5G Non-Coherent Transceiver Design}

The transceiver design considerations for B5G communication scenarios involving massive connectivity in a distributed, heterogeneous, and non-cooperative context operating over a wide range of the spectrum, including radically high-frequency bands, may emerge as a vital challenge for the future. The coherent transceiver designs are conventionally preferred over their non-coherent counterparts for their advantage of high bit error rate (BER) performance. On the other hand, this advantage has an associated cost of complex additional circuitry needed for carrier acquisition or synchronization. Also, the imbalance between the in-phase and quadrature-phase branches of a coherent transceiver leads to an imperfect image-frequency filtration. This performance-limiting factor becomes more noticeable for high carrier frequency communications, e.g, mmWave, sub-teraHz, and teraHz bands.

Moreover, the performance of coherent information detection methods heavily depends on the availability of a reliable estimate of the channel state information (CSI). Obtaining an accurate channel estimate for a massive number of devices employing distributed multi-antenna systems with the non-orthogonal resource allocation may emerge as a challenging task for B5G networks. In addition, the application scenarios requiring a simple transceiver design usually prefer non-coherent detection methods by sacrificing at the BER performance. The massive connectivity in B5G communications context, e.g., in mMTCs, the devices may only need occasional communication of small data. The suitability of non-coherent transceiver design for 6G communication applications requires a thorough investigation, which is one of the focuses of this paper.

\begin{table*}[t]
  \centering
  \caption{State-of-the-art of Potential 6G Technologies (\cmark -- represents clearly suggested/discussed, \xmark -- represents  not suggested/discussed)}
  \resizebox{\textwidth}{!}{
    \begin{tabular}{|p{6.5em}|c|c|c|c|c|c|c|c|c|c|c|c|c|c|c|c|}
    \hline
    6G Proposals &
      \multicolumn{16}{c|}{Potential 6G Technologies}
      \bigstrut\\
\cline{2-17}   (2019 - 2020)  &
      \multicolumn{4}{c|}{Spectrum} &
      \multicolumn{3}{p{8.5em}|}{Cellular or Cell-free 3-D Networks} &
      \multicolumn{3}{p{8.6em}|}{RF, Modulation, and Detection} &
      \multicolumn{2}{p{4.2em}|}{Energy Efficiency} &
      \multicolumn{4}{p{10.5em}|}{Data analytic, security, and computing}
      \bigstrut\\
\cline{2-17}     &
      \multicolumn{1}{p{1.8em}|}{VLC} &
      \multicolumn{1}{p{3.5em}|}{mm-Wave, Sub-teraHz, \& teraHz} &
      \multicolumn{1}{p{2.7em}|}{Full-Duplex} &
      \multicolumn{1}{p{3.3em}|}{Dynamic Sharing} &
      \multicolumn{1}{p{2.8em}|}{Distr-ibuted Systems} &
      \multicolumn{1}{p{2.3em}|}{UAVs, HAP, LAP} &
      \multicolumn{1}{p{2.2em}|}{Small- \& Tiny- Cells} &
      \multicolumn{1}{p{2.3em}|}{Fluid Ante-nnas} &
      \multicolumn{1}{p{2em}|}{OAM} &
      \multicolumn{1}{p{2em}|}{Non-Cohe-rent} &
      \multicolumn{1}{p{1.7em}|}{LISs} &
      \multicolumn{1}{p{2.8em}|}{BsC/ EH/ SWIPT} &
      \multicolumn{1}{p{3.7em}|}{Quantum Technologies} &
      \multicolumn{1}{p{1.8em}|}{AI/ ML} &
      \multicolumn{1}{p{2.25em}|}{Block- chain} &
      \multicolumn{1}{p{3em}|}{Edge computing \& caching}
      \bigstrut\\
    \hline\hline
    Nawaz et al. \cite{QML_6G_Junaid}
    & \cmark & \cmark & \cmark & \cmark & \cmark & \cmark & \cmark & \cmark & \cmark & \xmark & \cmark & \xmark & \cmark & \cmark & \xmark & \cmark
\bigstrut \\ \hline
    Tariq et al. \cite{Tariq6G_2019}
    & \cmark & \cmark & \xmark & \xmark & \cmark & \cmark & \xmark & \cmark & \cmark & \xmark & \cmark & \xmark & \cmark & \cmark & \cmark & \cmark
\bigstrut\\ \hline
    David et al. \cite{8412482}
    & \cmark & \cmark & \xmark & \cmark & \xmark & \xmark & \xmark & \xmark & \xmark & \xmark & \xmark & \cmark & \xmark & \cmark & \xmark & \xmark
\bigstrut \\ \hline
    Rappaport et al. \cite{8732419}
    & \xmark & \cmark & \xmark & \xmark & \xmark & \cmark & \xmark & \xmark & \xmark & \xmark & \xmark & \xmark & \xmark & \xmark & \xmark & \xmark
\bigstrut \\ \hline
    Corre et al. \cite{corre2019sub}
    & \xmark & \cmark & \xmark & \xmark & \xmark & \xmark & \xmark & \xmark & \xmark & \cmark & \xmark & \xmark & \xmark & \xmark & \xmark & \xmark
\bigstrut \\ \hline
    Saad et al. \cite{Saad_6G_2019}
    & \cmark & \cmark & \xmark & \xmark & \cmark & \xmark & \cmark & \xmark & \xmark & \xmark & \cmark & \cmark & \cmark & \cmark & \cmark & \cmark
\bigstrut \\ \hline
    Strinati et al. \cite{strinati20196g}
    & \cmark & \cmark & \xmark & \xmark & \cmark & \cmark & \xmark & \xmark & \xmark & \xmark & \xmark & \cmark & \xmark & \cmark & \cmark & \cmark
\bigstrut \\ \hline
    Letaief et al. \cite{letaief2019roadmap}
    & \xmark  & \cmark & \xmark & \cmark & \cmark & \cmark & \xmark & \xmark & \xmark & \xmark & \cmark & \xmark & \xmark & \cmark & \xmark & \cmark
\bigstrut \\ \hline
    Giordani et al. \cite{giordani2019towards}
    & \cmark & \cmark & \cmark & \cmark & \cmark & \cmark & \cmark & \xmark & \xmark & \xmark & \xmark & \cmark & \xmark & \cmark & \xmark & \cmark
\bigstrut \\ \hline
    Yaacoub et al. \cite{yaacoub2019key}
    & \cmark & \cmark & \xmark & \cmark & \cmark & \cmark & \cmark & \xmark & \xmark & \xmark & \xmark & \xmark & \xmark & \cmark & \cmark & \cmark
\bigstrut\\ \hline
    Yang et al. \cite{8782879}
    & \cmark & \cmark & \xmark & \cmark & \cmark & \cmark & \xmark & \xmark  & \cmark & \xmark & \xmark & \cmark & \cmark & \cmark & \cmark  & \xmark
\bigstrut\\ \hline
    Baiqing et al. \cite{8760275}
    & \cmark & \cmark & \xmark & \cmark & \cmark & \cmark & \xmark & \xmark & \xmark & \xmark & \cmark & \xmark & \xmark & \cmark & \xmark & \xmark
\bigstrut \\ \hline
    Zhang et al. \cite{8766143}
    & \cmark & \cmark & \xmark & \cmark & \cmark & \cmark & \xmark & \xmark & \cmark & \xmark & \cmark & \xmark & \cmark & \cmark & \cmark & \cmark
\bigstrut \\ \hline
    Raghavan et al. \cite{8603730}
    & \xmark & \cmark & \xmark & \xmark & \xmark & \cmark & \cmark & \xmark & \xmark & \xmark & \xmark & \cmark & \xmark & \cmark & \xmark & \cmark
\bigstrut \\ \hline
    Viswanathan et al. \cite{9040431}
    & \cmark & \cmark & \cmark & \cmark & \cmark & \cmark & \xmark & \xmark & \xmark & \xmark & \cmark & \cmark & \cmark & \cmark & \xmark & \cmark
\bigstrut \\ \hline
    Rajatheva et al. \cite{rajatheva2020white}
    & \cmark & \cmark & \cmark & \xmark & \cmark & \cmark & \cmark & \xmark & \cmark & \cmark & \cmark & \cmark & \xmark & \cmark & \xmark & \cmark
\bigstrut \\ \hline
You et al. \cite{You2020towards}
    & \cmark & \cmark & \xmark & \cmark & \cmark & \cmark & \cmark & \xmark & \cmark & \xmark & \cmark & \cmark & \cmark & \cmark & \cmark & \cmark
\bigstrut \\ \hline
    \end{tabular}
    }
  \label{tab_6G_Technologies_Comparison}
\end{table*}

\subsection{Contributions and Organization}

The advantages of hardware-simplicity and low-cost offered by non-coherent communications and energy-, spectral-, and cost-efficiency offered by BsC, a non-coherent BsC-based framework for 6G e-mMTC services is proposed. The main contributions of this work are explicitly described in as follows,

\begin{itemize}
  \item The need for 6G wireless networks is motivated by conducting a comprehensive survey of the 5G technologies and their shortcomings anticipated to emerge in B5G era. This work also summarizes the requirements and potential candidate technologies for 6G wireless networks, which have recently appeared in the literature.
  \item The potential of BsC technology for supporting the anticipated ultra-massive connectivity requirements in 6G networks is motivated along with a comprehensive review of the related state-of-the-art. Compared to some notable existing survey articles on BsC, e.g., \cite{8368232,memon2019backscatter}, our work additionally extensively discusses various new associated technologies which are not thoroughly reviewed in the existing literature, e.g., quantum BsC, quantum ambient BsC, visible-light BsC, acoustic BsC, and SR, etc. The potential of other emerging radio reflective communication technologies, such as LIS-assisted communications, for achieving ultra-high energy- and spectral-efficiency in massively connected 6G wireless networks is also reviewed.
  \item In the light of the 6G networks requirements and services, a thorough review of non-coherent communication concept is conducted. A comparative review of coherent and non-coherent communication methods for different futuristic use-cases and application scenarios is also conducted. The scope of employing advanced non-coherent transceivers in the context of distributed massive-MIMO (cell-free), NOMA, vehicular communications, and massive internet-of-things (IoT) technologies is also reviewed. Moreover, various open research challenges for non-coherent transceiver design are identified and discussed.
  \item An enabling framework for e-mMTC services delivery in 6G wireless networks is proposed, which utilizes two emerging technologies, namely, non-coherent communications and BsC. To this end, the joint applications and use-cases of non-coherent and BsC are thoroughly reviewed. Furthermore, in the context of proposed framework, the open research challenges and potential future research directions are also highlighted.
\end{itemize}

 The remainder of this paper is organized as follows. Section II discusses the 5G technology innovations, 5G target services, and the challenges expected to emerge in B5G era. Section III comprehensively reviews the state-of-the-art of non-coherent communications. Section IV discusses reflective radio technologies, primarily BsCs, for meeting the ultra-high energy efficiency demands of 6G wireless networks. Section V presents the proposed framework for 6G e-mMTC services along with highlights on open research challenges. Finally, Section VI concludes the paper.


\begin{figure*}[t]
  \centering
  \includegraphics[width=\textwidth]{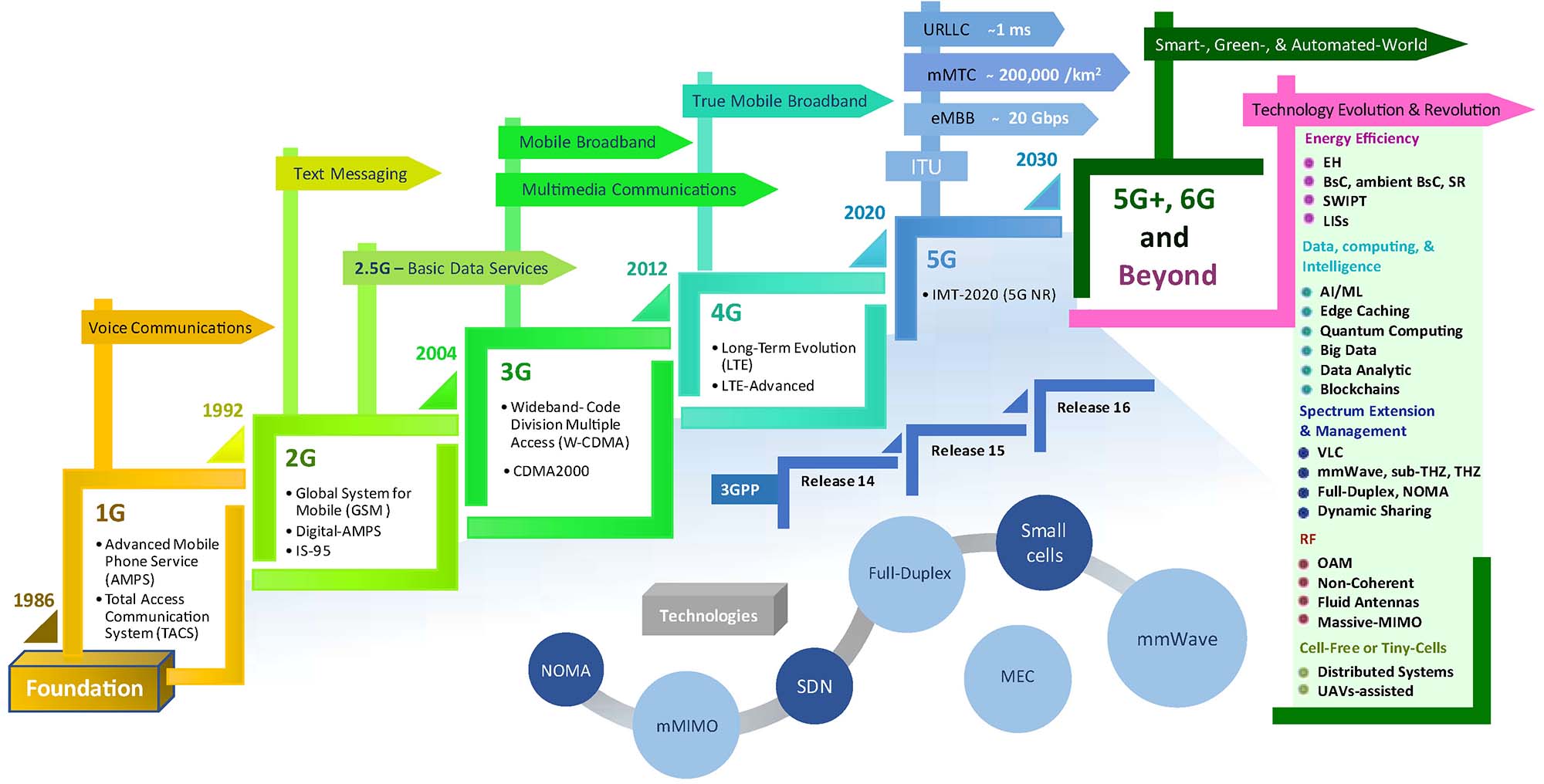}
  \caption{Evolution in Generations of Mobile Wireless Networks }\label{Fig_Evolution}
\end{figure*}

\section{Beyond 5G Wireless Era}

 The rapid growth of mobile technologies over the past two decades has completely transformed the way people live. Technological advancements in mobile phones have given rise to a variety of data-hungry applications resulting in ever-increasing demands for high data rate, low latency, and high reliability of wireless connectivity. To meet these demands, several innovative technologies are originated by 5G such as mMIMO, NOMA, mmWave, full-duplex, ultra-dense networks (UDN), network slicing, network function virtualization, SDNs, etc. The services offered by 5G can be broadly categorized into mMTC, eMBB, and URLLC service types \cite{Sharma2020COMST}. In Fig. \ref{Fig_Evolution}, a brief history of evolution and revolution in technologies and services of mobile networks is illustrated \cite{imtvision2020}, where some potential technology innovations expected in the B5G era are also indicated. The key requirements of the B5G wireless era include, but are not limited to, manifold improvement in network capacity, spectral and energy efficiency, and latency performance compared to its predecessor technologies. In the following subsections, a brief overview of 5G innovative services and technologies is presented followed by the discussion on the need for B5G networks.

\begin{figure*}[t]
  \centering
  \includegraphics[width=0.9\textwidth]{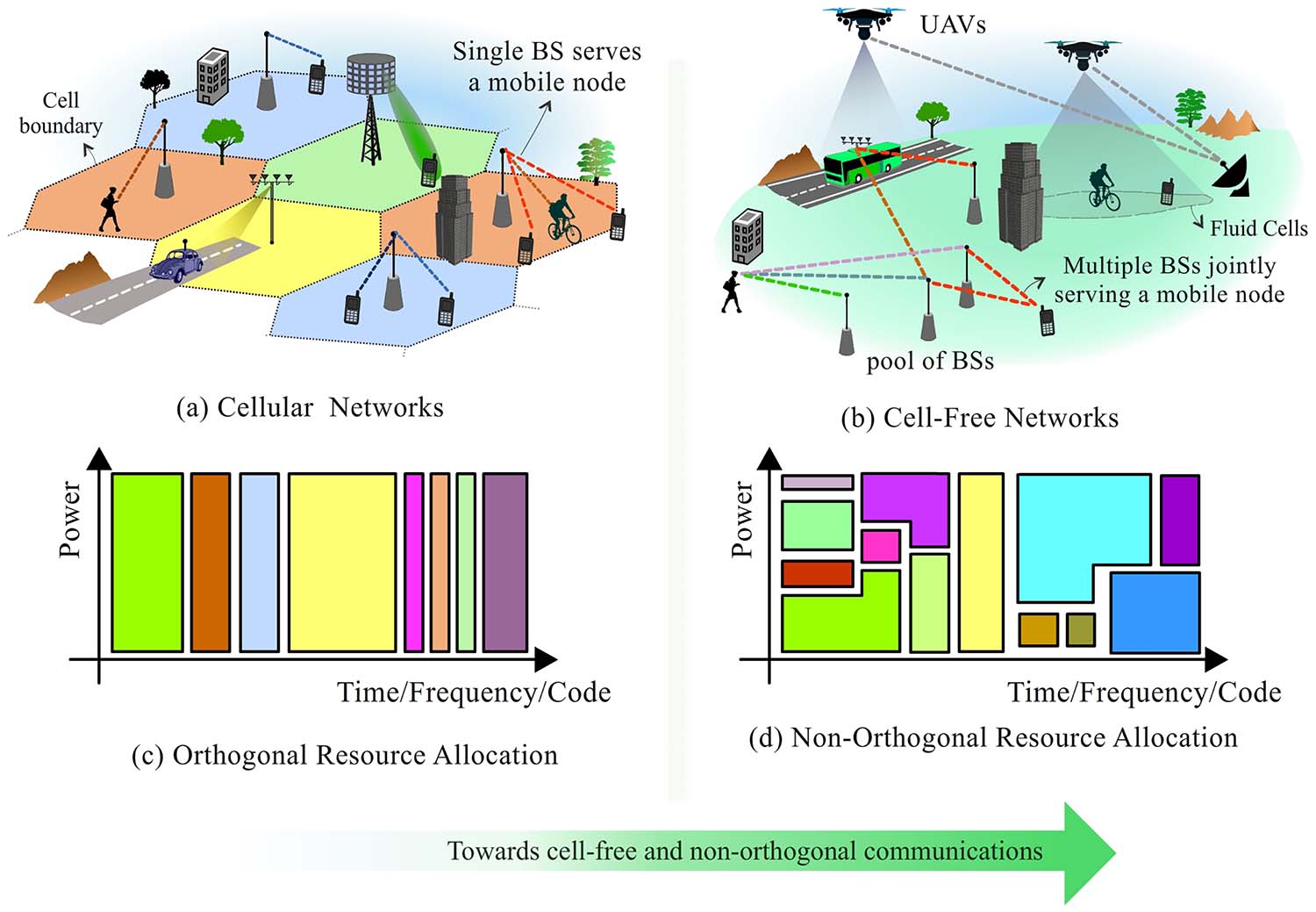}
  \caption{Towards cell-free and non-orthogonal communications.}
  \label{Fig_Block_Diagram}
\end{figure*}

\subsection{5G Target Services}

The key 5G services can be categorized as follows:

\subsubsection{Enhanced Mobile Broadband}
A primary use-case of 5G is the eMBB which is an extended and enhanced version of the current mobile broadband communication scenario. The notion of eMBB is to meet the demands of bandwidth hungry applications such as ultra high definition video streaming, virtual and augmented reality, and autonomous driving, by maximizing the data rates with high link reliability and by providing seamless coverage and high quality-of-service (QoS).

As per vision 2020 of ITU, the target peak data rates of $20$ Gbps in the downlink and $10$ Gbps in the uplink, the user experienced data rates of $100$ Mbps in the downlink and $50$ Mbps in the uplink, and mobility support of velocity up to $500$ km/h are defined for 5G. By taking these requirements into account, the key performance indicators for eMBB include the spectral efficiency and user experienced data rates. To meet these tight requirements, authors in \cite{7888974} have proposed a 3D mMIMO system, where it has been shown that the proposed system is capable to achieve the targeted spectral efficiency for 5G eMBB. By conducting field trials of the devised system, it is demonstrated that for full traffic load scenario, the proposed 3D mMIMO  enhances the throughput of the cell by a factor of $4\sim 6.7$. In order to fulfill the requirements for eMBB, some of the main enabling technologies include mMIMO, NOMA, mmWave based communications, and SDN.

\subsubsection{Massive Machine Type Communications}
``Everything connected to everything" is the slogan of emerging communication networks. The technological innovations have revolutionized not only the ways of communications but almost every walk of life such as homes, agriculture, transportation, and the industrial sector. The emerging concept of smart cities, autonomous cars, e-healthcare, utilities, and smart sensors has led to a huge number of device connections. Such devices possess the capability to measure, analyze, and transmit the information related to their surroundings.

The advent of mMTC in 5G is expected to drastically increase the number of network devices in the B5G era.
The anticipated increase in the number of devices, which sporadically communicate with each other, will require support for massive connectivity with efficient resource utilization in B5G networks \cite{Sharma2020COMST}. Due to infrequent and occasional network access requests generation nature of machine-type devices, only a subset of such devices will be randomly active at a given time. This necessitates the formulation of new efficient random network access protocols and control signaling mechanisms.
Moreover, due to the long-term battery-powered nature of such devices, a major challenge is to devise energy-efficient access and communication protocols.
In this context, authors in \cite{8360103} have discussed several medium access protocols for massive access attempts in the mMTC scenario. Also, authors in \cite{7405724} have proposed a scalable and adaptive network access protocol that exploits the broadcast information from neighboring devices in the 5G mMTC scenario. In addition, authors in \cite{7937902} have exploited data aggregation approach for resource scheduling in the mMTC scenario. Towards meeting the challenges of supporting mMTC in the 5G and beyond systems, some of the key enabling technologies include mMIMO, MEC, and SDN along with NFV.

\subsubsection{Ultra-Reliable Low-Latency Communications}

Another important use-case for 5G and beyond communications is URLLC. Several IoT-based application areas such as autonomous cars, remote surgery, machine-to-machine communications, smart grids, factory automation, and tactile internet (i.e., teleoperation systems, wireless virtual reality, wireless augmented reality \cite{TIAccess2020}) require support for high link reliability, very small packet errors, and very low latency.
To deliver the requirements of such applications in B5G networks, real-time data communication, processing, and decision making will be required. The goal 5G URLLC services is to achieve $1-10^{-5}$ success probability of transmitting a layer 2 protocol data unit (PDU) of 32 bytes within 1ms to the edge users in urban radio propagation environments \cite{3GPPURLLC}. The provision of URLLC is a challenging task due to conflicting requirements of low latency and high reliability.

In the above context, authors in \cite{8638959} have discussed several resource allocation and re-transmission schemes for URLLC scenario in 5G systems. In \cite{8782873}, authors have proposed ML and fountains code-based hybrid technique for spectrum access in mmWave range for URLLC scenario. The authors in \cite{8329618} have discussed the challenges faced in achieving URLLC while presenting some key technological enhancements for attaining the set targets in the URLLC scenario. Furthermore, in \cite{8329622}, authors have proposed a make-before-break handover mechanism for URLLC scenario. The fundamental trade-offs along with some fundamental rules that constitute radio access in URLLC have been discussed by the authors in \cite{8705373}. In this context, authors have considered mMIMO and massive-connectivity from the technological perspective. In order to achieve the targets set for URLLC, the key 5G technological enablers include full duplex, MEC, SDN, NOMA and mMIMO.

\subsection{Beyond 5G: Challenges and Requirements}
While the deployment of 5G is under-way, with some early versions already being rolled out in some countries, the debate for drawing a vision of B5G communication networks has also started in the research community. Though it is too early to clearly visualize the nature of B5G or 6G communication networks, speculative analyses can be conducted to initiate discussions and to set long-term goals. In 5G and beyond era, the shift in conventional cellular and orthogonal communications concepts towards cell-free and non-orthogonal communications, respectively, has strongly emerged for various different services and use-cases. In Fig. \ref{Fig_Block_Diagram}, a comparative illustration of cellular and orthogonal resources allocation setting with cell-free and non-orthogonal setting is provided.
Alongside, the emergence of new service areas and applications has led to an enormous growth in number of connected devices and global mobile data traffic.

It is predicted by the ITU that the global mobile data traffic will reach to an astonishing amount of about 5 zettabytes per month by the year 2030. Due to such an exponential growth in global data traffic, it is believed that the future mobile networks will be data driven rather than model driven \cite{8663550}.
Artificial intelligence (AI) and its sub-disciplines, such as ML and deep learning, are expected to play a vital role in the architecture, implementation, and management of B5G communication networks.
It is strongly believed that the AI empowered intelligent mobile networks will possess the capability to be autonomously modified and dynamically optimized as per users' experience. In this context, authors in \cite{8759894} have discussed the implementation of AI-based autonomous networks with knowledge-driven operations. Also, authors have also elaborated on both supervised and unsupervised learning methods of AI in the context of future wireless networks.

Furthermore, authors in \cite{8412482} have envisioned the requirements and services leading to communication networks B5G. Authors, in \cite{8663550}, have advocated to exploit the teraHz and sub-teraHz frequency spectrum in 6G to meet the demands of future wireless services. Moreover, authors have also discussed the technological enablers in terms of antennas and modulators, which could be used in teraHz range. In addition, authors in \cite{8760401} have presented the concept of integration of terrestrial and space networks in future communications for critical service areas such as defense, agriculture, and mining. a novel architecture for such a network by employing civil aircrafts leading to implement a low-cost airborne network is also discussed. In \cite{7073483}, submarine optical fiber cables based airborne Internet access architecture has been proposed to facilitate the use of high bandwidth to the elevated remote wireless nodes.

The emergence of several new services and applications has given rise to several new requirements and challenges for 5G communication networks. A few of such open challenges are highlighted in the following subsections.

\subsubsection{Network Capacity}

One of the key innovative techniques to meet the stringent requirements of 5G cellular communications is the densification of cellular networks by using the concept of small cells, having radii on the order of 100 to 10 meters, with the resulting network termed as ultra-dense networks (UDNs) \cite{6525591,6824752}. Such densification of cellular networks allows more aggressive frequency re-use, which helps in increasing the network capacity, improving spectral and energy efficiency, enhancing data rates, and providing seamless coverage with improved user experience. While on the other hand, the hyper densification of networks imposes an increase in challenges regarding network interference, mobility management, and optimal resource allocation. With the emergence of IoT and mMTC; resulting in several new applications and services, the deployment of small- and pico-cells will eventually reach its practical limitations in terms of inter-cell interference management, deployment cost and geographical area constraints. The 6G target connectivity density is defined as $10^7$ devices/km$^2$ with area traffic capacity of 1 Gbps/m$^2$ \cite{You2020towards}. To meet the stringent network capacity requirements of 6G wireless networks, there is a need for the exploration of the new frequency spectrum, and the development of novel network architectures and technologies.

\subsubsection{Energy Efficiency}
Energy efficiency is of paramount importance due to its significant economical and ecological effects. In order to provide seamless coverage and connectivity in a particular geographical location, e.g., a metropolitan area, 5G utilizes the concept of deploying a large number of small cells. Although mMIMO achieves significant energy efficiency, however, increasing the number of BSs also increases the energy consumption in terms of computational power. Moreover, billions of battery powered devices/sensors world-wide will be connected to the network posing serious challenges for improving the energy efficient in communication networks. Consequently, coping the strict energy efficiency requirements remains a significant challenge in B5G era.
The anticipated target of 6G wireless networks is to provide connectivity density of $10^7$ devices/km$^2$ with volumetric energy-spectral efficiency gain of $1000\times$ in terms of bps/Hz/m$^3$/Joule \cite{You2020towards}.

\begin{figure*}[t]
  \centering
  \includegraphics[width=0.9\textwidth]{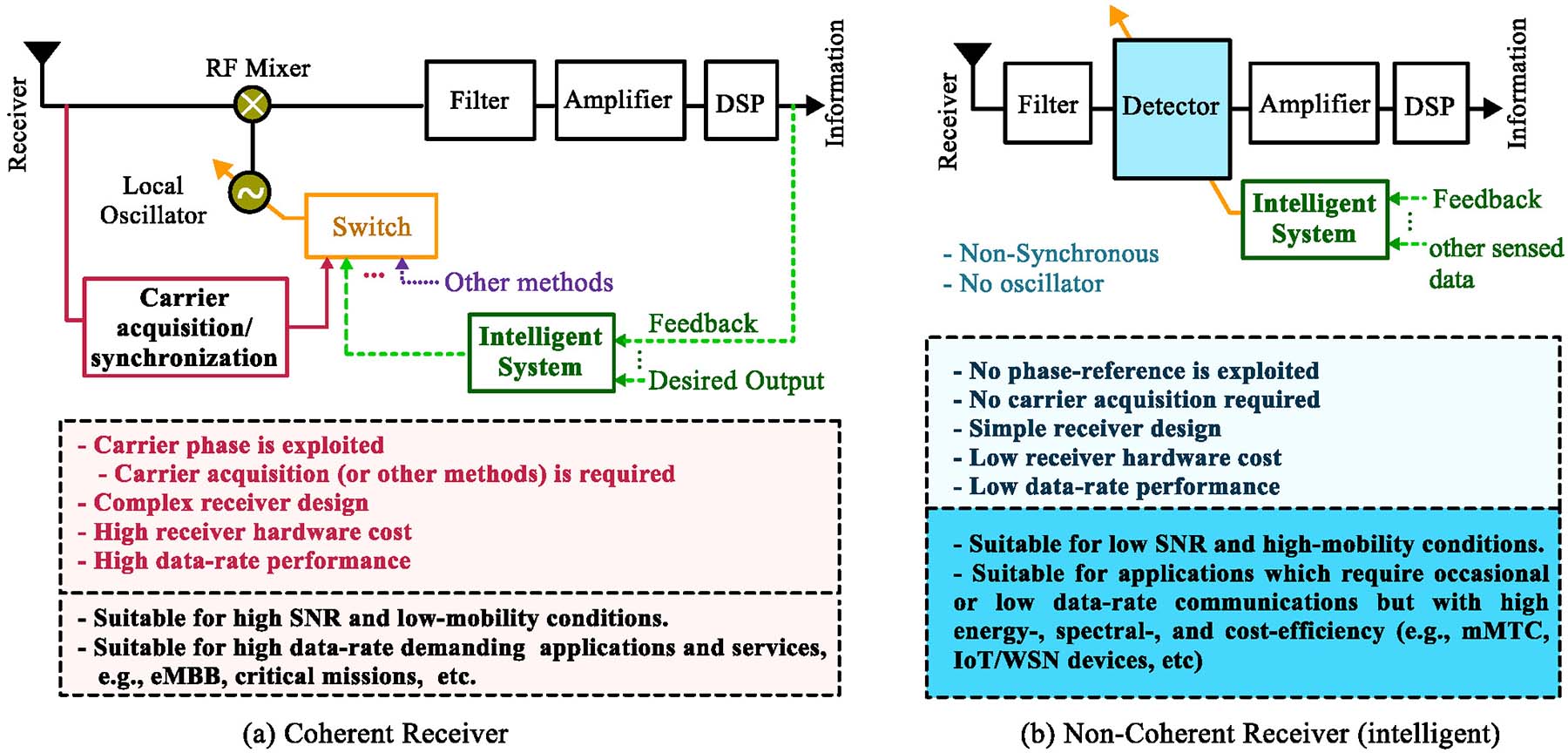}
  \caption{Coherent and non-coherent receiver design.}
  \label{Fig_coherent_vs_noncoherent}
\end{figure*}

\subsubsection{Network Security}
With the emergence of IoT and tactile Internet, billions of devices around the globe, generating a huge amount of data, will be connected to the communication networks.  Moreover, the 5G umbrella of services covers almost every walk of life- ranging from personal to organizational, health, industrial, and government levels. Consequently, the security risk factors of cyber attacks, intrusions, cyber warfare, and breach of data privacy will raise to new levels. Furthermore, the increased involvement of ML/AI and introduction of mobility at network-edge (e.g., UAVs-based BSs) has raised new data security and privacy concerns.
The provision of promising security solutions for communication networks is a key challenge in the B5G era which requires exploration of novel methods and new metrics (e.g., security capacity). To this end, the exploration of classical cryptography, physical layer, quantum, blockchain, and ML-based methods are the potential research directions.

\subsubsection{Network Throughput}
The advent of new applications and services such as augmented and virtual reality, ultra-high-definition television, self-driving cars, e-health, and tactile Internet, demands higher network throughput with lower end-to-end latency. Meeting these demands is a key research challenge in B5G wireless era, where exploration of new frequency spectrum in sub-teraHz, teraHz and visible light-range is the key research direction.

The target 6G goal for peak data rate, experienced data rate, and area traffic capacity is indicated as $>1$ Tbps, $1$ Gbps, and $1$ Gbps/m$^2$, respectively \cite{You2020towards}, for supportable mobility conditions of $>1000$ km/h. In 6G wireless networks, the evolution of URLLC services to enhances-URLLC (e-URLLC) is anticipated to promise end-to-end delay, radio-interface delay, and reliability requirements of $<1$ms, $10$ns, and $99.99999\%$ \cite{8766143,You2020towards,INGR_testbed}, respectively.
This anticipated target service of 6G networks holds the potential of revolutionizing many aspects of our lives, e.g., self-driving vehicles, processor-less consumer electronics devices, etc \cite{syed_sharma_patwary_asaduzzaman_2020}.

%
%
%
%
%
%
%
%
%


\section{Non-Coherent Communications} \label{Sec_nCoh_cFree}

This section reviews the significance, need, and recent trends of non-coherent wireless communication techniques.

\subsection{Significance and Fundamentals}
A brief comparative illustration of \emph{coherent} and \emph{non-coherent} receiver designs is provided in Fig. \ref{Fig_coherent_vs_noncoherent}, where their notable features are also highlighted. Non-coherent receiver design avoids the need for carrier recovery at the receiver. For example, for analog communications, instead of exploiting the frequency-mixing property of Fourier transform for performing the demodulation operation (shifting from passband to baseband) at the receiver, a simple envelope-detector circuit (involving a diode, a resistor, and a capacitor) can directly demodulate the information by sensing the envelope fluctuations in the received signal. As the carrier phase is not exploited in the non-coherent system designs, no carrier phase recovery stage is required at the receiver, i.e., the need for carrier acquisition or synchronization, which usually involves phase locked loop (PLL) and other steps, is avoided. For sampling, timing recovery is another critical step at the receiver that requires the alignment of receiver-clock with the transmitter-clock in order to avoid inter-symbol interference (ISI) and accurately recover the transmitted symbols.
\\ \\
Coherent receiver design offers high data rate performance or low detection error probability, however, it requires complex and expensive receiver circuit design. On the other hand, non-coherent methods offer simplicity in the receiver design, i.e., low hardware cost, however, it has an associated cost of reduced spectral efficiency or high detection error probability.
The sub-optimal non-coherent receiver designs can be considered suitable for low-power, low-data rate, and low-cost communication applications such as IoT, WSNs and mMTCs.
\\ \\
Differential encoding (and detection) is regarded as an important breakthrough and so differential non-coherent detectors are widely used in practice. This is because differential non-coherent receiver design offers low design complexity with the improved performance when compared to that offered by its coherent receiver counterpart. This differential encoding and non-coherent detection based system design helps in avoiding the influence of the carrier phase during symbol detection but it enhances the noise strength; therefore, the offered advantages are only limited to some specific low-order constellations. Furthermore, the employment of adaptive signal processing and learning methods holds the potential to enhance the performance of non-coherent detection methods.
\\ \\
The mobility of nodes in wireless networks usually imposes various diverse types of challenges (e.g., Doppler spread, handover, etc). Coherent receiver designs are usually considered suitable for low mobility conditions. This is because high mobility imposes a large Doppler spread (that causes time-variability in the channel) which leads to the requirement of rapid carrier phase tracking. Instead of striving to track the rapid carrier phase fluctuations in such high mobility conditions, non-coherent receivers can be used. To this end, non-coherent receiver designs are regarded as better in high-mobility (large Doppler spread) conditions. In \cite{8922634}, sixty years history of coherent and non-coherent communication systems is discussed in the context of 5G wireless communication applications.
\\ \\
Table \ref{tab: NoncoherentClassi} classifies the notable existing literature under the following themes along with the related sub-topics: (i) Characterization of non-coherent channels, (ii) Non-coherent modulation and detection, (iii) Non-coherent single-input multiple-output (SIMO)/MIMO systems, (iv) Non-coherent signal processing, and (v) Quantum-assisted non-coherent communications. These themes of non-coherent communication and associated subtopics are thoroughly reviewed in the following subsections.


\subsection{Modeling and Characterization of Non-Coherent Channels}


Accurate modeling and characterization of wireless channels is of prime importance in studying and designing of communication systems. Along with the degree of offered accuracy by a channel model, its mathematical simplicity is also important.
Fading, attenuation, interference, and noise are the prime phenomena that represent the behaviour of a wireless channel and affect the signals transmitted over the channel. The channel capacity metric indicates the reliably achievable maximum amount of information exchange through a certain channel.
Some existing works have analyzed capacity of non-coherent fading channels in different settings. For example, authors in \cite{Mezghani2008analysis} analyzed the capacity of limit of one-bit quantized non-coherent Rayleigh fading channel by assuming no knowledge of channel coefficients at both the transmitter and receiver in a single-antenna setting. It is observed that the channel capacity becomes zero if the channel coherence-time is in the order of symbol rate. Moreover, the on-off quadrature phase shift keying (QPSK) scheme can achieve the channel capacity for the slower channels with the on-off probability dependent on the channel coherence time.
Furthermore, authors in \cite{Vu2018optimalsignaling} investigated the optimal signalling schemes and capacity of non-coherent Rician fading channels by assuming no knowledge of fading coefficients at both the transmitter and receiver. For this analysis, low-resolution output quantization case with 1-bit analog-to-digital converter (ADC) was considered and the impact of line-of-sight (LoS) component of Rician fading channel on the channel capacity was analyzed in detail. It was shown that $\pi/2$ circularly symmetric is the optimal signalling input to achieve channel capacity in a Rician fading non-coherent channel.

It is of significant interest to understand the fundamental limits of non-coherent channels due to the consideration of non-coherent communication design in the upcoming wireless networks. Following the analysis of the capacity analysis of multi-antenna communications with Rayleigh fading channels in  \cite{Marzetta1999capacity}, authors in \cite{Zheng2000info} studied the fundamental information theoretic limits of non-coherent channels in multi-antenna settings without considering the knowledge of fading coefficients. The asymptotic capacity of such non-coherent multi-antenna channels was derived in the high signal-to-noise ratio (SNR) regime in terms of the number of transmit ($M$) and receiver ($N$) antennas and coherence interval ($T$). As compared the capacity gain of coherent multi-antenna channel, which is given by $\mathrm{min}\{M,N\}$ for every 3dB increase in the SNR, the capacity gain of non-coherent channel is given by; $M'(1-M'/T)$ bps/Hz, with $M'=\min\{M,N,LT/2\}$ \cite{Zheng2000info}.

Authors in \cite{Barletta2018dof} analyzed the effects of oversampling on the capacity of oversampled non-coherent communication channels impacted by the phase noise for the scenario with the coherence time of the phase process being much smaller than the sampling time interval. Furthermore, authors developed the Generalized Degrees of Freedom (GDoF) for such an oversampled non-coherent channel by considering the case with the oversampling factor growing to infinity with the transmit power. Furthermore, the work in \cite{Pikus2016discrete} studied the application of discrete signalling schemes, which have better power efficiency that Gaussian i.i.d. signals in mid-to-low SNR region, for non-coherent single-input single-output Rayleigh block-fading channels. The mutual information of the proposed discrete signalling schemes has been computed and compared with the existing signalling schemes in the literature.

In the wideband channels including the case of mmWave channels which have very high bandwidth, the capacity regime is highly impacted by the channel uncertainty in the absence of CSI at the transmitter or receiver. Furthermore, it becomes challenging to implement coherent detection since it becomes difficult to estimate channel coefficients with the required sufficient precision due to the constrained power in the wideband channel \cite{Cuba2015bandwidth}.  Therefore, it is important to investigate suitable non-coherent detection for wideband fading channels. Moreover, energy is limiting resource than spectrum in non-coherent wideband fading channels. Towards the analysis of non-coherent wideband channels, authors in \cite{Cuba2015bandwidth,Cuba2017unified} considered the bandwidth occupancy (average bandwidth utilization over time) as a performance metric and performed the information theoretic analysis of wideband fading channels by considering both the peaky and non-peaky signals. It was shown that  the achievable rates for all types of signalling techniques with the identical bandwidth occupancy  approach to the wideband additive white Gaussian noise (AWGN) capacity within the same gap at the critical value of bandwidth occupancy and subsequently decreases to zero when the bandwidth occupancy tends to infinity.


The conventional coherent detection techniques are designed based on the CSI knowledge to mitigate the harmful effects of fading in communication channels. In multi-antenna relaying systems, it may be challenging to estimate all the involved channels due to power-hungry and complex channel estimation techniques. For example, a MIMO system having $M$ transmit antennas and $N$ receive antennas requires to estimate the $M \times N$ channels. Also, in relaying systems, it may not be possible for the resource-constrained relay nodes to allocate resources for estimating source-to-relay channel required for coherent detection. To address these issues, non-coherent detection seems highly advantageous. In this regard, some existing works have studied the design and performance analysis of relay channels in non-coherent scenarios. For example, authors in  \cite{Li2015relaying} provided a survey of non-coherent successive relaying methods while considering multi-user wireless systems. The discussed successive relaying techniques can perform the function of full-duplex relaying towards compensating the $50$\% throughput loss by a half-duplex transmit/receive of the widely-used transceivers.

In air traffic management, one of the current issues is that the widely-used wireless communication technologies are not suitable for aeronautical scenarios with high Doppler. As an example, the existing LTE-A based systems are able to supper users at a high speed velocity of 500 km/h (train) but the speed of aircraft may reach upto 1080 km/h. The Rician fading channels in high-mobility aeronautical scenarios are characterized by the high frequency offset and normalized maximum Doppler frequency and the uncertainty of phase rotation of a random channel \cite{8501993}.

In the literature, mostly training-based (pilot-based) channel estimation methods have been studied for the coherent channels while the channel estimation techniques for non-coherent channels are based on the modifications of conventional differential detection \cite{8501993}. Both of these approaches suffer from the irreducible error flow while assuming constant CSI. Also, current differential phase shift keying (DPSK) based non-coherent detectors used for Ricean fading channels demand for significant enhancements. To this end, authors in \cite{8501993} proposed adaptive switching between non-coherent and coherent schemes at high and low Doppler frequency, respectively for aeronautical communications networks. Also, the comparison of various non-coherent techniques to their coherent counterparts is presented in a wide-range of channel-coded scenarios.

\begin{table*}[t]
\caption{\small{Classification of existing works on non-coherent communications}}
\centering
\begin{tabular}{|l|l|l|}
\hline
\textbf{Main Theme} & \textbf{Sub-topics} & \textbf{References}  \\
\hline
Characterization of non-coherent channels & Capacity analysis and fundamental limits & \cite{Mezghani2008analysis,Vu2018optimalsignaling,Zheng2000info} \\
                                          & Signalling schemes & \cite{Vu2018optimalsignaling,Pikus2016discrete} \\
                                          & Effects of oversampling & \cite{Barletta2018dof} \\
                                          & Wideband non-coherent channels & \cite{Cuba2015bandwidth,Cuba2017unified} \\
                                          & Non-coherent successive relaying & \cite{Li2015relaying} \\
                                          & High mobility aeronautical communications & \cite{8501993} \\ \hline
Non-coherent modulation and detection & Non-coherent spatial modulation (SM) & \cite{Xuadaptive2019,xu2019adaptive}  \\
                                      & Non-coherent index modulation (IM) & \cite{Choi2018noncoherent,Hanif2019noncoherent} \\
                                      & Non-coherent detection & \cite{Jamali2018detection,Safdari2016IWCMC} \\ \hline
Non-coherent SIMO/MIMO systems & Non-coherent SIMO & \cite{Xie2019SIMO,Ngo2018multipleaccess} \\
                                      & Non-coherent MISO & \cite{Vu2019correlatedMISO} \\
                                      & Non-coherent MIMO & \cite{roger2014non} \\
                                      & Temporally-Correlated non-coherent MIMO Channels & \cite{Cabrejas2016openloop} \\
                                      & Space-time codebook design & \cite{4359531} \\ \hline
Non-coherent signal processing & Non-coherent SWIPT systems  & \cite{Chowdhury2016scaling,Jing2016design} \\
& Non-coherent AoA estimation and beamforming  & \cite{8453887,Rieken2004} \\
& Jamming attack detection &  \cite{Xuattack2019}  \\ \hline
Quantum-assisted non-coherent communications & Quantum search algorithms for non-coherent receivers/detectors & \cite{8010959,Botsinis2015noncoherent}   \\
\hline
\end{tabular}
\vspace{-15 pt}
\label{tab: NoncoherentClassi}
\end{table*}

\subsection{Non-Coherent Modulation and Detection}
As compared to the conventional digital modulation schemes which are based on the modulation of frequency/phase/amplitude of a carrier signal, there exist other modulation schemes such as spatial modulation (SM), permutation modulation (PeM), and index modulation (IM) \cite{Ishikawa2018permutation}. SM is considered as one promising candidate technology for future wireless networks since it can exploit the spatial dimension (i.e., antennas) as an additional degree of freedom and can efficiently operate in a variety of MIMO configurations \cite{Yang2015spatial}. In the SM scheme, information bits are transmitted by using the combination of two schemes, namely, amplitude-phase modulation (APM) and a space-shift keying (SSK), which correspond to the modulation and space domains, respectively.  To decode these transmitted bits, the receiver needs to have the CSI knowledge, which is difficult to acquire in practice with the desired accuracy. To address this, one possible solution could be the combination of SM with the differential modulation (named as differential SM) \cite{Bian2015differential} to address the complexity of symbol detection and channel estimation overhead in the conventional coherent SM. To further enhance the performance of SM, multi-ring amplitude phase shifting keying (APSK)-assisted differential SM has been proposed in the literature \cite{Liu2017highrate}. However, this scheme has the drawback of limited flexibility for high-dimensional constellations with the scattered constellation points in the outer circle and intensive points on the inner circle.
To address this issue, authors in \cite{Xuadaptive2019} recently proposed the concept of non-coherent SM which employs a non-coherent detection at the receiver. Unlike the differential operation in the differential SM, the  detection process in this approach relies only on the previous demodulated transmission block.

Furthermore, in the context of SM-aided unmanned aircraft systems, authors in \cite{xu2019adaptive} suggested to switch between coherent and non-coherent SM to enhance the QoS performance. Towards this, the authors proposed a three-fold adaptive approach to perform the following adaptations: (i) to switch between coherent and non-coherent techniques based on the Doppler frequency, (ii) to reconfigure the system in its own between single transmit antenna and multiple transmit antenna systems based on channel coding, and (iii) to switch between high diversity and high spectral efficiency multiple transmit antenna techniques based on the modulation throughput.

PeM concept was introduced by Slepian in 1965. This concept offers various desirable features and its use-cases span from data storage (steganography, holographic, flash memories, solid-state drive memories) to information modulation (wired and wireless communications) applications. This concept aims at compressing the input bits by selecting a permutation of a set of sequences, where the sequences are composed of  a finite number of states (e.g., two states, i.e., `on' and `off'). By reducing the number of occurrences of a certain state in a physical material, the storage capacity can be enhanced while maintaining the latency and complexity considerations for read and write operations. By representing the signals in 3-D space representing spatial, temporal, and frequency parameters, the hybrid PeM concept can be deduced to the SM concept. Different PeM coding schemes offer different performance attributes when compared to other known digital modulation schemes. In \cite{Ishikawa2018permutation}, a comprehensive review of PeM along with SM and IM schemes is provided.

Another promising modulation technique, IM, utilizes a binary permutation vector and maps the information bits by activating different parameters of the transmission medium such as sub-carriers, transmit antennas, relays, time slots, precoding matrices, signal powers and spreading codes \cite{Basarindexmodualtion}. The conventional IM technique can be utilized to carry additional information by incorporating information bits into the activated indexes, however, the demodulation of this information requires the CSI knowledge at the receiver side \cite{Hanif2019noncoherent}. Also, the receiver needs to estimate channel coefficients for all the indexes since the subset of active indexes may vary according to the message. For example, in the conventional orthogonal frequency division multiplexing (OFDM)-IM, the overhead caused by the pilot transmissions via all subcarriers may become very high, especially in fast fading environment,  since only a fraction of subcarriers are active \cite{Choi2018noncoherent}.  In conclusion, the conventional coherent-IM may lead to a very high channel estimation burden and also high energy consumption, thus making it not suitable for resource-constrained wireless sensors/devices.

To address the aforementioned issues of the conventional IM, the concept of non-coherent IM is emerging \cite{Choi2018noncoherent,Hanif2019noncoherent}. In this approach, the information bits are transmitted with only index selection and without the need of channel estimation. In this direction, optimal maximum-likelihood (MLh) detector for non-coherent IM has been studied in different settings \cite{Choi2018noncoherent,Hanif2019noncoherent}.  The authors in \cite{Choi2018noncoherent} carried out the analysis of optimal MLh receivers for non-coherent IM under fast fading environment. However, the assumption in this study was that signal belonging to each index experiences independent fading and the optimally seems to cease in the case when signals on individual channels experience the identical fading. This study has been extended to the case of complex-valued channels in \cite{Hanif2019noncoherent}, with the consideration of the signals on different indexes suffering from the same amount of fading, which could be the  case in narrow-band wireless sensor networks.

In the context of ultra-wideband Direct Chaotic Communication (DCC), the authors in \cite{Mesloub2017chip} proposed a non-coherent modulation scheme, which is based on the adaptive threshold value. As compared to the conventional modulation techniques where the bit energy per given symbol is fixed, the bit energy of each symbol is time-varying due to the non-periodic chaotic signals. To address this issue,  differential chaos shift keying (DCSK) modulation and several its variants have been proposed in the literature \cite{Fang2016DCSK}. In the similar context, the non-coherent scheme in \cite{Mesloub2017chip} utilizes  chip averaging chaotic on-off keying (CA-COOK), which uses the concept of the Cell Averaging Constant False Alarm Rate paradigm utilized in radar systems.

Authors in \cite{Safdari2016IWCMC} studied the combination of differential binary phase shift keying (DBPSK) modulation and non-coherent detection at the relay and the destination while considering auto-regressive time series model for the time-varying Rayleigh fading channel. The employed non-coherent detection doesn't need the knowledge of instantaneous CSI at both the relay and destination to obtain average BER. Three-node relaying with decode-and-forward (DF) strategy was employed with selection combining (SC) at the destination in the following phases: (i) First phase: the source transmits the DBPSK signal, and relay and destination operate in the listening mode, and (ii) Second phase: the relay decodes the received signal in a non-coherent manner and re-transmits to the destination. Via numerical results, it was shown that the performance of SC with the considered non-coherent DF relaying framework is closer to the semi-maximum ratio combining (MRC), which doesn't need the second-order statistics of the channels.

In recently emerging molecular communications \cite{Farsad2016molecular}, the CSI may vary with time due to the variations in the flow velocity as well as the distance between the receiver and transmitter. This requires the need of tracking the CSI variations repeatedly for the CSI acquisition \cite{Jamali2018detection}. The traditional detection techniques, which requires the knowledge of CSI, can be employed only in the scenarios where the molecular channel is slowly varying with the time and the coherence time is sufficiently large. In dynamic molecular scenarios, the CSI estimation process requires a huge overhead and detecting data-symbols directly without going through the CSI acquisition process seems attractive, leading to the utilization of non-coherent detection. To this end, authors in \cite{Jamali2018detection} studied the optimal non-coherent Multiple Symbol (MS) and Symbol-by-Symbol (SS) detectors, as well as a non-coherent decision feedback detector, which do not need the knowledge of instantaneous CSI. An analytical expression for the BER of the optimal non-coherent SS detector as well as the lower and the upper bounds for the BER of the proposed optimal non-coherent MS detector were derived. Via simulation results, the performance of the optimal MS detector was shown to outperform the sub-optimal blind detector, mainly for the case of small detection window sizes, and the performance of both the optimal and sub-optimal detectors was found to converge to that of the coherent MLh detector, which needs the knowledge of perfect CSI.

\subsection{Non-Coherent SIMO/MIMO}
As highlighted earlier, the acquisition of accurate CSI in a timely manner with large-antenna arrays in mMIMO systems is a challenging task. Also, the estimated channel coefficients may be corrupted due to the aggressive spatial reuse of pilots (i.e., pilot contamination). To address these issues, various approaches including blind or semi-blind, precoding and protocol-based techniques have been suggested in the literature \cite{Elijah2016massiveMIMO}. Furthermore, non-coherent communication systems based on energy detection can be utilized as an alternative as it does not require the knowledge of CSI. Besides, as compared to the coherent receivers, non-coherent receivers provide the advantages of low power consumption, low complexity and simple structures at the cost of sub-optimal performance \cite{Xie2019SIMO}. By considering the non-coherent reception mode in single-input multiple-output systems, authors in \cite{Xie2019SIMO} analyzed the effect of correlated Rayleigh fading on the system performance and also derived the theoretical expressions for achievable rate, symbol error rate and outage probability.
Authors in \cite{Vu2019correlatedMISO} carried out a comprehensive theoretical analysis of optimal signaling schemes, and also studied the capacity of non-coherent multiple-input single-output (MISO) Rayleigh fading channels under various constraints including the per-antenna power constraints and the joint per-antenna and sum-power constraints. The capacity gain of MISO under both the scenarios with per-antenna power constraints, and joint sum-power and per-antenna constraints is compared with that of the single-input single-output (SISO) systems by utilizing the finiteness and discreteness of the optimal inputs obtained from the optimization problems.

As compared to the conventional training-based methods, non-coherent signaling can provide several advantages in wireless systems including the following \cite{8602450}: (i) non-coherent signalling can avoid the cost of training, which consumes about 14.3\% of resources in multi-antenna systems \cite{3GP_selfevaluation}; (ii) non-coherent techniques can provide higher spectral efficiencies, and (iii)  pilot contamination issue in mMIMO system can be largely mitigated.

The conventional coherent receivers need the perfect knowledge of the instantaneous channel, and may result in significant signalling overhead while employing coordinated multipoint (CoMP) transmission or reception in MIMO systems in the presence of higher number of transmission points or in the presence of rapidly varying wireless channels \cite{roger2014non}. The widely used pilot-based channel estimation techniques demand for high receiver complexity and signaling overhead in the aforementioned CoMP scenarios. Also, in fast fading scenarios, coherent systems become almost impractical to implement due to very short coherence time. Moreover, channel estimation errors in coherent communications may cause significant degradation of the system performance, leading to the need of non-coherent communications, in which the receivers can perform data detection without requiring the knowledge of channel coefficients.

In the above context, authors in \cite{roger2014non} have reviewed non-coherent techniques for MIMO communication systems for both the slowly-varying and block-fading channels. In slow-fading scenarios of non-coherent MIMO systems, differential transmission-based techniques such as differential unitary space-time modulation seems promising \cite{Hochwald2000}, which is high dimensional version of the standard DPSK modulation. On the other hand, for non-coherent MIMO systems with block fading channels, codebooks composed of unitary matrices which are distributed isotropically on the Grassmann manifold seem promising \cite{Gohary209noncoherent}.  Despite some existing research in this domain, the performance of non-coherent schemes under realistic cellular scenarios requires significant further research. In this regard,  \cite{roger2014non} has presented some preliminary results regarding the performance of non-systematic Grassmannian constellations (GC) in two practical cases including a spatial correlation in a MIMO system and unbalanced transmission points in a CoMP system.

As highlighted earlier, in fast fading coherent scenarios, the channel coefficient changes so rapidly that a new and independent channel coefficient may appear before it is being estimated by the receiver by using training signals. Also, the presence of multiple antennas gives rise to the problem of estimating higher number of parameters, thus leading to the need of non-coherent detection techniques. To this end, authors in \cite{4359531} studied the problem of space-time codebook design for non-coherent multi-antenna systems, in which the channel matrix was modeled as an unknown deterministic parameter at both the transmitter and receiver and a generalized likelihood ratio test principle is employed for the operation of a non-coherent receiver. By considering the detection error probability as the code design criterion at the high-SNR region, a space-time code design methodology was suggested in the considered non-coherent set-up. Subsequently, a two-phase method was proposed to solve a non-linear, non-smooth and high-dimensional underlying optimization problem, with first phase involving a sequence of convex semidefinite programming (SDP) and the second phase using a descent optimization algorithm.

The article \cite{Cabrejas2016openloop} studied three non-coherent MIMO schemes including Grassmannian signaling, differential space-time block code and differential unitary space-time modulation by considering temporally-correlated Rayleigh fading channels. The performance of these techniques has been compared with two Alamouti and Golden code-based coherent schemes via numerical results, and it has been shown that these non-coherent methods provide significant advantages over coherent methods in medium to high mobility scenarios, especially, when the number of transmit antennas increases.


Regarding multiple access schemes for non-coherent systems, authors in \cite{Ngo2018multipleaccess} proposed a Grassmannian constellation based non-coherent multiple access by considering uplink SIMO communications. In this approach, an individual codebook was constructed for each user by using bijective mapping for the encoder, and a greedy approach was suggested at the receiver by exploiting a geometric separation among the code-books of the users to extract the signals from different users.
In \cite{9049073,9115900}, an efficient message passing algorithm for non-coherent multi-user (soft and hard) detection scheme based on expectation propagation is proposed.
In \cite{ngo2020joint}, a joint constellation design to minimize the detection error for two-user MAC is investigated.

\subsection{Non-Coherent Signal Processing}
Although non-coherent systems can provide the advantages of low cost and low power consumption, in general, they suffer from performance loss as compared to the coherent systems. However, the performance of power detection based non-coherent modulation is shown to be similar as that of the coherent modulation scheme in different settings \cite{Chowdhury2016scaling,Jing2016design}. Furthermore, the power detection-based non-coherent receivers find importance in simultaneous wireless information and power transfer (SWIPT) systems since the SWIPT receiver may employ a coherent and an energy receiver either in a power-splitting or a time-switching mode \cite{Perera2019challenges}. To this end, authors in  \cite{Jointcoherent2017} proposed a splitting receiver by utilizing the joint coherent and non-coherent processing, which can provide better performance gain than those of the separate non-coherent and coherent receivers in terms of symbol error rate (SER) and mutual information.

The angle-of-arrival (AoA) estimation is needed in several applications including beamforming in wireless communications, radar and sonar, and there exist several AoA estimation algorithms in the literature including multiple signal classification (MUSIC), root-MUSIC and estimation of signal parameters via rotational invariance techniques (ESPRIT). Most of these algorithms perform coherent processing and require the knowledge of covariance matrix of the entire array along with the inter-subarray covariance matrices. The main drawback with this coherent processing is the huge communication overhead involved in sending the raw measurements from the sub-arrays to the central processing unit, and high computational load at the central processing unit \cite{8453887}. To address these drawbacks, non-coherent processing can be employed, which is dependent only on the sub-array covariance matrices unlike the covariance matrix corresponding to whole array aperture required in coherent processing \cite{Wax1985trans}. To this end, authors in \cite{Rieken2004} generalized the MUSIC and minimum variance distortionless response beamformers for non-coherent arrays, in which sub-arrays estimate noise sub-spaces locally and forward the estimates to the processing center. Furthermore, the authors in \cite{Suleiman2014ICASSP} generalized the root-MUSIC algorithm for non-coherent processing in which the sub-arrays calculate the polynomial coefficients of root-MUSIC locally and then forward these coefficients to the processing center. Moreover, authors in \cite{8453887} considered the AoA estimation problem for partly calibrated arrays by utilizing non-coherent processing, by considering the case that the sub-arrays are not able to locally identify the sources.


Like coherent systems, non-coherent systems are also susceptible to jamming attacks. It is critical to investigate and develop suitable jamming attack detection and prevention schemes for the non-coherent systems. In this regard, authors in \cite{Xuattack2019} proposed a jamming detection method based on the normalization of variance and standard variance by utilizing the the principle of likelihood ratio test. Via numerical results, it has been shown that the detection probability increases with the increase in the number of receive antennas and gets converged quickly, however, the performance is affected by the channel statistics from the jammer to the receiver.


\subsection{Quantum-Assisted Non-Coherent Communications}

The integration of quantum computing and wireless communications can be a promising platform for enabling future low-complexity and near-capacity wireless systems.  It should be noted that on one hand, Shannon's channel capacity did not consider aspect the complexity needed to approach the capacity of a wireless channel, while on the other hand, the information to be transmitted over a wireless channel can be encoded with the polarization or the spins of electrons besides only their charge and number. The classical communication principles can be adapted for quantum-assisted communications \cite{Imre2013quantum,QML_6G_Junaid} with the help of quantum computing algorithms and processes enabled by quantum chips.

In multi-user detection (MUD) problem in a  synchronous NOMA system, one of the issues for multi-level detector is to identify a specific symbol based on the received signal power, the set of legitimate symbols with (in case of coherent) or without the knowledge of channel states (in case of non-coherent) \cite{Botsinis2015noncoherent}. Since the channel state information is not available to the non-coherent receiver at the base station, two users need to be separated in code domain or frequency domain or in time domain during different time-slots. Therefore, the multi-user detector can be represented by simple correlation filters while matching the spreading code of each user. In contrast to the coherent receivers, the complexity introduced by the channel estimation process is avoided in non-coherent receivers but they may need higher bandwidth or experience performance degradation \cite{8010959}.

From the practical implementation perspective, the complexity of MUD becomes the main issue. For example, for supporting $P$ number if users with the same code and frequency resources, the optimal maximum a posteriori probability (MAP) with M-ary modulation in a coherent receiver requires $M^P$ number of cost function evaluations \cite{8010959}. Similarly, the non-coherent receiver with MAP and the MS differential detector (MSDD) needs to search in $N_w$ symbol windows in the time domain independently for each user and the MSDD for an M-ary modulation technique requires $M^{N_w-1}$ cost function evaluations to search among all possible multi-level symbol combinations of each user. To reduce such  symbol detection complexity, the quantum computing assisted MUD can be utilized to reduce the complexity of symbol detection processing while also achieving the near-optimal performance. To this end, authors in \cite{8010959} investigated various quantum search algorithms and proposed a methodology to select a suitable quantum search algorithm based on the nature of the search problem and operation.

In emerging large-dimensional wireless systems including mmWave systems, mMIMO systems and cooperative multicell processing/coordinated beamforming, it becomes highly challenging and complex to acquire the channel gains of all the links to carry out coherent detection. To address this, non-coherent detection based differential modulation could be an alternative to avoid the need of channel estimation. For the non-coherent receivers, the classical MLh MSDD \cite{Botsinis2014} is considered as optimal but suffers from high complexity. Towards reducing the detection complexity in non-coherent communications, other detectors including MS differential sphere detector (MSDSD) \cite{Xu2011DDST} and Decision-Feedback Differential Detector (DFDD) \cite{Schober2000decision} have been proposed in the literature, which can provide near-optimal performance with regard to the MSD with the reduced complexity.

As highlighted earlier, quantum computing can be utilized to reduce the number of cost function evaluations needed for MSDDs. In this regard, authors in \cite{Botsinis2015noncoherent} have discussed various Quantum Search Algorithms (QSAs) including Grover's QSA, Durr-Hoyer algorithm, and Boyer, Beassard, Hoyer and Tapp. Furthermore, Durr-Hoyer algorithm-based Quantum-assisted symbol differential detection was proposed for reducing the complexity of conventional differential detector and its performance was analyzed via numerical results.

\section{Backscatter Communications for B5G/6G} 

In the provision of uninterrupted power supply and ultra-reliable connectivity to the massive number of nodes envisaged in 6G and beyond wireless networks, the designing of ultra-power-efficient wireless links, harvesting of energy from various diverse sources, designing of reconfigurable intelligent reflective surfaces are among the prime research trends.
Today's battery-powered wireless devices are vulnerable to various performance constraints imposed by limited lifetime of the nodes due to limited batteries-life. The wireless transfer of power to recharge the batteries of remote devices is also an emerging paradigm. The wireless powered communication networks are comprehensively reviewed in \cite{7984754}, where significant developments and open research challenges are highlighted. Various interesting technologies for enabling self-sustainable communications in future networks are recommended with strong focus on BsC systems.
BsC concept has a strong potential in enhancing the uplink (i.e., devices to reader) energy- and spectral efficiency of communication nodes, which ultimately vastly influences the overall energy and spectrum requirements of ultra-massively connected wireless networks.
The proliferation in the number of devices envisioned in the emerging paradigm of massive-IoT networks necessitates the conduction of thorough studies for designing energy-efficient wireless network with wireless-powered battery-assisted or battery-less devices.

In the emerging paradigm of green communications with the requirements of ultra-dense connectivity in 6G and beyond wireless networks, BsC can play a vital role in delivering the emerging services, such as massive-IoT, further enhanced broadband, and various other new types of services. Transceiver design with ultra-high power efficiency for future IoT devices is a basic necessity in this context, where the harvesting of energy from radio frequency (RF) sources is one of the robust potential solutions. This section starts with discussing the fundamentals and design aspects of BsC and then proceeds with reviewing its state-of-the-art. Moreover, the section also highlights various interesting future communication trends, e.g., quantum BsC, mmWave and teraHz BsC, energy-efficient intelligent Surfaces -- to name a few.

\begin{table*}[t]
  \centering
  \caption{Characterization of the literature review on BsC Systems.}
    \begin{tabular}{|l|c|c|}
    \hline
    \textbf{Main Theme} &
      \textbf{Sub-topics} &
      \textbf{References}
      \bigstrut\\
    \hline
    \textbf{Fundamental Aspects} &
      \multicolumn{1}{p{25em}|}{System configuration (Ambient, MonoStatic, Bistatic)} &
      \cite{ruttik2018does,7948789,memon2019backscatter,8253544}
      \bigstrut\\
\cline{2-3}    \multicolumn{1}{|c|}{} &
      \multicolumn{1}{p{25em}|}{Cooperative (Symbiotic Radio)} &
      \cite{long2018symbiotic,8746225,8692391,8274950,8665892,8638762,8636518}
      \bigstrut\\
\cline{2-3}    \multicolumn{1}{|c|}{} &
      \multicolumn{1}{p{25em}|}{Wireless Channel (modeling, characterization, and Estimation)} &
      \cite{8665892,8171133,8345348,jameel2019simultaneous,8320359,8641171,zhao2019channel,duan2018multi,jameel2019simultaneous,8171133}
      \bigstrut\\
\cline{2-3}    \multicolumn{1}{|c|}{} &
      \multicolumn{1}{p{25em}|}{Multiplexing, Multiple Access, and Duplexing} &
      \cite{li2019secure,8647245,8103807,8299928,8422524,8399824,8439079,8491248,8636518,7913737,8700623,7913737,8700623,8636518,8638762,8642363}
      \bigstrut\\
\cline{2-3}    \multicolumn{1}{|c|}{} &
      \multicolumn{1}{p{25em}|}{Modulation, Transmission, and Detection} &
      \cite{harms2017modulation,8644139,7551180,8474331,nguyen2019signal,8445952,8423609,8672080,han2019joint,8534458,8672080,8534458}
      \bigstrut\\
    \hline
    \textbf{Scope with other technologies} &
      \multicolumn{1}{p{25em}|}{Quantum Technologies} &
      \cite{8269081,lanzagorta2016improving,inomata2016single,di2018quantum,8970171}
      \bigstrut\\
\cline{2-3}    \multicolumn{1}{|c|}{} &
      \multicolumn{1}{p{20em}|}{Multi-Antenna Systems} &
      \cite{7274644,shen2019ambient,8254467,wu2019beamforming,8446004,8302460,8468064,wyner1966bounds,7948789,7997479,6685977,li2019secure}
      \bigstrut\\
\cline{2-3}    \multicolumn{1}{|c|}{} &
      \multicolumn{1}{p{25em}|}{Non-Coherent} &
      \cite{7769255,8399824,8642363,8474331,8269048}
      \bigstrut\\
\cline{2-3}    \multicolumn{1}{|c|}{} &
      \multicolumn{1}{p{25em}|}{UAV-Assisted } &
      \cite{8762096,memon2019backscatter}
      \bigstrut\\
\cline{2-3}    \multicolumn{1}{|c|}{} &
      \multicolumn{1}{p{20em}|}{Machine Learning} &
      \cite{9024401,jameel2020machine}
      \bigstrut\\
    \hline
    \multirow{3}[6]{*}{\textbf{Spectrum}} &
      \multicolumn{1}{p{25em}|}{mmWave, Sub-teraHz, teraHz} &
      \cite{digiovanni2014surface,mou2014backscattering,8594703,4631469,6697382,8319409,8058702,4405373,mou2014backscattering,mizojiri2018wireless,mizojiri2018wireless}
      \bigstrut\\
\cline{2-3}    \multicolumn{1}{|c|}{} &
      \multicolumn{1}{p{25em}|}{VLC} &
      \cite{7752834,xu2017passivevlc}
      \bigstrut\\
\cline{2-3}    \multicolumn{1}{|c|}{} &
      \multicolumn{1}{p{25em}|}{Acoustic (underwater)} &
      \cite{jang2019underwater}
      \bigstrut\\
    \hline
    \textbf{Applications and other aspects} &
      \multicolumn{1}{p{20em}|}{Massive IoT} &
      \cite{8692391,memon2019ambient,kim2018novel,8115787,6613706,liu2019next,jameel2020machine,long2018symbiotic}
      \bigstrut\\
\cline{2-3}    \multicolumn{1}{|c|}{} &
      \multicolumn{1}{p{25em}|}{Biomedical/ molecular communications/ eHealth} &
      \cite{grosinger2012evaluating,grosinger2013feasibility,thomas2012modulated,zhang2016enabling,ciuonzodecision,8746225,Li2016localconvexity,Jamali2018detection,vasisht2018body}
      \bigstrut\\
\cline{2-3}    \multicolumn{1}{|c|}{} &
      \multicolumn{1}{p{25em}|}{Decision Fusion} &
      \cite{ciuonzodecision}
      \bigstrut\\
\cline{2-3}    \multicolumn{1}{|c|}{} &
      \multicolumn{1}{p{25em}|}{Cognitive Networks} &
      \cite{8738811,8700623}
      \bigstrut\\
\cline{2-3}    \multicolumn{1}{|c|}{} &
      \multicolumn{1}{p{25em}|}{Localization and Tracking} &
      \cite{kotaru2017localizing,niu2012target}
      \bigstrut\\
\cline{2-3}    \multicolumn{1}{|c|}{} &
      \multicolumn{1}{p{25em}|}{Security and Privacy} &
      \cite{8970171,li2019secure}
      \bigstrut\\
\cline{2-3}    \multicolumn{1}{|c|}{} &
      \multicolumn{1}{p{25em}|}{Blockchain} &
      \cite{8731639}
      \bigstrut\\
\cline{2-3}    \multicolumn{1}{|c|}{} &
      \multicolumn{1}{p{25em}|}{Routing in B2B networks} &
      \cite{8253954,8737551,8364537,8170764}
      \bigstrut\\
\cline{2-3}    \multicolumn{1}{|c|}{} &
      \multicolumn{1}{p{25em}|}{Cost (Hardware, Deployment, and Operational )} &
      \cite{liu2019next}
      \bigstrut\\
    \hline
    \multicolumn{2}{|p{26.215em}|}{\textbf{Survey Articles and Book Chapters}} &
      \cite{8368232,memon2019backscatter,liu2019next,memon2019ambient,jameel2020machine,8454398}
      \bigstrut\\
    \hline
    \end{tabular}%
  \label{Literature_Summary_BsC}%
\end{table*}%

\subsection{Fundamentals and State-of-the-Art}

The RF signals received at a device can be potentially manipulated as an energy source.
The fundamental concept of BsC is to utilize the incident RF signals (e.g., modulated or unmodulated), instead of utilizing a locally generating new carrier, by reflecting it back after modulating (new) information over it in an over-the-air modulation fashion. This concept of BsC offers the advantage of reliving the burden of system complexity and power cost from the remote devices (e.g., sensing devices) by transferring them to the serving stations (e.g., base station).
The roots of this concept trace back to 1948 when Harry Stockman demonstrated the concept of point-to-point communications by reflecting the received RF signals after modulating information over it \cite{1697527}. RF identification (RFID) is a typical example of BsC, in which, a reader device generates an unmodulated carrier, which is utilized by the passive tags to transmit it back with the information modulated over it. Another notable example of the systems obeying the fundamental principles of BsC is passive radars (a special case of bistatic radars). Such radars detect and track the target objects by processing and analyzing the radio-scene generated through the radio illumination of the environment from existing non-cooperative radiation sources (e.g., commercial radio broadcast signals, land-mobile radio communication signals, etc).

In \cite{memon2019backscatter}, BsC is presented as a practical solution to the limited battery life problem for emerging heterogeneous networks (cellular, IoT, etc). A thorough survey on different types, modes, modulations, and architectures of BsC systems is presented. Moreover, various important aspects of BsC are discussed, e.g., reliability, security, range, etc.
In \cite{liu2019next}, a thorough review of BsC as a potential solution to the problem of limited lifetime of massive-IoT networks is conducted. Moreover, the limitations and open challenges of BsC are also highlighted, e.g., short transmission range, limited data rate, etc. Another comprehensive literature review on fundamentals and state-of-the-art of ambient BsC is presented in \cite{8368232}. The scope of ambient BsC for IoT, wireless body area networks (BAN), and mmWave range is investigated; where various designs and solutions available in the literature are thoroughly reviewed. In \cite{liu2013ambient}, an implementation of BsC is introduced along with detailed analysis for its different proof of concept applications. Various important aspect at physical layer and link layer are highlighted and prototype implementation is presented. The notable existing literature on BsC systems is classified into different themes and sub-topics in Table \ref{Literature_Summary_BsC}.

\begin{figure*}
  \centering
  \includegraphics[width=\textwidth]{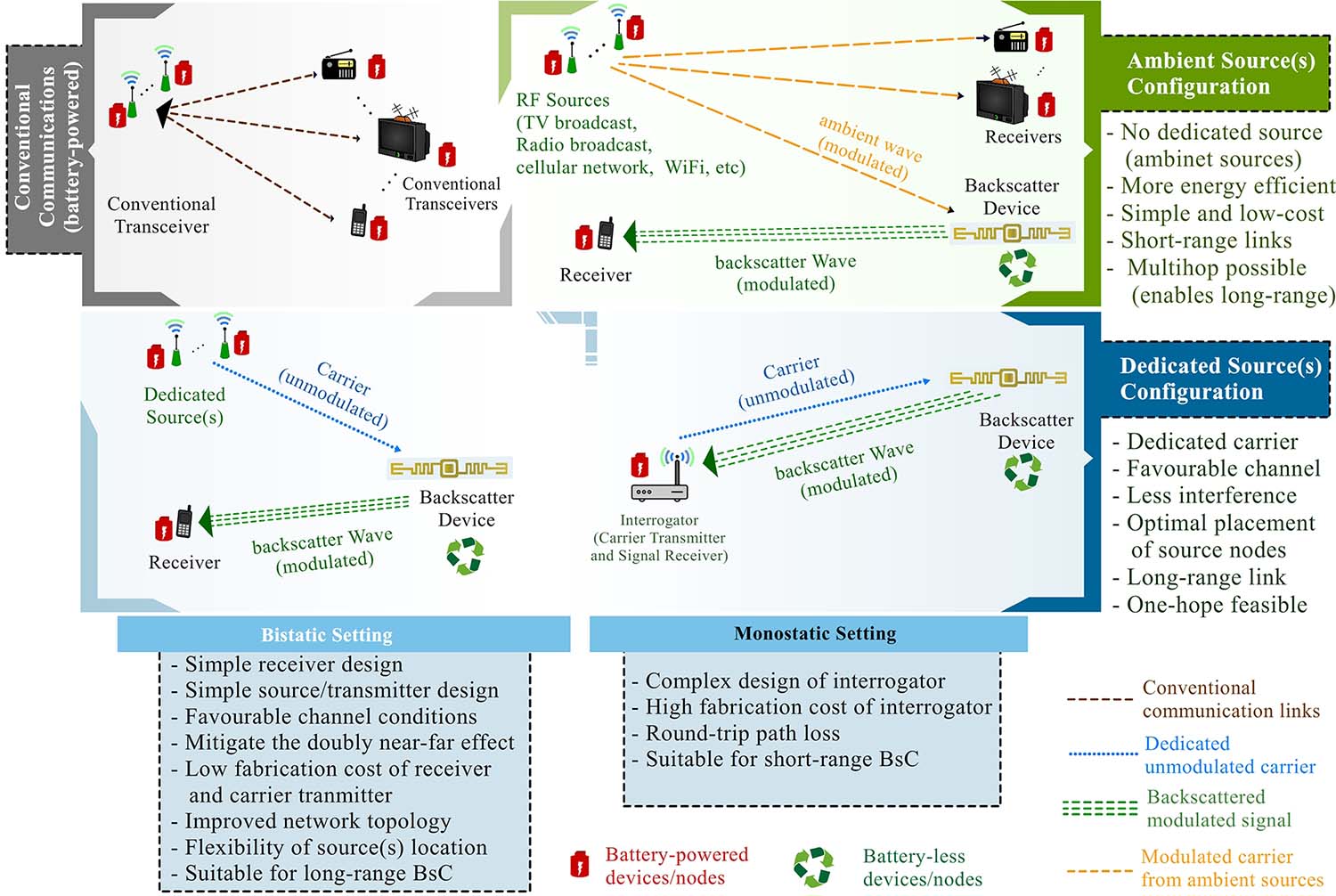}
  \caption{Comparative illustration of the conventional and BsC Systems; configuration of ambient, monstatic, and bistatic BsC systems are also indicated.}\label{Fig_BsC_Diagram}
\end{figure*}

\paragraph{System Configuration}
The configurations of BsC systems can be classified into dedicated and ambient source based configuration, while the dedicate source configuration can further be subcategorized into monostatic and bistatic settings. An illustration of different system configurations for BsCs is provided in Fig. \ref{Fig_BsC_Diagram}, where different salient features associated to different system configurations are also highlighted.

In the system configuration involving \emph{dedicated RF source}, a dedicate unmodulated carrier is transmitted by a carrier source which the backscatter device receives and backscatters it after modulating its information over it. In a \emph{monostatic} setting, the carrier source and the intended receiver of backscatter device is the same node. For example, in communication setup of RFID-tag to reader, the  reader node provides the carrier signal to the tag (backscatter device) and the same reader node itself is the intended receiver of the backscattered signal from tag.
In a \emph{bistatic} setting, the source and receiver are different nodes. A separated source node supplies the carrier signal to the backscatter device through an independent wireless link while the intended receiver of backscattered signal is a different node.
Such setting makes the design of both receiver and RF source nodes simpler and cheaper. Also separating the carrier source from the receiver enables the deployment of multiple RF sources in mMTC scenario. This further helps in not only creating favourable channel conditions but also in optimizing the network topology.

The \emph{Ambient configuration (or ambient BsC)} is a type of BsC in which the RF signals from nearby/ambient signals from non-dedicated sources (e.g., televisions, wireless fidelity (WiFi) APs, cell phones, etc) are utilized to power the communication devices. An illustration of monostatic, bistatic, ambient, dedicated, and hybrid scatter configurations is presented in Fig. \ref{Fig_BsC_Diagram}.
The ambient BsC configuration can also be interpreted as a spectrum sharing method, as it reutilizes the spectrum of existing broadcast systems. The spectrum regulation aspects for ambient BsC spectrum sharing are discussed in  \cite{ruttik2018does}, where it is suggested that the the ambient BsC can share the spectrum in coexistence with digital broadcast systems without imposing any significant amount of interference.
In \cite{3384419}, an ambient LoRa backscatter system design employing on-off keying is proposed which is named Aloba. The proposed backscatter device design can sense low-power ambient LoRa transmissions and modulate its information over the LoRa transmissions using on-off keying, where a promising trade-off between backscatter link distance (50-200 m) and throughput (199.4-39.5 Kbps) is demonstrated.

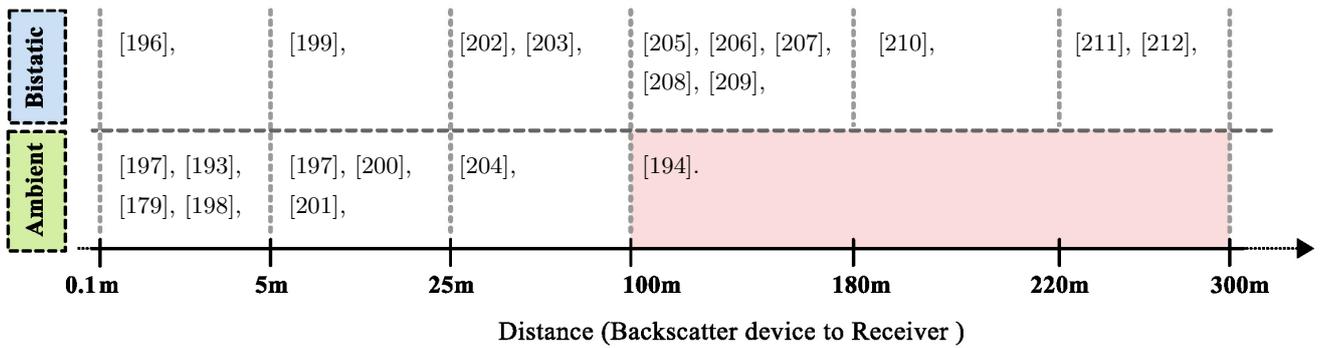
\begin{figure*}[t]
  \centering
\begin{overpic}
  [width=0.95\textwidth]{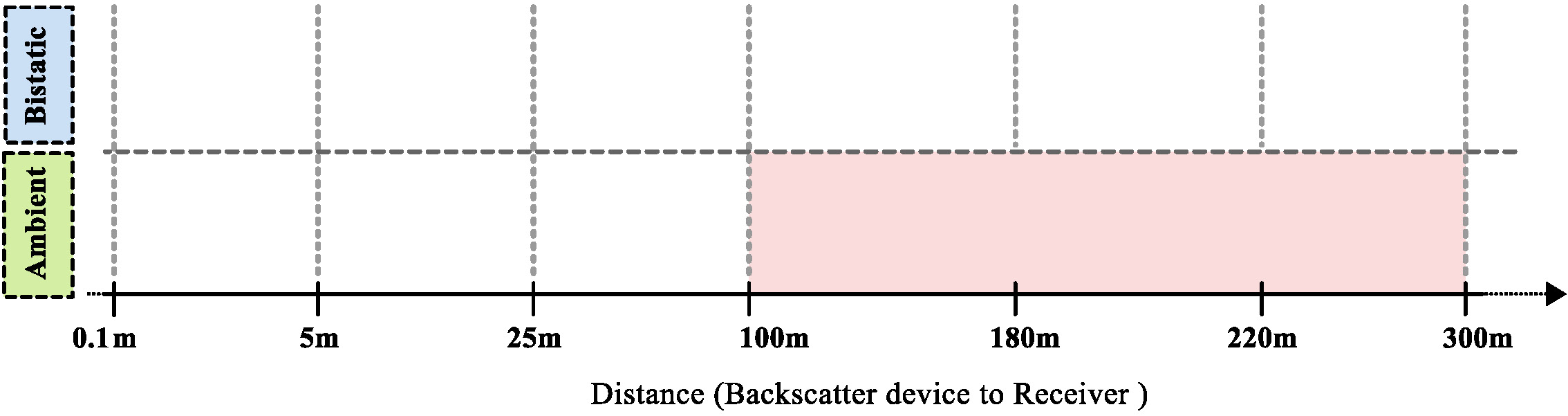}
  \put (8,23) {\small\cite{kellogg2014wi},}
  \put (8,13.5) {\small\cite{parks2014turbocharging}, \cite{liu2013ambient},}
  \put (8,10.5) {\small\cite{zhang2016enabling}, \cite{7175077},}
  \put (21,23) {\small\cite{kellogg2016passive},}
  \put (21,13.5) {\small\cite{parks2014turbocharging}, \cite{bharadia2015backfi},}
  \put (21,10.5) {\small\cite{wang2017fm},
  }
  \put (34,23) {\small\cite{6404546}, \cite{zhang2016hitchhike},}
  \put (34,13.5) {\small\cite{6942497},}
  \put (48,23) {\small\cite{6742719}, \cite{7412542}, \cite{6934197},}
  \put (48,13.5) {\small \cite{3384419}.}
  \put (48,20) {\small\cite{7059230}, \cite{6838975}, }
  \put (66,23) {\small\cite{varshney2016lorea},}
  \put (81,23) {\small\cite{7498843},\cite{7490345},}
\end{overpic}
  \caption{Classification of the literature demonstrating experimental results for link distance (i.e., from backscatter device to receiver) w.r.t. different system configurations.}
  \label{fig_distance_characterization}
\end{figure*}

In Fig. \ref{fig_distance_characterization}, the literature demonstrating experimental results for different system configurations of BsCs is classification with respect to backscatter-link coverage distance (i.e., from backscatter device to receiver). Ambient configuration is more energy efficient, whereas, it is more suitable for short range communications. However, long-range communications can be enabled in ambient configuration through establishing multihop backscatter-device-to-backscatter-device (B2B) (also referred as tag-to-tag) networks.

\paragraph{Multi-antenna BsC Systems}

The use of multiple antennas for transmitting and receiving the radio waves has shown various primary advantages in the overall efficiency of the communication systems. The advantages in communication and energy efficiency of the systems are achieved through the exploitation of angular domain, which is an extra degree of offered leverage, for diversity, multiplexing, multiple access, and/or beamforming, etc. Multiple antennas can also be used for wireless energy transfer through the concentration of energy into narrow beams radiated in the desired physical directions; which is also referred to as \emph{energy beamforming} \cite{7274644}. In \cite{shen2019ambient}, a system design with $16$ antenna ports is presented for the harvesting of energy from different frequency bands. In \cite{9249377}, a beamforming scheme for simultaneous information and power transfer exploiting estimated CSI of backscatter channel is proposed for multiuser wirelessly powered BsC, where achievable energy harvesting rate and ergodic rate are investigated. Deploying multiple antennas for energy beamforming can also help in increasing the range of ambient BsC links, which otherwise can only cover a very short range.

A multi-antenna energy beamforming technique is proposed in \cite{7274644}, where a multi-antenna energy transmitter serves multiple energy receivers. An algorithm for the optimization of performance trade-off between different energy transfer efficiency parameters (e.g., weighted sum energy,  proportional-fair-energy, etc) is also proposed. The availability and accuracy of the CSI is an essential requirement for such energy transfer methods; which can be obtained through different channel estimation methods, e.g., pilots based etc, as discussed in next subsection.

The combination of energy beamforming and BsC has a strong potential to support battery-less devices for the future wireless networks.
In \cite{8254467}, a beamforming system for ambient BsC is proposed, which optimizes the beamforming vector to enhance the spectrum sharing efficiency of the system. In the considered system, a cooperative receiver decodes the information from two transmitters, i.e., one multi-antenna primary and one single-antenna secondary. Moreover, a new transmit beamforming structure for enhancing the computational complexity required for the beamforming vector optimization is also proposed. Through numerical analysis, it is concluded in \cite{8254467} that the proposed beamforming optimization based ambient BsC system outperforms the conventional systems in terms of sum-rate.
The beamforming design optimization for multi-antenna SR BsC system to maximize the achievable transmission rate of the backscatter link under the given constraint of primary link's achievable rate is investigated in \cite{wu2019beamforming}.
In \cite{8446004}, an interesting multi-antenna receiver design for ambient BsC systems is proposed which does not require any knowledge of channel or noise statistics. Moreover, a new method for detection of ambient backscatter OFDM symbols is proposed. Another wireless energy beamforming assisted BsC system is proposed in \cite{8302460}, which reveals that the cooperative relay based BsC strategy can significantly enhance the overall system throughput.
The harvesting of energy from ambient RF signals assists in operating the backscatter circuit, however insufficient availability of the harvested power to the circuit can lead towards the suspension of BsC. In \cite{8468064}, an opportunistic scheme for a battery-assisted system (with multi-antennas at the receiver) for exploiting the residual and full battery power of the users for insufficient harvested power is proposed to ensure the availability of backscatter signals at all times. MRC along with successive interference cancellation (SIC) is suggested. The proposed scheme is demonstrated to be superior than its counterparts in the literature through a conducted simulation analysis, for high SNR and large number of antennas.

Theoretical bounds on the capability of coding and modulation schemes in different channel and noise conditions is of high importance in designing and studying of the systems \cite{wyner1966bounds}. The achievable sum-rate of bistatic ambient BsC is studied in \cite{7948789}, where a legacy MIMO system in co-existence of a multi-antenna node constitute the system model. The ambient BsC system is found to impose only a nominal limit on the achievable sum-rate, i.e., it causes only minimal interference. Moreover, it can also offer an additional advantage of assisting the legacy system as in the role of bring a passive relay. In cooperative spectrum sharing setup, the interference analysis and management is of high significance. In \cite{7997479}, a multi-antenna receiver design with SIC algorithm for cooperatively decoding the information from both BsC and RF sources is proposed. For the considered system model, closed-form analytical relationships for BER performance of the MLh and SIC based detectors are derived. The conducted numerical performance analysis has suggested that the SIC-based detector can nearly achieve the same performance as that of MLh detector.
The principles of BsC from the perspective of communication systems theoretic is thoroughly reviewed in \cite{6685977}. The performance of both energy harvesting and modulation are linked to each other in BsC. The joint investigation of both energy and spectral efficiency in the energy-constrained applications of multi-antenna BsC systems is also strongly advocated for redirecting the focus of the researchers towards the devising of coded modulation schemes. The statistical behavior of the fading channels in such multi-antenna BsC systems may be vastly different from those encountered in conventional MIMO communication systems; therefore it is pivotal to study the channel statistics and the multi-antenna associated multiplexing and diversity gains for the BsC systems.

\paragraph{Backscatter Wireless Channels (Modeling, Characterization, and Estimation)}\label{SubSec_BsC_Channel}

In realistic radio propagation scenarios, various different physical phenomenon result in arrival of multiple copies of the signal. Each multipath signal may experience single or multiple bounces causing different amount of shift in delay, frequency, and angle of each arriving multipath depending upon the nature of the propagation environment and mobility conditions of the nodes \cite{nawaz2017tunable}. These shifts further cause the signal to spread in delay, frequency, and angular domain imposing frequency-, time-, and spacial-selectivity in the channel characteristics, respectively. The use of multiple antennas with directed radiation patterns can reduce the angular spread, which can further help in reducing the Doppler and root mean square (RMS) delay-spread \cite{5462882}. Accurate modeling and characterization of radio propagation channels is of vital importance in designing, studying, and capacity enhancement of the communication systems.

The capacity of  ambient BsC channel between a reader and a passive tag is studied in \cite{8171133,8345348} by using Gaussian channel model.
In \cite{8171133,8345348}, it is suggested that the Gaussian channel capacity can be obtained for the scenario when the RF signals are  backscattered by the tag with an unequal probability. Moreover, for the considered setup, it is suggested that the channel capacity with complex Gaussian RF signals is not exactly the twice of the the capacity with real signals.
In \cite{jameel2019simultaneous}, ambient BsC under Rayleigh fading channel conditions is studied for simultaneous energy harvesting and transmission. For achieving optimal energy harvesting and spectral efficiency from BsC, ambient BsC, and cooperative ambient BsC systems, accurate understanding of the channel parameters is of vital importance. Particularly in SR systems, the estimate of the channel parameters is critical in effectively removing the interference from the primary transmission for detecting and decoding the backscatter transmission. Moreover, in muti-antenna BsC systems, effective exploitation of angular domain for energy beamforming is only possible when accurate understanding of the channel statistics is available.

The estimate of the channel can be obtained through the exploitation of known pilots, blindly (without pilots), semi-blindly, and superimposed-pilots based methods. In most of the existing studies on ambient BsC systems, the perfect CSI is assumed to be available, while practically obtaining a reasonable CSI estimate is a challenging task.
In ambient BsC systems, as the environmental RF signals are unknown to the receiver, therefore the pilots based estimation methods may not be a suitable choice.
This lack of availability of the dedicated pilot signals makes the CSI estimation problem in ambient BsC systems as vastly different from that in the conventional systems. In \cite{8320359}, an expectation-maximization algorithm based blind channel estimation method for ambient BsC is proposed, and Cram\'er-Rao bound of the proposed estimator is derived. Channels estimation in multi-antenna readers based ambient BsC systems is an open research problem.
In \cite{8641171}, a blind channel estimation method for multi-antenna ambient BsC systems is proposed, which is based on eigen value decomposition (EVD) of received signal covariance matrix.
Further in \cite{zhao2019channel}, a blind channel estimation method for large-scale multi-antenna ambient BsC systems is proposed. A large-scale uniform linear antenna array is employed at the reader, a method of estimating AoA and channels' gain is proposed for which  Cram\'er-Rao lower bound is also derived.
The multi-bounce effect between ambient BsC devices is mostly ignored in the literature on ambient BsC systems. The work in \cite{duan2018multi} highlights it by considering a modified multiplicative reverberant channel model and binary Phase Shift Keying (BPSK) signals. The presented results demonstrate the importance of not ignoring the multi-bounce effect to achieve optimal rates.

\paragraph{Multiplexing, Multiple-Access, and Duplexing in BsC}

The BsC links and networks has received significant attentions over the last decade. Different methods for multiplexing in the primary and backscatter links are proposed in the literature. In \cite{8647245},  spread-spectrum and multi-carrier multiplexing perspective for ambient BsC systems are presented. A method for exploitation of in the air ambient OFDM carriers is proposed. The cancellation of interference from the primary link while decoding the backscatter link, can be carried out by utilizing the repeating cyclic-prefix structure of ambient OFDM signals. In such multiplexing setting, the symbol duration of backscatter device is designed as larger (integer multiple) than the symbol duration of primary link; which along the cyclic-prefix helps in the cancellation of the interference in decoding of backscatter symbols. In \cite{8103807}, design for different symbol waveforms for the ambient OFDM backscatter systems is proposed.  The waveform for bit `0' maintains the same state and the waveform for bit `1' goes through a state transition in the middle of each OFDM symbol duration within a larger duration symbol of backscatter link. For the proposed setting, the detector only requires the channel gain, which simplifies the system implementation. Moreover, for a multi-antenna receiver, a new method which linearly combines the test statistics of different antennas is also proposed.

Investigating the scope of BsC in integrated large scale systems, which the wireless access is required to be provided to a huge number of devices (sensor, IoT devices, etc) is critical in designing of various emerging applications like massive-IoT, BAN, smart-industries, etc. For example, the provision of simultaneous access to a huge number of devices through BsC links in a massive-IoT network is another interesting research direction.
A multiple access technique named as Multiple Subcarrier Multiple Access (MSMA) is proposed in  \cite{8299928}.
The optimal allocation of subcarrier frequencies and unavoidable harmonics among different adjacent subcarriers are the two primary research challenges indicated for the proposed MSMA method. The random assignment of the subcarrier may lead to the degradation of the system performance, as the inter-carrier mutual interference may be unevenly spread over the contemplated frequency band. The proposed heuristic approach compared to random sub-carrier allocation can provide significant enhancement in system performance through effective suppression of the interference.
A multiplicative ambient backscatter multiple-access method is proposed in \cite{8422524}. The new multiplicative multiple-access method for ambient BsC is comprehensively presented in \cite{8399824}, where the receiver is designed to simultaneously detect the signals from the direct transmitter and its multiplicative backscattered copy containing the additional information of tag. For this new multiplicative backscattering, which is different from the conventional additive backscattering, a new channel model is also proposed. The achievable rate of the proposed multiplicative multiple-access method is found to be higher than the conventional time-division multiple access (TDMA) scheme for the cases when SNR is high and direct channel is stronger than the backscatter channel.

The demands of delivering multiple-access requirements for enabling massive connectivity have escalated with the emergence of applications like massive-IoT. NOMA has a recognized potential in accommodating multiple users within a single resource block for addressing the demands of future generation multi-access requirements.
A NOMA enhanced BsC system design is proposed in \cite{8439079}.
Another study on backscatter-NOMA systems is presented in \cite{8491248}.
A BsC system with multiple backscatter nodes being served by power-domain NOMA is considered. It is suggested that using different backscattering coefficient on different multiplexed backscatter devices can help in optimal exploitation of power-domain NOMA. A cooperative bakscatter-NOMA system is proposed in \cite{8636518}, in which donwlink NOMA system is incorporated with backscatter device. Considering the high energy and spectral efficiency performance of NOMA, ambient BsC, and full-duplex systems, these technologies can make a natural alliance to address the demands of future massive-IoT and cellular networks.

The two way communication in a full-duplex system is accommodated within a single time-frequency resource, while the interference is suppressed through different signal processing methods.
Full-duplex BsC links are thoroughly studied and encouraged in the literature, see e.g., \cite{7913737,8700623}.
A full-duplex backscatter system for wireless information and energy transfer is proposed in \cite{7913737}. In \cite{8700623}, simultaneous transmission of signals from primary cellular system and receiving of backscatter signals to enhance the overall system throughput is strongly encouraged.

\paragraph{Modulation, Transmission, and Detection in BsC}

The optimal setting for modulation and transmission parameters to ensure accurate signal detection in BsC systems is a challenging task, especially in ambient BsC systems. This is because of the absence of any prior training/pilot data availability for estimating the channel parameters, as the ambient BsC systems utilize the signals from different unknown sources.
Another critical difficulty in the detection of backscatter signals is its mixing with the direct signal received from the primary sources; as the primary signal causing interference is always stronger than the desired backscatter signal. The inaccurate estimate of the channel and high interference from the primary sources limit the overall rate of the BsC link.

A new simple method of detection is proposed in \cite{8644139}, where the on-off keying is used to modulate information over spontaneous parametric down conversion based generated signals and the receiver exploits the non-classical Hong-ou-Mandel effect.
In \cite{7551180}, a theoretical model for transmission and detection in ambient BsC systems is proposed which adopts different encoding method at the tag-side in order to avoid the need for estimating the channels. Moreover, analytical expressions to characterize the minumum-BER and optimal threshold detection methods are derived. Another learning based method which avoids the need for estimating the CSI at the reader-side is proposed in \cite{8474331}. In \cite{nguyen2019signal}, repeating structure and cyclic-prefix of ambient OFDM signals are utilized to assist in the cancellation of the interference from the primary link. Moreover, the optimal detection threshold and power order for the proposed system are also derived. The signal detection in multi antenna devices based systems is more challenging, as the number of channels to be estimated are directly proportional to the number of antennas and only a limited transmission can be supported due to the power constraints on backscatter devices. A design for detector and antenna selection for multiple antenna backscatter devices is proposed in \cite{8445952}, which exploits blind channel estimation methods.

Due to energy constraints and unavailability of active signal source at backscatter nodes, the optimal decision of the choice of modulation is a challenging task. The work in \cite{harms2017modulation} studies the extent of achievable data rate and link-distance for different modulation schemes in combination with off-the-shelf hardware in BsC links. The performance of on-off keying (OOK) and frequency-shift keying (FSK) for backscattering television broadcast signals and a constant carrier signal is investigated. For the case of backscattering the ambient television broadcast signals, for these modulation schemes, the link-distance of over a meter can be achieved with a maximum signal strength of -70dBm. However, for the case of a direct constant carrier signal, BsC over a link-distance of up to 225m and 30m can be achieved in a LoS and non-LoS (nLoS) environment, respectively. The two state amplitude shift keying (ASK) or phase shift keying (PSK) may help in more accurate detection of the symbols at the receiver in high SNR conditions, but it imposes reduction in the link data rate. In \cite{8423609}, a high order PSK, i.e., $M$--PSK modulation scheme is adopted for BsC links, and a design for optimal multilevel energy detector is proposed. Closed-form expression for symbol error rate (SER) is also derived. Hardware prototype for 4--PSK ambient BsC systems is designed, where it is shown to achieve a data rate of 20kbps. This high data rate is seen to be achievable for 98.7\% of the time compared to the two-state modulations, while to distinguishable symbols mean number is observed to be 3.66.

In \cite{8672080}, a multiple frequency shift keying (MFSK) modulation scheme for BsC links and corresponding MLh and direct energy detectors are proposed. It is suggested that MFSK outperforms the OOK modulation in terms of SER performance. Moreover, the impact of the order of modulation on SER performance is primarily dependant on the operating bit SNR, and reducing the link-distance can help in minimizing the outage probability.
An application of BsC for localization in vehicular communication context is presented in \cite{han2019joint}, where backscatter tags  deployed along the vehicles' routes (roads, rail-track, etc) assist in the the localization operation. Moreover, a new waveform design named as ``Joint Frequency-and-Phase Modulation" is proposed to exploit maximum degree-of-freedom of backscatter channel. The frequency and phase are are proposed to be jointly modulated in order to estimate the distance and to facilitate the differentiation operation (required for demodulation) at the receiver, respectively.

The authors in \cite{8534458} proposed 4-pulse amplitude modulation (4-PAM) to represent the backscatter device information over the ambient FM signal to be communicated through the backscatter link. An ultra-low-power microcontroller controlled single transistor RF front-end of the tag.  In an indoor environment, when the bit rate was taken as 345bps, the power consumption of 27$\mu$W,  the primary source link was a real FM station at 34.5Km, the distance of backscatter link was taken as 1m; the energy spent for modulation was calculated to be 78.2 nJ/b. Moreover, for a bit rate of 10.2 Kbps, the energy spent on modulation was calculated to be 27.7 nJ/b.

\subsection{Cooperative Ambient BsC}

Cooperation between primary transmitter and backscatter nodes can potentially help in minimizing the interference caused by the primary transmissions for the backscatter links.
Cooperative BsC concept is regarded as a promising solution to the challenges in enhancing the transmission rate and detection accuracy of BsC. In cooperative BsC concept, the primary signal causing the interference in the backscatter signal can jointly be decoded and removed from the received composite signal through the sharing of some prior information between primary and backscatter transmissions.
SR is an emerging paradigm in which both the radio-spectrum and radio-source are shared by the backscatter and primary transmission devices.
The transmitter of primary source is particularly designed to assist the transmission of both primary and backscatter device, and the primary receiver is designed to assist detection and decoding of information from both primary and backscatter device.
In \cite{long2018symbiotic}, a MISO SR system is proposed, where beamforming optimization and achievable rates of both primary and backscatter device are derived. The symbol duration of backscatter transmission is suggested to be devised as either the same or much longer than that of in primary transmission, which leads to a parasitic or commensal relationship between the backscatter and primary transmissions. The appropriate designing of backscatter symbol duration can significantly help in enabling the opportunistic transmission for the backscatter link as well as in enhancing the transmission rate of the primary link. This enhancement is achieved through the exploitation of the additional signal path from the backscatter link.
In \cite{8692391}, three transmission methods for cooperative BsCs are proposed, and analysis for different fading conditions is conducted.
The authors in \cite{8274950} proposed a spread spectrum and spectrum share conception based model for cooperative detection of signals and decoding of information from the primary and backscatter devices. MLh detection, linear detection, and SIC methods are investigated. Detection schemes for both frequency-flat and -selective channels are proposed and closed-form analytical expressions for BER are derived.

The authors in \cite{8665892} study the challenges in resource allocation in SR systems, where the fading imposed by the wireless channels is also considered into the account. The optimal setting of transmit power of primary transmission and the reflection-coefficient of backscatter device is also investigated in order to maximize the ergodic weighted sum rate of both the links.

Full-duplexing in SR systems is studied in \cite{8638762}, where a passive backscatter device is parasitic to primary transmission. A multi-antenna primary transmission link broadcasts common messages to both primary and backscatter receivers, while it also supports passive BsC. The backscatter device utilizes a fraction of the received power to decode the common messages while it simultaneously uses the remaining power to backscatter with its information towards the primary receiver in a full-duplex fashion. The conceivable rate of the full-duplex SR system with both Gaussian and QAM schemes is also investigated. It is also established that the full-duplex system outperforms the time-division-multiplexing based half-duplex method.

\subsection{Non-Coherent BsC}

A thorough literature survey on non-coherent detection and modulation schemes has been presented in Sec. \ref{Sec_nCoh_cFree}. This subsection briefly revisits the concept and significance of non-coherent detection in BsC systems. Both coherent and non-coherent BsC systems have been proposed in the literature, see e.g., \cite{6942497,7059230} and \cite{7769255,8642363}, respectively. Acquiring the synchronized carrier signal and accurate CSI in ambient BsC receivers for decoding information is a challenging task, therefore non-coherent detection with BsC can potentially deliver promising trade-off.
In \cite{7769255}, the fundamentals of non-coherent detection in ambient BsC systems are studied when no information of CSI is available. A MLh detector and closed-form expression for BER based outage probability of the system are derived.
In \cite{8399824}, a performance analysis of coherent and non-coherent modulation methods used by primary transmitter and tag, respectively, is presented for cooperative multiple-access based ambient BsC systems. In \cite{8642363}, a non-coherent energy detection based modulation scheme for the BsC systems exploiting legacy OFDM signals is proposed. Binary and higher-order modulation schemes are investigated. Exact and approximate expressions for the probability of error for the binary and M-ary cases are derived. Based on conducted Monte-carlo simulations, it has been established that the proposed non-coherent method is superior to other BsC methods exploiting legacy OFDM signals.

\subsection{Machine Learning (ML) for Detection in BsC}
The ML methods have recently gained notable attention from the research community as an alternative to the conventional model-based algorithmic solutions. The potential of ML methods in addressing various challenges in delivering B5G wireless networks is thoroughly studied in \cite{QML_6G_Junaid}.
Numerous applications of ML at different layers of wireless communications networks, spanning from network layer management to autoencoding of end-to-end systems, are highlighted. Given the availability of sufficient data and computing capability, ML has a recognized potential in the optimization of modulation, transmission, and detection operations. High computational capability may not be available at Tag side of BsC systems, however, it can be provided at the receiver side for detection of primary and backscatter signals.
A label-assisted transmission framework is developed in \cite{8474331}, in which two known labels are transmitted from the tag before sending the actual information. An algorithm for expectation maximization of the corresponding received label signals at the reader is proposed. As the scope of ML methods for clustering and detecting operations is well established, the work in \cite{8474331} performs the signal detection through the learning and classification of labels' constellation into clusters and categories.
An unsupervised learning-based signal detection method for ambient BsC systems is proposed in \cite{8269048}, which learns the behavior of the received signal through clustering the received symbols into different groups.
Due to the spectrum sharing nature and the difficulties associated with the channel estimation in ambient BsC systems, ML has a strong potential in interference suppression and signal detection.


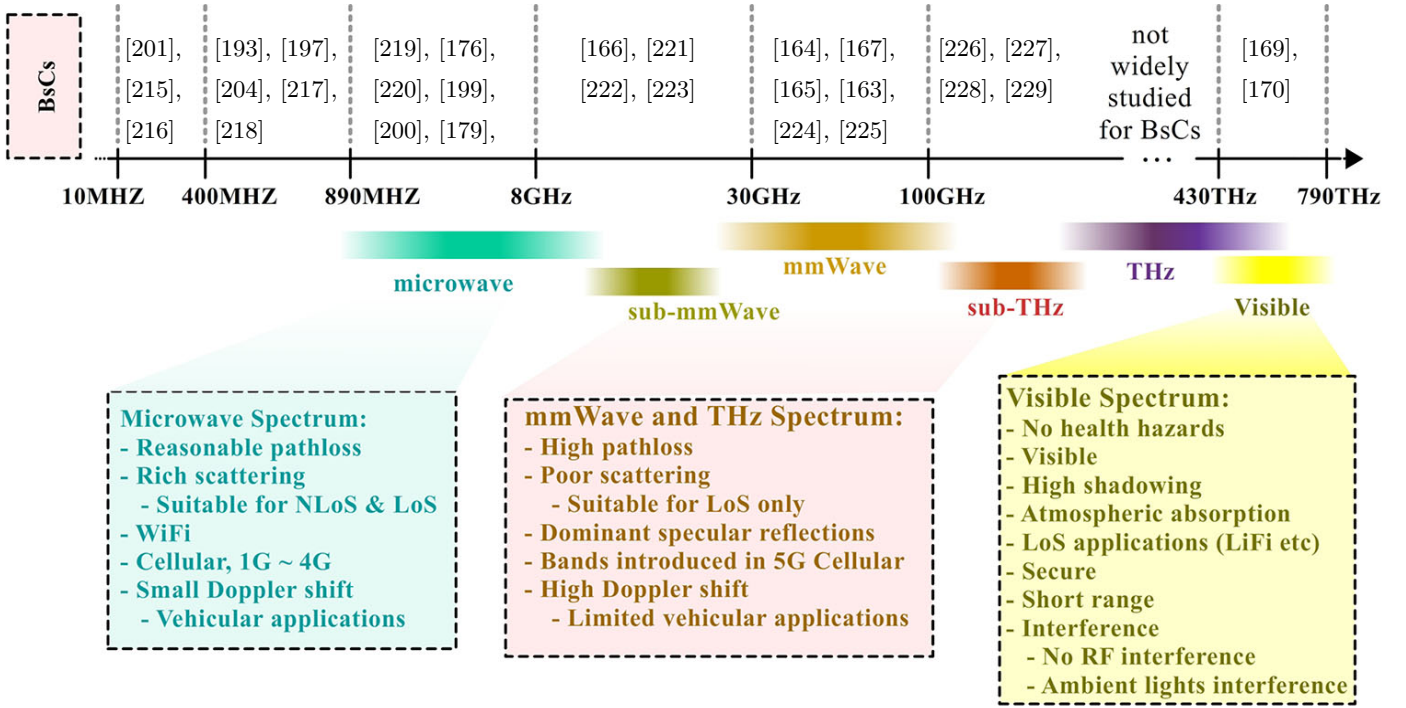
\begin{figure*}[t]
  \centering
\begin{overpic}
  [width=\textwidth]{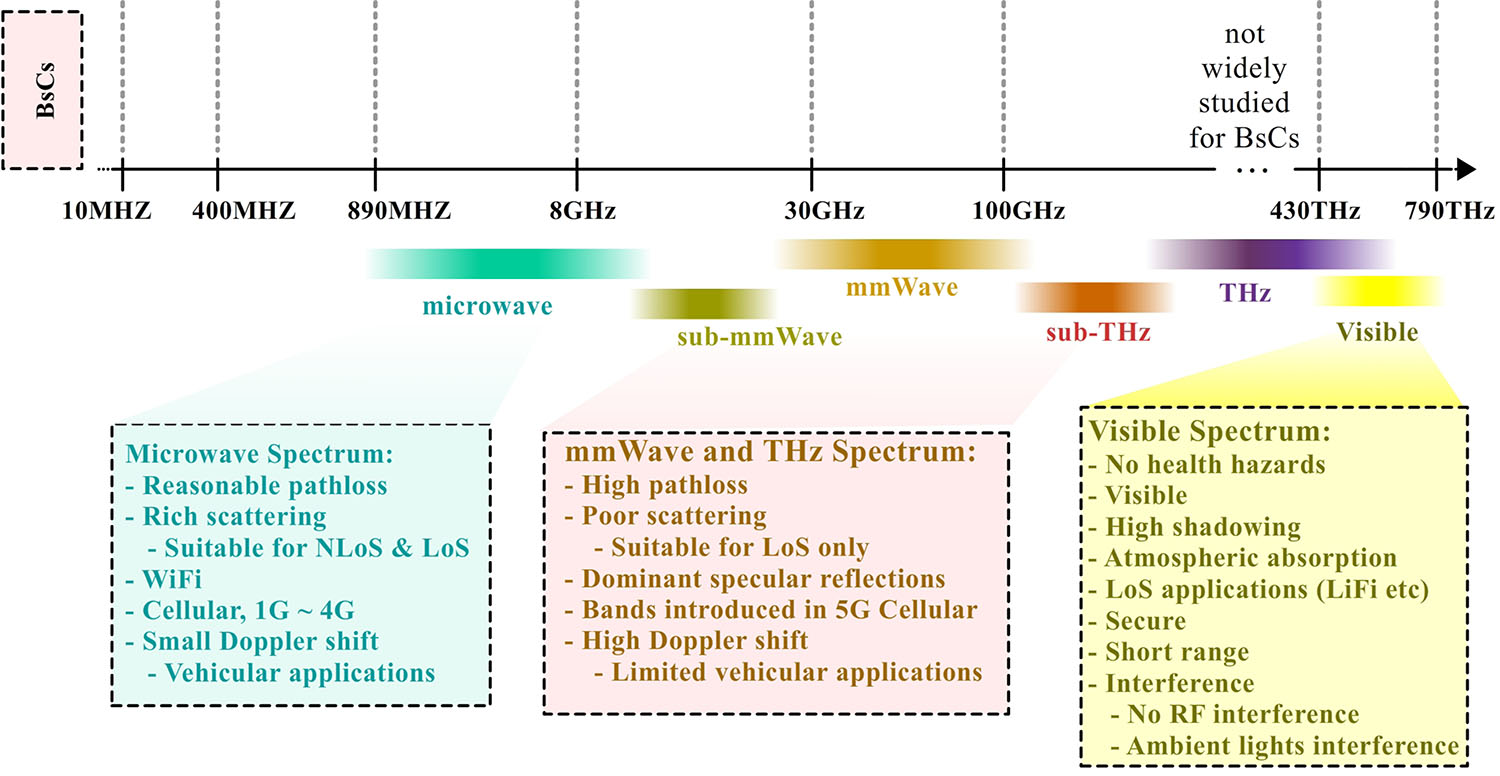}
  \put (8,48) {\small\cite{wang2017fm},}
  \put (8,45) {\small\cite{8103031},}
  \put (8,42) {\small\cite{8471572}}
  \put (14.5,48) {\small\cite{liu2013ambient}, \cite{parks2014turbocharging},}
  \put (14.5,45) {\small\cite{6942497}, \cite{munir2017low},}
  \put (14.5,42) {\small\cite{8757120}}
  \put (26,48) {\small\cite{7419631}, \cite{grosinger2012evaluating}, }
  \put (26,45) {\small\cite{liu2014enabling}, \cite{kellogg2016passive},}
  \put (26,42) {\small\cite{bharadia2015backfi}, \cite{zhang2016enabling},}
  \put (41,48) {\small\cite{8058702}, \cite{8058927}}
  \put (41,45) {\small\cite{7488012}, \cite{8815450}}
  \put (55,48) {\small\cite{6697382}, \cite{4405373},}
  \put (55,45) {\small\cite{8319409}, \cite{4631469}, }
  \put (55,42) {\small\cite{abbas2019thz}, \cite{7927322}}
  \put (67,48) {\small\cite{7750893}, \cite{8454696},}
  \put (67,45) {\small \cite{8823734}, \cite{6548163}}
  \put (89,48) {\small\cite{7752834},}
  \put (89,45) {\small\cite{xu2017passivevlc}}
\end{overpic}
  \caption{Classification of notable literature w.r.t. utilized frequency spectrum.}
  \label{fig_spectrum}
\end{figure*}

\subsection{Radio Spectrum for BsC}

The scarcity of usable electromagnetic spectrum is a major bottleneck in meeting the data rate demands of future networks. The ultra-high capacity and energy efficiency demands of communication networks has led to the exploration of new usable frequency spectrum beyond the microwave band.
This has motivated the research community to explore the large chunk of underutilized mmWave range based electromagnetic spectrum from $30-300$ GHz, and sub-mm range such as $28$ GHz. The 3GPP Release-15 has introduced the mmWave communications as a key technology for 5G wireless networks \cite{3GPP_r15}.
The ongoing thorough research work on the frequency spectrum beyond the mmWave bands makes the adoption of such bands (some) in future generation of communication networks  highly likely, i.e., 6G and Beyond. For example, the researchers around the globe have started exploring and characterizing the mmWave, sub-teraHz, tera-Hz, visible light, and laser bands for their applications in communication networks, see e.g., \cite{ju2019scattering,digiovanni2014surface,mou2014backscattering}.

Some example research articles on BsCs characterized into different frequency spectrum ranges are indicated in Fig. \ref{fig_spectrum}. Some salient radio channel characteristics associated to frequency ranges are also indicated in the figure.
The design of BsC systems for the microwave frequency bands may be favorable because of the low losses in the active components and transmission lines. However, along with the advantage of large bandwidth offered by the higher frequency bands, these bands also offer the miniaturized design of high-gain antennas elements and antenna arrays. This advantage has the potential to enable the integration of antenna-arrays into various portable and sensing devices. This opportunity can open various new horizons of applications and services through joint exploitation of spatio-temporal statistics of the radio propagation environments for optimized services delivery to a massive number of communicating nodes in an ultra densely connected world of the future.

The behavior of radio propagation channels at higher frequency bands compared to the conventional microwave band is vastly different in terms of the pathloss, shadowing, specular reflections, and multipath scattering phenomena.
The propagation characteristics of mmWave frequencies make them more suitable for frequent frequency re-use in small and tiny (e.g., inter-chip communications, etc) cells.
Moreover, the polarization in mmWave, due to smaller wavelengths, is exploited for aggressive spatial multiplexing by using mMIMO and adaptive beamforming \cite{6375940}, leading to the increased densification of small cells in urban areas.
The mobility of communicating nodes (in vehicular communication scenarios) at such high-frequency bands imposes a large Doppler shift, which is because of the carrier frequency being a linear scaling factor in equating the Doppler shift with direction and velocity of mobility. In a multipath propagation environment, this high Doppler shift associated with each multipath leads to a high Doppler spread, which further leads to a reduced coherence-time causing increased rate of fluctuations in the envelopes. There is a need to accurately characterization the fading statistics of radio propagation channels at mmWave, sub-teraHz, and teraHz bands for their maximal utilization in B5G/6G wireless networks \cite{8594703}.


The conventional ultra high frequency (UHF) based RFID concept is extended to mmWave based identification concept in \cite{4631469}; where the advantages and limitations of the system are also highlighted. The feasibility of the mmWave based identification concept for 60GHz based downlink and backscatter-uplink is studied, where various applications of the idea are also suggested.
A 60GHz based semi-passive design for tags identification in high data rate applications is proposed in \cite{6697382}, where an SNR of over 20dB for a link-distance of 30cm is observed.
In \cite{8319409}, the scope of utilizing highly directive mmWave communication links for passive RFID based BsC is investigated. For the considered setup, it is suggested that the identification and localization of tags can be performed with reasonable accuracy at the 60GHz band by only utilizing a single reader. The energy requirements in the proposed setup for identification and localization of tags at 60GHz are observed to be less stringent compared to the microwave band.
In \cite{7010535}, architecture of mmWave-based UDN for 5G communication networks is discussed. Also, authors in \cite{7306533} have discussed an mmWave mMIMO based wireless backhaul for the 5G UDN.
A high-speed backscatter transmitter for the 24GHz band is presented in \cite{8058702} for RF sensing and multi-gigabit communications. The design is implemented through inkjet printing on flexible substrates that can be easily integrated into different devices, e.g., IoT, wearable, etc. The design is capable of sensing the deviation of up to 4GHz in the carrier frequency. The energy consumption of transmission is recorded as 0.15 pJ/b at the communication speed of 4 Gb/s.
A receiver design for 77GHz band for identification systems is presented in \cite{4405373}.

Researchers have recently begun to explore a few frequency bands beyond the mmWave range to establish their characteristics and usability in different communication applications for meeting the anticipated enormous capacity and energy efficiency demands in 6G and beyond communication networks.
The propagation behavior at sub-teraHz and teraHz band being vastly different from the microwave band, the scattering behavior of 100 GHz, 160 GHz, 240 GHz, and 1.55 teraHz for different materials (metallic, homogeneous dielectric, and inhomogeneous dielectric surfaces) is investigated in \cite{digiovanni2014surface}. The co-polarization backscattering coefficient in an indoor environment is analyzed for different elevation angles (5 to 75 degrees). Another study on the backscattering behavior of different teraHz and laser bands for different materials is presented in \cite{mou2014backscattering}. These investigations provides a detailed insight into the coherent and non-coherent surface and volumetric scattering behavior of different types of materials at sub-teraHz, teraHZ, and laser bands, which is of high significance in designing communication systems and integrated circuits for teraHz and sub-teraHz applications. A through survey on design aspects for wireless power transfer via mmWave and sub-teraHz wave is provided in \cite{mizojiri2018wireless}, where 303 GHz is the highest examined frequency band. Rectenna design aspects for wireless power transfer via mmWave and sub-terHz is investigated.
The DC output power and the RF-DC conversion efficiency associated with 303GHz band was found to be 17.1 mW and 2.17\%, respectively. It is suggested that thorough research work on the development of efficient and high-breakdown voltage diodes (e.g., Gallium Nitride (GaN) diodes) is of vital need to improve the efficiency of wireless power transfer at sub-teraHz bands.
Several examples of the chipless RFID concept extended to teraHz frequency bands, also refereed as teraHz identification (THID) and RF bar codes, are reported in the literature, see e.g., \cite{7750893,8454696,8823734,6548163}. Such low-cost substrate and all-passive structure based tags with added sensing capability operating at teraHz bands can play a vital role in the evolution of IoT paradigm for achieving the goal of green- and automated-World.

\subsection{BsC Application Scenarios}

BsC is a new paradigm for enabling massive connectivity in the future wireless networks with stringent spectrum and energy efficiency requirements. Various applications of BsC are reported in the literature from enabling battery-free ad-hoc wireless networks to the assistance in cognitive radios. Massive-IoT and bio-medical communications are among the most promising applications of BsCs. This subsection briefly discusses the recent developments and applications of BsC in enabling B5G and 6G wireless networks.

\paragraph{Massive-IoT}

Ensuring uninterrupted power supply to a huge number of small devices in a massive-IoT network is a challenging task. The provision of power through batteries to these wireless devices has associated tedious tasks of recharging, replacing, and maintaining the batteries. Ambient BsC has recently emerged as a promising technology to realize the ambition of battery-less passive devices operating in massive-IoT networks. The authors in \cite{memon2019ambient} conducted a survey on the different types of ambient RF signals and different bands of frequencies that can be utilized as sources of energy to energize the passive IoT devices. Maintaining a desirable balance between wireless energy and information transfer is another area of research.

The inherent sparsity in ambient energy harvesting supported BsC signals can be exploited to design sparse codes for non-orthogonal signaling and detection. In \cite{kim2018novel}, a sparse-coded ambient BsC based non-orthogonal signaling method is proposed to assist NOMA and M-ary modulation with an extra degree of diversity for cooperative backscatter transmissions. Moreover, dyadic message passing and channel estimation algorithms are proposed.

The benefits offered by backscatter wireless communication links at the physical layer can be extended to design the optimal higher layers operations for future IoT networks. In \cite{8115787}, a medium access control (MAC) protocol to support BsC enabled IoT networks in the coexistence of a primary Wi-Fi network is proposed. The Wi-Fi access links are designed to operate in full-duplex while the backscatter devices operate in a half-duplex fashion. New control frames to provide the necessary support in the protocol at Wi-Fi access points (APs) are introduced for enabling the suppression of uplink transmission while simultaneously transmitting the downlink data for BsC links. The mean of overhead time and downlink utilization factor are reported to be 121 $\mu$s and 80\%, respectively. The average network throughput is said to be 63.5 Mbps and 59 Mbps for data delivery frequency of 2 and 8 times per backscatter-nodes, respectively. These observations are obtained for the setup containing 30 backscatter- and 30 Wi-Fi client-nodes; while the average throughput performance is observed to degrade with an increase in the number of backscatter-nodes.

The minimum power required as input for performing the sensing operations was reported to be -18dBm for ambient RF energy harvesting sensing and communication platform developed in \cite{6613706}. A high sensitivity RF harvester with 6dBi receive antenna was shown to operate up to a distance of 10.4 km and 200m over 1 MW UHF television broadcast transmitter and a cellular BS transceiver, respectively.

BsC holds various attractive applications in the automation of homes, offices, and industries, etc. The low manufacturing cost of backscatter nodes facilitates in maintaining the deployment cost to minimal for the application scenarios involving a massive number of nodes, e.g., in industrial automation. \emph{Logistics management} is another attractive application of massive-IoT where both non-coherent and BsC offer various fundamental advantages in designing of their communication system,i.e., low power consumption, spectrum sharing, and low device cost capabilities.
The deployment of BsC in automation and logistics can assist in reducing the operational costs, improving the production quality, and accelerating the production process \cite{liu2019next}.


\begin{figure*}[t]
  \centering
  \includegraphics[width=0.85\textwidth]{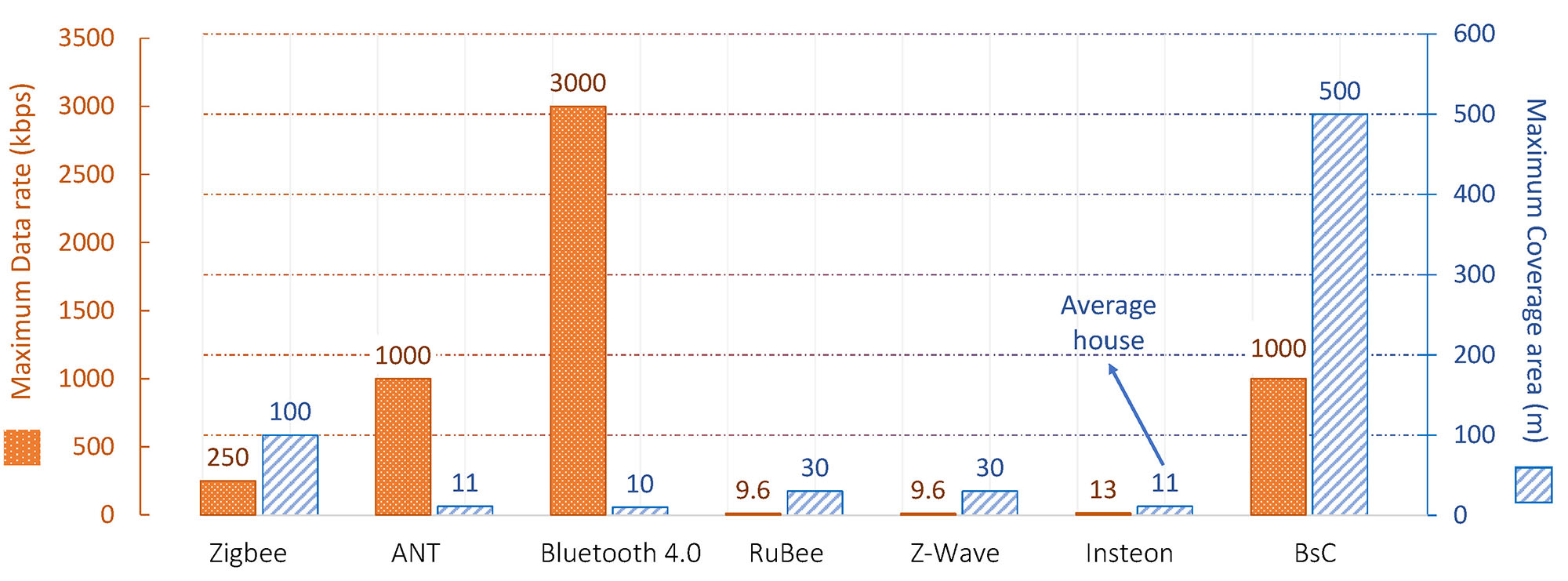}
  \caption{Maximum achievable data rate and coverage distance of different technologies \cite{zhang2016enabling}.}
  \label{Fig_technologies_bsc_comp}
\end{figure*}

\paragraph{Biomedical Communications}

The on-body and in-body wireless devices are usually constrained with limited power and bandwidth resources. Among many attractive applications of BsC, reinvigorating the healthcare systems is a notable application. The need for enabling battery-less communications with on-body and in-body sensors and small embedded devices makes the BsC an important candidate technology.
The feasibility of on-body (Human) BsC for 900 MHz and 2.45G Hz bands is studied in \cite{grosinger2012evaluating,grosinger2013feasibility}.  Realistic on-body channel measurements are conducted, which the channel response is utilized to study the sufficiency of available power for reliable BsC. Outage analysis for forward and BsC links is conducted, where the BsC is seen to demonstrate robust performance for 900 MHz monopole antenna configuration.
In \cite{jameel2019applications}, BsC for healthcare applications are reviewed. Measurements campaign for BsC at 590 MHz in different indoor radio propagation environments is conducted, where the link budget results are seen as promising for healthcare networks.
In \cite{thomas2012modulated}, an ultra-low-power biomedical telemetry system that exploits near- and far-field propagation through modulated BsCs is proposed. An example of far-field telemetry for EMG/neural signals of flying dragonflies through multiple channels by a device of mass 38mg, at a range of 5m, and a data rate of 5 Mbps is demonstrated. Another example of near-field telemetry for sensors implants in mice with an implant depth of 6cm and data rate of 30 Mbps is shown. The animal-side device power cost for dragonfly and mice examples is claimed to be 4.9 pJ/b and 16.4 pJ/b, respectively.
In \cite{zhang2016enabling}, a taxonomy of applications of BsC in healthcare systems is presented, viz: patients' registration, monitoring and evolution of body conditions, routine health monitoring, patients' lab data analysis, to name a few.
The impact of variations in different network parameters on the performance of BsC based healthcare systems is presented, where the link-budget based analysis established BsC a promising solution for indoor healthcare applications.
In \cite{jameel2019applications}, applications of BsCs for healthcare networks are discussed.
There exist various promising platforms which can be considered as competitor platforms of BsC, e.g., Zigbee ( IEEE 802.15.4-based specification), Bluetooth 4.0, Adaptive network topology (ANT), RuBee (IEEE standard 1902.1), Z-Wave, and Insteon. A comparison of these platforms with BsCs for healthcare applications context is presented in \cite{zhang2016enabling,jameel2019applications}. The summary of provided maximum achievable data rate and coverage-distance of different competitor platforms is illustrated in Fig. \ref{Fig_technologies_bsc_comp}. It can be observed that BsC can be seen as a promising platform which provides a good trade-off between coverage-area and data rate performance.

\paragraph{Decision Fusion}

BsC based decision fusion is another potential application of BsC. Decision fusion is a type of data fusion in which the decisions from multiple classifiers are combined to make a common decision about the activity under investigation.
In \cite{8493480}, a collaborative weighted data fusion scheme based on evidence theory for distributed target classification in IoT networks is proposed.
In \cite{8305509}, blind hard decision fusion rules are proposed which are based on mean-value interpretation of secondary user signal characteristics for resource-constrained distributed networks.
In \cite{7331232}, an energy-efficient and reliable decision exchange for intelligent spectrum sensing and sharing in industrial IoT networks is proposed.
In \cite{6327686}, the optimality of the received-energy test is demonstrated for decision fusion over non-coherent diversity multiple access Rayleigh fading channels with conditionally mutually independent and identically distributed sensor decisions.
In \cite{ciuonzodecision}, single antenna backscatter devices perform decision fusion in a multiple-access scenario over fading wireless channels with multiple-antennas employed at the fusion-center reader. A set of fusion rules for the fusion-center reader are derived and computer simulations based investigations are conducted.

\paragraph{D2D and Cooperative Communications}

Device-to-device (D2D) communications is believed to have a strong potential in offering various types of assistance to the central cellular network, e.g., coverage extension, spectral efficiency, etc.
D2D communication can be seen as an active application scenario for BsC. In \cite{8746225}, a cooperative communication application for BsC assisted relaying is studied.  A wirelessly powered two user communication network is considered, in which a hybrid AP broadcasts wireless energy. The user with a stronger channel helps the user with the weaker channel by harvesting energy and receiving the information from the backscatter channel in return during the wireless energy transfer phase. The multi-objective optimization problem of time-slot and power allocation for joint energy and information transmissions is also investigated. In joint relaying of information and energy, BsC can make a robust application for overall enhancement of energy and spectral efficiency in D2D and cooperative communication context.

\paragraph{UAV-Assisted BsC}

UAVs are believed to play a vital role in future wireless networks to meet the explosive demands of high network capacity, energy efficiency, and data rate.
The integration of BsC with UAVs-assisted networks is an interesting proposal for efficient utilization of network resources in various futuristic application scenarios (e.g., smart cities, remote sensing and coverage, etc). In \cite{3414045}, performance analysis of UAV-enabled multi-node BsC network is conducted by modeling with wireless channels as $\kappa-\mu$ shadowed wireless fading channels. In \cite{9222571}, a UAV powered BsC network is proposed. The ground backscatter devices are illuminated through ground carrier sources, while the devices backscatter the signal with modulated information towards a flying UAV in a TDMA fashion. Subject to throughput and other important system constraints, the joint optimization of backscatter devices scheduling, backscatter reflection coefficient, UAV trajectory, and carrier sources transmit power is performed to counter the critical issue of overall energy efficiency. A communicate-while-fly method is proposed, which is shown to provide significant gain in overall energy efficiency compared to the benchmark hover-and-fly method.
In \cite{8781883}, energy efficiency optimization aspects of UAV-assisted BsCs are studied in the context of a remote sensing application. In \cite{9238842}, a UAV-assisted wireless powered IoT network is studied. The UAV performs both sensory data collection and energy signal broadcasting operations for IoT devices. ML for UAVs flight control and trajectory planning is applied.

\paragraph{Cognitive BsC Networks}

Cognitive radio refers to the radio systems in which any available underutilized radio channels are intelligently and adaptively sensed and utilized by the secondary users (low-priority) to improve the overall radio operating behavior and spectral efficiency.
Cognitive radio is regarded as one of the potential technologies to resolve the spectrum congestion in future massively connected networks. Employing BsC to support cognitive networks is an interesting research proposal.
In \cite{8738811}, a BsC-assisted RF-powered cognitive network is investigated. The selection of APs and services in the network is studied by formulating an evolutionary game. Based on the requirements of the use, an algorithm for performing the secondary transmissions by acting as players and adjusting the selections of APs is proposed.
In \cite{8700623}, a full-duplex ambient BsC links based cognitive network is studied. In the considered setup, the primary cellular network is underlaid by ambient BsC network, where the APs can perform primary transmission full-duplexed with the backscatter receiving. An iterative method exploiting the block coordinated decent is proposed for throughput maximization, which conducts the tweaking of reflection-coefficient, transmit-power, and time-schedule parameters. Moreover, the time-scheduling and the joint reflection-coefficient and transmit-power adjustment problems, convex and a sequence of convex optimization problems are formulated, respectively. The proposed system setting is shown to significantly increase the system throughput with a fast convergence rate.

\begin{figure*}[t]
  \centering
  \includegraphics[width=\textwidth]{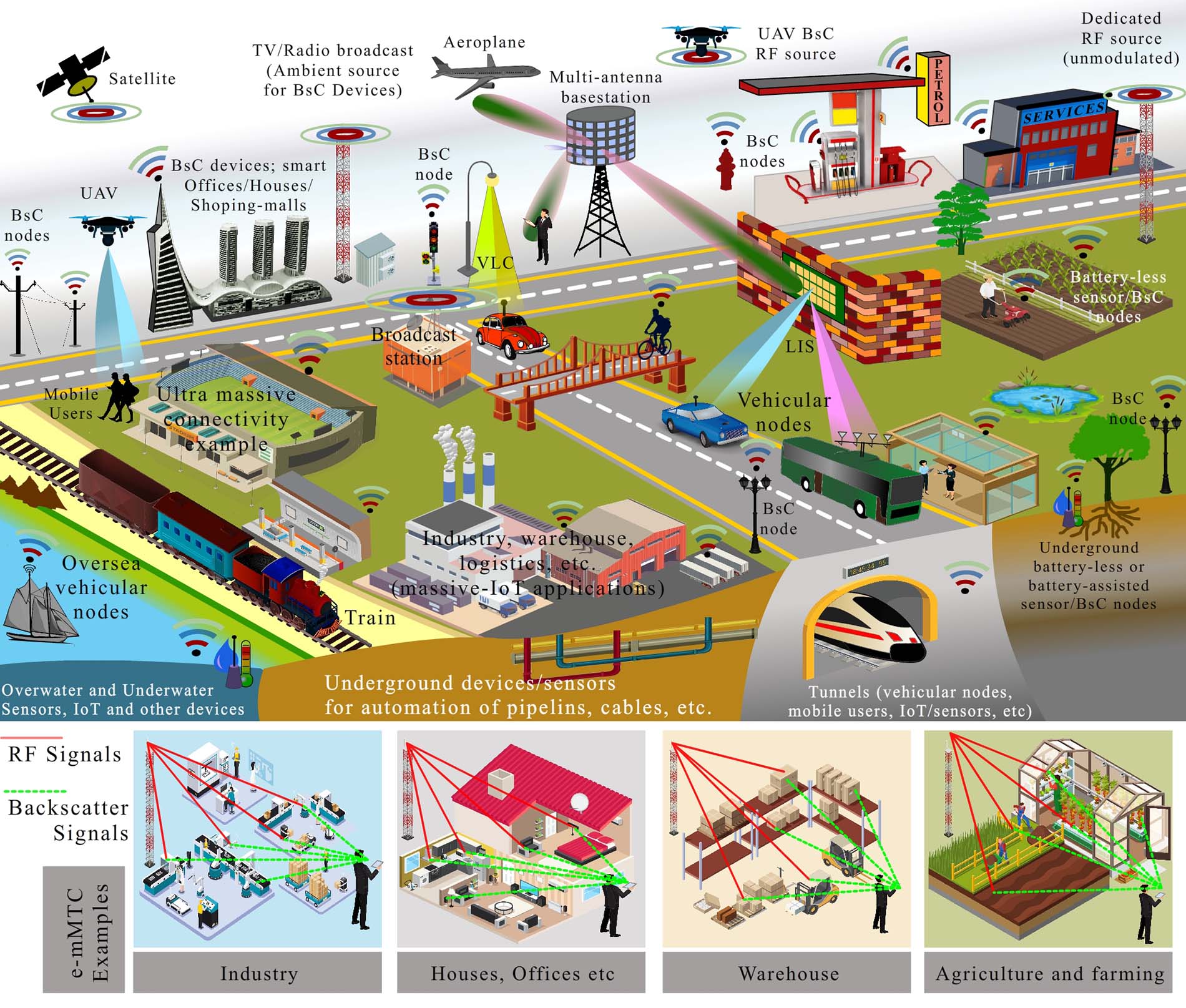}
  \caption{Applications and use-cases of the proposed framework for ultra-massive connectivity in B5G/6G wireless networks.}
  \label{Fig_framework}
\end{figure*}


\begin{table*}[t]
  \centering
  \caption{The scope of non-coherent communications, BsC, non-coherent BsC, and LIS for performance optimization, services delivery, and different types of device nodes in B5G and 6G communication networks.}
    \begin{tabular}{|p{12em}|p{15em}|c|c|c|}
    \hline
    \textbf{6G Networks: Requirements Domains} &
      \textbf{Parameters} &
      \multicolumn{3}{c|}{\multirow{2}{25em}{\centering\textbf{Scope of Reviewed Technologies} \\ \textbf{(categorized as Primary, Partial, or Marginal)}}}
      \\
\cline{3-5} &
      \multicolumn{1}{c|}{} &
      \textbf{Non-Coherent} &
      \textbf{BsC} &
      \textbf{Non-Coherent BsC}
      \bigstrut\\
    \hline
    \multicolumn{1}{|c|}{\multirow{3}[6]{*}{\textbf{Optimization}}} &
      \textbf{Energy} &
      Marginal &
      Primary &
      Primary
      \bigstrut\\
\cline{2-5}     &
      \textbf{Spectrum} &
      Marginal &
      Primary &
      Primary
      \bigstrut\\
\cline{2-5}     &
      \textbf{Cost} &
      Primary &
      Primary &
      Marginal
      \bigstrut\\
    \hline
    \multicolumn{1}{|c|}{\multirow{3}[6]{*}{\textbf{Service types}}} &
      \textbf{e-mMTC} &
      Primary &
      Primary &
      Primary
      \bigstrut\\
\cline{2-5}     &
      \textbf{e-URLLC} &
      Marginal &
      Marginal &
      Marginal
      \bigstrut\\
\cline{2-5}     &
      \textbf{se-MBB} &
      Marginal &
      Partial &
      Partial
      \bigstrut\\
    \hline
    \multicolumn{1}{|c|}{\multirow{4}[8]{*}{\textbf{Devices}}} &
      \textbf{Smart-phones, Laptops, computers, etc} &
      Partial &
      Partial &
      Partial
      \bigstrut\\
\cline{2-5}     &
      \textbf{Sensors, IoT devices, etc} &
      Primary &
      Primary &
      Primary
      \bigstrut\\
\cline{2-5}     &
      \textbf{Drones (UAVs, HAP, LAP, etc)} &
      Partial &
      Primary &
      Primary
      \bigstrut\\
\cline{2-5}     &
      \textbf{Body implant, skin patch, textile integration, human-robot interface devices, etc} &
      Primary &
      Primary &
      Primary
      \bigstrut\\
    \hline
    \end{tabular}%
  \label{table_Opt_Ser_dev}%
\end{table*}%

\section{Proposed Framework, Potential Enablers, and Open Challenges}

Ensuring successful progression of 5G \emph{service types} (i.e., mMTC, URLLC, and e-MBB) towards enhanced services (e.g., e-mMTC, e-URLLC, and se-MBB) in 6G wireless networks is an emerging critical research challenge. The overall connectivity-density of $10^7$ devices/km$^2$ is projected in 6G networks with a requirement of 1000$\times$ gain in volumetric energy-spectral efficiency (bps/Hz/m$^3$/Joule) \cite{You2020towards}.
The anticipated ultra-massive connection-density, traffic-capacity, spectral-efficiency, cost-efficiency, and energy-efficiency requirements to deliver different 6G service types require extensive advancements in the communication framework. To support this, new technological revolutions along with the evolution of existing technologies is also required.
\\ \\
A revolution in \emph{devices-types} in the B5G era is also contemplated with the advent of 5G services. For example, new device types may include robot-human interaction, body implants, skin patch, textile integrated devices, etc. From the experiences learned from 5G development, it can be established that not a single standalone technology can promise the delivery of all the anticipated dynamic services for diverse types of devices in future wireless networks, and the resolution lies in the appropriate amalgamation of multiple technologies. The battery-less operational capability with only about sub-milliWatt of power consumption by the BsC devices makes it a strong candidate for the provision of connectivity to a massive number of devices (passive) in the future networks. The critical requirement of acquiring precise channel knowledge and phase estimate at the network nodes in such ultra-massively connected future networks, especially for less-privileged devices (in terms of computation, energy, etc.), will be a challenging issue. Since non-coherent detection schemes do not require phase or channel state information, they can be considered as a potential enabler in such scenarios. Furthermore, non-coherent communication and BsC together can serve to deliver the \emph{optimization} of resources (energy, spectrum, etc.), enhancement in the delivery of different \emph{service types}, and efficient device design for different \emph{device types}. The scope of BsC and non-coherent communications for these aspects are summarized in Table \ref{table_Opt_Ser_dev}.
\\ \ \\
To deliver the anticipated stringent requirements of 6G e-mMTC services, a non-coherent communication and BsC technologies based framework is proposed in this section. As discussed, non-coherent communication methods provide the advantages of hardware simplicity and cost efficiency, which can be combined with the advantages of energy-, spectral-, and cost-efficiency offered by BsC to constitute an operative framework for 6G e-mMTC services.
Fig. \ref{Fig_framework} illustrates a contemplative scenario of envisioned future wireless networks. The notable application scenarios and use-cases of 6G e-mMTC services, where the proposed framework can take a primary role, are highlighted in the figure, e.g., industries, homes, offices, public services, warehouses, parking spaces, and agricultural sites automation --  to name a few. Various configurations of BsC, wireless sensor networks (WSN), massive-IoT, VLC, LIS, multi-antenna systems, UAVs-assisted networks, and airborne internet applications are also highlighted in the figure. The scope of the proposed framework for healthcare services is demonstrated in Fig. \ref{Fig_BAN}. The proposed framework, associated open research topics, and future recommendations are discussed in the following subsections.

\subsection{Multi-antenna Systems}

In order to achieve the much anticipated gains in spectral and energy efficiency for 5G networks, mMIMO \cite{5595728} is one of the potential enabling technologies. By virtue of employing a very large number of antennas at the BS, mMIMO provides orders of magnitude increase in the spectral and energy efficiency by incorporating horizontal and vertical beamforming to serve several users simultaneously using the same time/frequency resources \cite{6375940}. With the help of massive number of antennas, the mMIMO BS focuses energy for the desired user only by producing sharp beams, thus, reducing the transmit power and minimizing the interference from undesired users. The large-scale multi-antenna systems has a recognized role in providing the capability and support for all the three basic types of 5G services (i.e., eMBB, URLLC and mMTC). The multiplexing of multiple users can help in achieving high spectral efficiency for the eMBB services. Channel hardening through increase in the number of antennas helps in reducing the small-scale fading severity making the links more reliable for achieving low-latency for URLLC services. Directed radiations help in extending the coverage and making the communication power efficient to support mMTC services.


In frequency division duplex (FDD) based mMIMO systems, both up- and down-link operate at distinct frequencies and therefore both the channels need to be estimated.
In TDD based mMIMO systems, owing to time reciprocity property of the channels, TDD operation helps in relaxing the burden of estimation task as limited to only uplink channels. Therefore, it relieves the user-side from the task of channel estimation. Moreover, it also eliminates the dependency of number of channels to be estimated from the number of antennas deployed at the BS. Along with these advantages offered by the TDD operation, time available for uplink data, downlink data, and pilot transmission gets limited by the channel coherence time. This limited time further imposes reduced number of available orthogonal pilots causing pilot contamination leading to inaccurate estimate of the CSI \cite{mansoor2017massive}. In high mobility conditions, further reduced channel coherence time further elevates the severity of pilot contamination problem. In such scenarios, unavailability of accurate CSI estimate makes the non-coherent signal processing methods as a suitable choice.

Due to asymptotic feature of large-scale multi-antenna receiver design, noise hardening makes the noise behaviour as predictable. Alongside, for such multi-antenna systems with large number of antenna elements, the channel hardening makes the system robust to inaccurate CSI. Coherent signal processing for multi-antenna receiver can be applied for low-mobility and high SNR conditions, while non-coherent signal processing is suitable for high-mobility and low SNR conditions.

\emph{Multi-antenna splitting receiver}: By utilizing the benefits of joint coherent and non-coherent processing as depicted in \cite{Jointcoherent2017}, it is an important future research direction to investigate multi-antenna splitting receiver for mMIMO systems. Furthermore, future research in this direction should investigate suitable constellation design and coding methods to facilitate the implementation of splitting receiver in future B5G non-coherent wireless systems.

Energy-efficient communications along with reducing the network operational cost also helps in achieving the goal of minimal carbon emissions. To this end, Massive MIMO is regarded as one of the prime 5G technologies which contribute to achieve higher energy efficiency. The anticipated more aggressive throughput and capacity demands in B5G era will require a matching increase in energy efficiency of the communication techniques, i.e., ensuring the delivery of high capacity and throughput demands with a reduced transmissions power level.
Multi-antenna systems along with BsC and LIS have a strong potential to meet these energy-efficiency demands. For example, directed energy radiations can be leveraged through mMIMO systems to power the massive number of passive (backscatter) devices while LISs can create favourable and energy-efficient propagation conditions.


\subsection{Localization and Tracking}

The awareness of nodes location in a wireless communication network has received a great interest in various application scenarios in order to improve the performance of wireless links and enable new types of services. Some examples in this regard include navigation services for robots and channel capacity enhancement in mmWave communications by steering the beams and nulls towards the desired and interference directions, respectively. Furthermore, various new unique demands and applications of tracking are also emerging in which accurate localization is of prime importance, such as tracking of wallet, keys, files, tools and pills box.

Moreover, appending battery-free tags to such objects and their localization through existing infrastructure is a challenging task and it is an emerging area of research. The distance of tags can be estimated on the basis of received backscattered power (e.g., WiFi AP), and by manipulating phase difference in multi-antenna systems, the AoA can also be estimated. Whereas,  the estimation of power, AoA, and other channel indicators is not a straight forward task. Exploiting the estimated distance and angle information and by simultaneously using multiple serving stations, various localization methods can be employed to estimate the spatial position of tags, e.g., triangulation method, ellipsoidal method, etc. For example, by utilizing the triangulation method in WiFi network setup, a strong potential has been demonstrated through an achieved  median localization error of about $0.92$m and $1.48$m in LoS and nLoS propagation scenarios \cite{kotaru2017localizing}.

Non-coherent localization and radar system design is believed to have a strong potential in reducing the cost and complexity of the system. Non-coherent signal processing, though sub-optimal, only involves the processing of signal envelope and avoids the use of any phase information which in turn has various practical advantages over the coherent processing. One of the prime advantages is the substantial relaxation in the dependence of localization algorithm on the radar's position. The other associated advantages include the reduction in cost and complexity of the system.

\begin{figure}[t]
  \centering
  \includegraphics[width=\columnwidth]{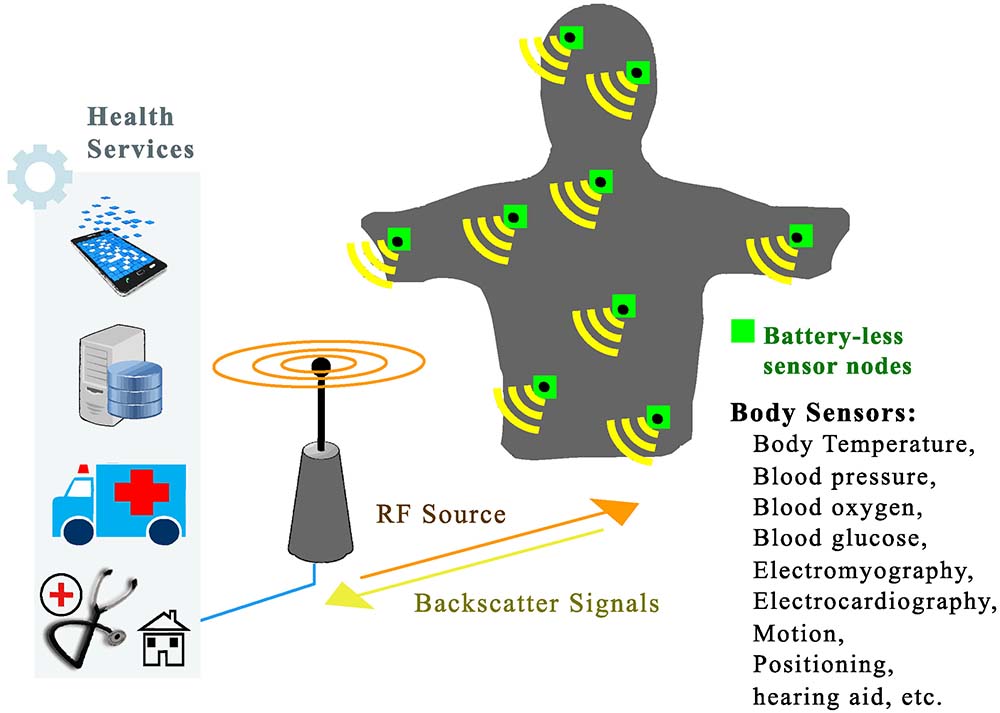}
  \caption{Application of proposed framework for health monitoring and services.}
  \label{Fig_BAN}
\end{figure}

In nLoS conditions (e.g., sensing through the wall), the detection and localization of objects (especially, multiple objects) is another prime area of research, where non-coherent signal processing has been throughly studied in the literature. In \cite{4107975}, a trilateration based non-coherent radar system for localization of multiple objects behind a wall is proposed. Three independent measurement points employing monostatic radar units (with various possible observation and inter-element distances settings) are used and  time-of-arrival (no phase coherence required) information obtained at the points is exploited to localize the target object. In \cite{niu2012target}, it has been established that a non-coherent multi-antenna radar with widely spaced antenna elements at both the nodes can achieve a significantly higher localization and tracking accuracy as compared to that of monostatic phased array radar.

\subsection{Learning Provisions and Edge Computing for BsC Devices}

The foundation of intelligent network operations has been laid with some basic provisions of AI operations in 5G networks. ML for delivering services-driven, fully intelligent, and self-configurable networks is rapidly becoming widespread. ML at the physical layer of the communication systems can  support various exciting applications and operations \cite{8054694} spanning from the estimation of parameters to the optimization of resources to meet the increasingly stringent performance requirements of future networks. Moreover, extending the provisions of intelligence to IoT devices is also well recognized in the literature, e.g.,  \cite{8291121}, for the realization of future smart cities with people-centric network services. However, the learning aspects for BsC devices and systems are still not well comprehended in the literature. A few recent research articles have discussed the need for leveraging learning capabilities to BsC devices \cite{jameel2020machine}.

The BsC devices are limited in available energy and computational resources, which makes them less-suitable to execute ML algorithms. However, now with the advent of edge computing in 5G, the computational tasks of such less privileged devices can be offloaded to the edge platforms and intelligence provisions can be extended. MEC is a promising 5G technology that helps to meet the challenging demands of latency-critical applications through the provision of computing and storage facilities deployed within a close vicinity of the devices/users (e.g. at the BS or APs).
The provision of data processing facilities at the network edge offers various advantages such as reduced the end-to-end latency and energy consumption for the devices/users. One promising technology enabler for MEC is collaborative edge-cloud processing \cite{LivedataIoT}, which can benefit from both the cloud computing (in terms of massive storage, huge processing capacity and global view) and edge computing (in terms of supporting applications demanding low-latency, high mobility and high QoS) paradigms. In \cite{9024401}, a reinforcement learning-based framework for BsC networks is proposed. A multicluster BsC model for short-range data sharing along with an intelligent power allocation algorithm for BsC devices for minimization of overall network interference are also proposed in \cite{9024401}. Considering the primary benefits of both ML and BsC in ensuring high efficiency of network resources (e.g., energy and spectrum etc) consumption in future massive-IoT networks, it is vital to comprehensively define and characterize the intelligence capability for BsC devices. We envisage that with a future evolution in edge computing platforms in the era beyond 5G, research work for the advancements in consumer electronics and IoT devices with added provisions of intelligence will be at the forefront.

\subsection{Biomedical Applications}
The globally increasing population, shortage of healthcare facilities, high cost of healthcare services, and rising chronic diseases demand a thorough revolutionization of healthcare practices. Recently, IoT-assisted solutions to these challenges have received notable attention. IoT-devices, which are tiny-sized and rely only on passive RF components, can be considered as best suitable for deploying in different biomedical applications, as there is no significant disadvantage on the subjects or organs being monitored. In this context, the role of IoT-assisted technologies has already emerged as dominant as cutting edge transformative frameworks to healthcare services.

\emph{Molecular communications} is an emerging branch of communications for nano-scale applications in which information is transmitted by embedding information data into the physical and chemical features of molecular messengers \cite{Li2016localconvexity}. The nature of molecular communication channels is more stochastic as compared to other wireless channels since the CSI of a molecular system relies on a number of parameters including the flow velocity, distance between receiver and transmitter, diffusion coefficient of information molecules and enzyme concentration \cite{Jamali2018detection,Li2016localconvexity}. Due to this, it is crucial to repeatedly track the CSI variations in the molecular channels, and coherent detection techniques are suitable only for the cases where the channel remains almost completely invariant  between two consecutive training intervals, do not seem attractive for molecular communications. Thus, it is highly important to investigate novel low-complexity non-coherent detection techniques for molecular communications, which do not need the knowledge of CSI and provide negligible inter-symbol interference.

BsC devices being completely battery-less are considered as highly suitable for different in-body communication applications \cite{vasisht2018body}. Illustration of BsC based health monitoring and services is shown in Fig. \ref{Fig_BAN}. However, there are various challenges associated with the application of BsC for in-body communications, e.g., the interference signals received as scattered from skin and other body tissues at a magnitude stronger than the desired backscattered information signal. Furthermore, as discussed before, the in-body electrical and propagation characteristics of signals are significantly different from those which are (well-established) for air and other mediums. Consequently, coherent  modulation  and  detection schemes for in-body BsC may not be suitable applications. Therefore, it is envisaged that the joint exploitation of the concepts of battery-less and CSI-free backscatter non-coherent communications for embedding, transmitting/backscattering, and decoding information through molecular signals can be seen as a potential enabler for various micro- and nano-scale in-body communication applications in 6G and beyond wireless networks.









\begin{figure}[t]
  \centering
  \includegraphics[width=\columnwidth]{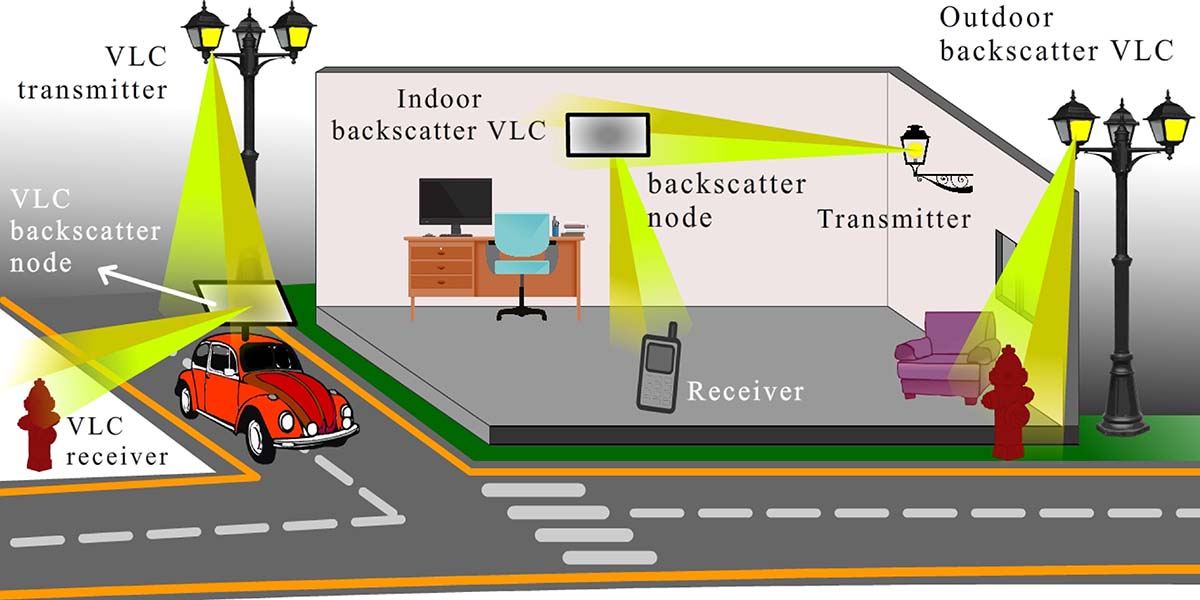}
  \caption{Example of indoor and outdoor VLC BsCs (bistatic).}
  \label{Fig_VLC}
\end{figure}

\subsection{Visible Light Communications (VLC)}
To meet the capacity demands of the future wireless networks, Visible Light Communications (VLCs) has recently received a great deal of interest from the research community. Example of indoor and outdoor VLC communication applications are illustrated in Fig. \ref{Fig_VLC}. A light emitting source along with the provision of light to the users can also extend the provisions of information exchange over the visible spectrum. This can help in relaxing the scarcity of radio spectrum resources. Alongside, it also offers the advantages of high information rate, no health hazards, and low power requirements.

Radio BsC has become a viable concept for enabling wireless connectivity to battery-less nodes (e.g., for IoT devices, etc). Similarly, backscattering of visible light has also been demonstrated as a potential foundation for enabling battery-less communications in the visible spectrum.
In \cite{7752834}, visible light BsC (VLBsC) is proposed as a potential wireless access technology for enabling connectivity of IoT devices. In \cite{xu2017passivevlc}, a practical VLBsC system is proposed for IoT applications. The proposed passive VLBsC system is implemented and evaluated, where it is demonstrated to achieve up to 1 Kbps of speed. Furthermore, it is established that the VLBsC is robust to ambient light conditions and is flexible to operate in different tag orientations. In the light of these demonstrated strong capabilities of VLBsC system, it can be seen as a potential enabler for massive-IoT applications in B5G and 6G wireless networks.

The fluctuations in the amplitude of the light signal with respect to the information signal at a rapid fluctuation rate enable the transmission of information through visible light spectrum without causing any observable light fluctuation to a human eye. Non-coherent detection of information from such amplitude modulated light signals is the a natural and practice detection solution. In this regard, VLC is also refereed to as non-coherent form of communications \cite{6685754}. Also, Optical discrete multi-tone modulation is regarded as a state-of-the-art non-coherent modulation scheme for VLC \cite{6747983}. The modulation concept for a multiple transmitter VLC environment is studied in \cite{6747983}.

The aforementioned discussion asserts that non-coherent BsVLC can be viewed as a potential enabler for massive-IoT in 6G and beyond wireless networks.

\subsection{Acoustic Underwater Communications}

Enabling BsC for underwater communication networks is an important research topic to eliminate the need of battery for underwater nodes (e.g., sensors). The underwater applications of BsC include long-term battery free WSNs for sensing sea conditions, for investigating habitat and migration patterns of marine-animals, military applications, etc. Typically, Underwater communication relies on acoustic communications (e.g., SONAR for submarines), which is mainly due to highly unfavorable propagation conditions for radio signals in the sea water. The need for battery-less wireless communication and limitation of relying on acoustic communications necessitates the need for exploring Acoustic BsC (AcBsC).

In \cite{jang2019underwater}, piezoelectric effect of materials is exploited to enable AcBsC. Piezoelectric ability of certain solid materials is usually utilized to produce electricity in response to an applied mechanical stress. This idea is exploited by utilizing the pressure of acoustic waves to induce strain on solid piezoelectric material to transform it into voltage. A throughput of 3kbps for a range of up to 10m is demonstrated for the designed AcBsC systems. Considering the potential and application significance of AcBsC, it can be regarded as an important future research direction.

The fast time-variability in the underwater acoustic channels statistics, which makes the robust phase tracking a difficult task. Therefore, non-coherent schemes for such channels are regarded as suitable choice. In \cite{8866887}, a non-coherent scheme for underwater acoustic communications is studied to improve energy-efficiency and link-reliability by keep the bandwidth consumption same. In \cite{wu2014signal}, channel characteristics for underwater acoustic non-coherent communications (in adverse conditions) is modeled as a phase-random Rayleigh fading channel and its capacity trend is determined. To achieve the channel capacity trend, the concatenated code of the nonbinary Low-density parity-check code and the constant weight code is suggested for such non-coherent channels.
Experimental investigations for deep-oceanic (5km) and shallow-lake (3km, with 50ms delay-spread) channels are also conducted over a 6kHz bandwidth. A promising performance of proposed non-coherent scheme is observed for SNR conditions as low as 2dB. OOK and MFSK and are two typical non-coherent modulation schemes. In \cite{yao2020efficient}, OOK for seafloor acoustic observation networks is studied, where based on the conducted simulations and experimental results, non-coherent acoustic communications scheme is observed superior over the conventional coherent communication schemes.

Considering the aforementioned promising features of non-coherent and BsC techniques for underwater communications, the joint application of both the techniques can help in materializing a holistic approach for enabling energy-efficient underwater communication networks.


\begin{figure}[t]
  \centering
  \includegraphics[width=\columnwidth]
  {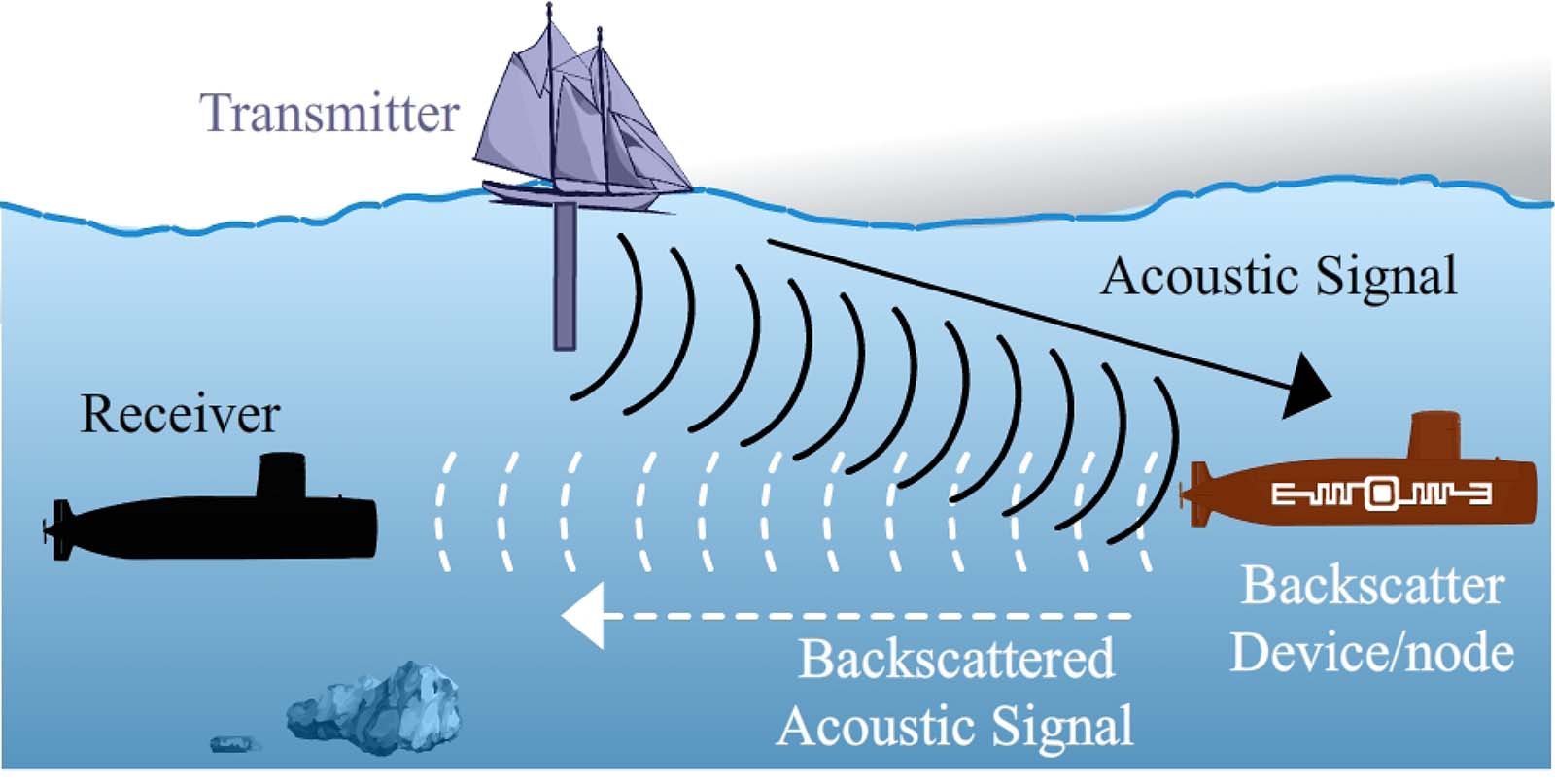}
  \caption{Underwater acoustics bistatic BsC example.}\label{Fig_acoustic_BsC}
\end{figure}

\subsection{UAVs-assisted Communications}

UAVs-assisted communications has received an overwhelming research attention in the recent years. Mobility of vehicular nodes in a dense scattering environment imposes Doppler spread which reduces the coherence time and causes difficulties in obtaining accurate estimate of CSI through pilot-based estimation techniques. Error in the estimate of such dynamic (fast time-varying) channels further leads to eroded log-likelihood ratio function. Consequently, in high mobility scenarios, the bandwidth and power efficiency of non-coherent communication schemes is seen better than that of coherent communication schemes \cite{8922634}. Therefore, coherent and non-coherent communications are recognized as suitable choice for low and high Doppler conditions, respectively. High mobility scenarios are also encountered in various other communication applications, e.g., vehicle-to-vehicle communications and airborne communications.

The notable applications of UAVs in communications include its role as a relay, as a BS, etc. UAVs for relaying of information from Tags to reader through backscattering of the signals is an attractive proposal for future low-power communication applications \cite{memon2019backscatter}. Moreover, BsC-enabled UAVs for assisting in various communication tasks (e.g., supplying ambient power to remote sensors) have been discussed in \cite{zeng2016wireless}. Furthermore, backscattering-enabled UAV nodes can also assist in creating favourable channel conditions for the user nodes which are under shadowing conditions. Exploration of non-coherent detection schemes together with BsC in  vehicular communication applications can be regarded as a potential research direction in designing suitable air-interfaces for 6G wireless networks.

\subsection{Backscatter-Device-to-Backscatter-Device (B2B) Communications}

The short-range of backscatter links, the volatility of RF environment, and weak link reliability are among the major limiting factors which restrict the deployment of BsC at a large scale. Direct B2B devices based networks (also usually referred as Tag-to-Tag (T2T) networks) can help to overcome these limitations; however, designing efficient and intelligent routing protocols for such networks is a crucial research challenge. Also, enabling communication between the backscatter devices opens the opportunity to perform different processing tasks collaboratively.
In \cite{8253954}, a robust cooperative routing protocol for B2B links based WSN is proposed. A collaborative framework to reduce the path failure probability along with a new routing quantifier for measuring and ensuring robustness in routing is proposed.
Joint optimization of routing stability, throughput performance, and the end-to-end delay has the potential to enable the scalability of such B2B networks.  In \cite{8737551}, a discrete components-based BsC transceiver design for the multi-hop B2B network is proposed. The muti-hopping is observed to be promising in extending the coverage of B2B networks and reducing the impact of dead spots. In \cite{8364537}, an innovative device architecture to support multi-hop B2B routing is proposed where a four-hop link is seen to be capable of communicating over 12m distance. Interference between multiple backscatter transmissions in such B2B networks is another attention-seeking factor in designing network architecture and routing protocols. In \cite{8170764}, a non-linear optimization based multi-hop routing protocol for B2B networks is proposed, which accounts for several network parameters to enhance the number of concurrent transmissions with minimized network interference.

In B2B networks, the unique phase cancellation problem is one of the prime performance-limiting factors. When multiple backscatter-devices directly communicate with each other, the on the air modulated and backscattered signal gets superimposed with the externally excited signal at the receiver device. These superimposed signals being different in phases (because of difference in delay) causes the ubiquitous self-interference problem, which further leads to significant performance degradation of non-coherent demodulators (e.g., envelope detector). To this end, various backscatter modulation techniques have been proposed in the literature, see e.g., multiphase backscattering technique in \cite{7419631}. Nevertheless, there is a scope to conduct research work for finding an optimal solution to this problem.

\subsection{LIS-Assisted Communications}

LISs (or intelligent reflective surfaces) assisted communications is another promising reflective radio communication technology besides BsCs. Employing LISs in a wireless communication environment provides an extra degree-of-freedom to the communicating nodes through leveraging of control over a few of the multipath. This makes the wireless channel behavior partially manipulatable, which was conventionally considered completely unmanipulable. The evolution in wireless networks is steering towards the paradigm of designing everything as reconfigurable through software definition. The applications of such reconfigurability include software-defined (SD) fluid antennas, RLISs, etc. The concept of SD-RLISs helps in controlling the behavior of radio propagation channels and thus achieving the optimal trade-off between different performance indicators, e.g., energy efficiency, coverage, reliability, data rate etc. This technique exploits the low-cost passive reflectors (EM materials) as coated over the surface of different environmental objects implemented over a large-scale. These surfaces coated with real-time programmable integrated electronic circuits manipulate the phase and/or amplitude shifts associated with each occurring physical phenomenon (e.g., reflection, refraction, scattering etc) during the radio propagation in order to achieve an overall desirable response from the wireless channel. The RLISs is also an attractive technology for amplifying-and-forwarding of the incoming signals through manipulations of suitable phase-shifts and without involving any power amplifiers. Moreover, through the assistance of RLISs in joint application scenarios with other attractive enabling technologies (e.g., cell-free massive-MIMO, mmWave, tiny-cells, etc), the interference aggravated from the ultra-dense deployment of APs and user nodes can also be effectively minimized. This new revolutionary technology has immense potential in achieving many-fold performance gains along with enabling massive connectivity in B5G communication networks.

In \cite{9149709}, a MISO system assisted by an LIS with power budget constraint is investigated. The LIS works in phases, i.e., it first harvests RF energy received from the beam of serving BS and then utilizes it to perform its phase manipulation and reflecting operations.
In \cite{huang2018large}, an LIS-based design for optimization of energy and spectral efficiency through tweaking of transmit power and coefficients of reflecting elements for multi-user multi-antenna communications is proposed. Gradient descent and fractional programming based methods are offered for optimizing LIS phase-shifts and transmit power, respectively. The conducted simulation analysis indicates that a gain of 300\% in energy-efficiency can be achieved through the exploitation of LISs in realistic outdoor environments compared to conventional relay-assisted systems. In \cite{Subrt2012IET}, the concept of an indoor intelligent wall is introduced. The wall is an autonomous part of a cognitive network with a ML-enabled cognitive engine. The surface of the wall is designed as an active frequency-selective surface and it is equipped with low-cost sensors. Such a system is seen to influence the overall system performance by intelligently responding to the immediate demands of the indoor wireless networks and controlling the radio coverage. The applications of such indoor intelligent walls include conference rooms, class room, corridors, etc.
A dynamic SM mechanism in the context of LIS-assisted communications in BsC environment is investigated in \cite{9171362}, where 3D selective radiation towards the receiver position is studied. Moreover, a time-varying LIS concept relying on delay adjustments between different elements of the LIS is also presented.
A review of RLIS technology is provided in \cite{8910627}, where its applications in wireless communications along with hardware architectures and signal model are discussed. A comparative analysis to demonstrate the advantages offered by RLISs over its counterparts is also conducted. Moreover, the critical design and development challenges in the implementation of RLISs are indicated. Numerical analysis for gauging the performance of the proposed RLIS system in creating hotspots and interference-free zones is conducted. It is demonstrated that the RLIS scheme, compared to the conventional methods, can substantially reduce the interference power by solely manipulating the RLISs phase-shifts. Moreover, it is also demonstrated that co-channel interference can also be effectively suppressed, primarily when the transmit/receive antennas are sufficiently large. A large number of antennas provide an additional degree of amplitude control, which further helps the RLISs to impose interference cancellation signals at the users and to achieve virtual interference-free zones.
In \cite{9122596}, a comprehensive survey of the recent advances on LIS-assisted wireless communications is conducted, and the design aspects of LISs are also thoroughly reviewed.

\subsection{Security and Privacy Considerations}
Ensuring secure wireless communications in densely connected wireless networks is one of today's most critical research challenges. Furthermore, in the future networks, where everything (that can benefit from) will be connected to the communication networks, securing the wireless links and ensuring the privacy (users data, network data, etc.) will become more challenging. When seeing the challenges in the provision of security and privacy demands together with the demands of massive connectivity (network capacity), power efficiency, and throughput enhancement in future networks, the attentions of the research community for designing of a comprehensive overall solution are required. Non-coherent schemes, together with BsC, as discussed earlier, has a strong potential in optimizing the power, capacity, cost, and throughput demands of the future networks. However, these potential schemes require comprehensive research to ensure the provisions of security and privacy in such networks.
The use of simple encoding and modulation methods in non-coherent BsC systems makes them vulnerable to security attacks.
The potential of physical layer security, quantum, and blockchain technologies for extending security and privacy provisions in such massively connected complex networks (e.g., IoT) are regarded as highly promising. The joint scope of these technologies together with non-coherent BsC systems for devising an IoT framework can be identified as a potential research direction for the future.

Security is conventionally ensured through computational security methods, which are deployed at higher layers of the protocol stack in wireless communication systems. Inherently, these methods rely on the assumption that the eavesdroppers are limited in their computational resources to decipher encrypted messages. However, the rapidly increasing computational capability of the systems offered through the advent of new technologies (e.g., Quantum technologies), this assumption of limited computational capability of eavesdroppers may not stand long. Furthermore, ensuring the distribution of the secret keys between the legitimate nodes in a massively connected wireless network is another challenging aspect of such conventional security provision methods. Distribution of secret keys at a very large scale (e.g., in massive-IoT application scenario) will need substantial infrastructural support.

In light of the aforementioned concerns, physical layer security (PLS) has emerged as a potential natural solution in which the physical layer characteristics of wireless propagation channels are exploited to generate the secret keys. This eliminates the need for a dedicated infrastructure for the distribution of secret keys among the legitimate nodes, as both the communicating nodes can independently generate the same secret key from the characteristics of the channel between the nodes, and the eavesdropper cannot measure their channel. This PLS concept has been independently employed for both non-coherent and BsC in the literature, see e.g., \cite{gu2020physical} and \cite{6836141}, respectively. However, there is a scope to investigate the potential of PLS solutions jointly for non-coherent BsC systems. Furthermore, quantum technologies are believed to have enormous potential for extending the provisions of security and data privacy in communication systems, which are highlighted in the context of non-coherent BsCs in the following subsection.

\subsection{Quantum Backscatter Communications (QBsC)}

Quantum radars are based on Quantum entanglement concept which are usually used for remote sensing applications \cite{QML_6G_Junaid}. S. Lloyd in 2008 introduced the Quantum Illumination (QI) concept \cite{lloyd2008enhanced}, which is a photonic quantum sensing method for enhancing the detection sensitivity in noisy and lossy environments. A particular target region containing bright thermal-noise can be illuminated through an optical transmitter, from where, the light received in return can be utilized to determine the absence or presence of the target objects (with low-reflectivity) in the region.
In QI concept a quantum illumination transmitter is utilized.
A signal with a plurality of entangled particles can be produced through an entangled quantum particle generator \cite{allen2008radar}. Subsequently, the signals returned from the target region can be processed to characterize the target by using the information of entangled particles.
The propagating quantum states of light can be displaced to generate entanglement for facilitating quantum communications. In \cite{fedorov2016displacement}, it is established that the entanglement generated through the displaced squeezed states remain constant over a wide range of displacement power. Moreover, it is demonstrated that there is no degradation of the squeezing level in the reconstruction of quantum states even for the displacements of strong amplitudes.
Two-mode squeezing is an exciting model of quantum entanglement for applications in quantum communications.
In  \cite{fedorov2018finite}, a theoretical model on the dephasing process of quantum cross-correlations in continuous variable propagating two-mode squeezed microwave states is proposed. Moreover, insight into finite-time entanglement theory and its limits are provided.
In \cite{tan2008quantum}, a performance comparison of the systems using quantum illumination transmitter and coherent-state transmitter is presented. Through the optimum joint measurements on the received light from the target illuminated region and the retained joint spontaneous parametric downconversion idler beam, the quantum illumination transmitter is observed to provide a 6dB gain in the detection error probability exponent compared to the coherent-state transmitter. Moreover, this advantage accrues despite there is an absence of entanglement between the light collected from the illuminated region and retained idler beam.
This QI concept lead to the development of a prototype for quantum radars in 2015 \cite{barzanjeh2015microwave}. Later, the use of QI technology for BsC was proposed in 2017 in \cite{8269081}.

In QBsC, a pair of entangled photons is generated by the transmitter, while only a single photon is transmitted and idler photon is retained at the receiver.
One of the primary research challenges in QBsC, which requires the attention of research community, is the designing of effective and fast generation of entangled photons.
The use of multi-antennas for encountering the problem of slow rate of generation of entangled modes in microwave band is also suggested in \cite{8269081}.
In \cite{lanzagorta2016improving}, a distributed design for synthetically increasing the distinguishable effective entangled modes is proposed.
Detection of single-photon is a vital step in quantum systems (e.g., for sensing, information processing, communications, etc) in both optical and microwave domains. The single-photon detection in the microwave domain is a particularly challenging task, as the energy of quanta in the microwave domain is usually 4 to 5 orders less than the optical domain \cite{inomata2016single}. An impedance-matched single microwave photon detection implementation for the propagation of photons through a waveguide is demonstrated in \cite{inomata2016single}, where the detection efficiency of $0.66\pm0.06$ with a reset time of $\sim400$ns is achieved. In \cite{di2018quantum}, the QI concept based novel QBsC protocol is proposed. The proposed QBsC protocol, using the sum-frequency-generation receiver design, has shown a gain of $6$dB, $6$dB, and $3$dB in the error exponent over its conventional counterparts for PAM, BPSK, and QPSK, respectively. The Quantum BsC (QBsC) concept has a strong potential in increasing the receiver sensitivity beyond the conceive able range of classical BsC systems.


Quantum technologies have recently attracted substantial research interest in redefining the conventional communication methods. Quantum communications, quantum computing assisted conventional communications, and quantum machine learning assisted communications are promising research directions for 6G and beyond 6G eras \cite{QML_6G_Junaid,8540839}.
For provision of security, quantum error correcting codes and quantum key distribution are among the rapidly emerging applications of quantum technologies in communication networks \cite{1668133,botsinis2016quantum}. In this regard, for BsCs,  in \cite{8970171}, a quantum-enhanced BsC system for microwave bands is proposed, which is observed to outperform the classical BsC systems in terms of receiver sensitivity as well as security. In \cite{8010959}, a thorough comparative analysis of quantum technologies assisted coherent and non-coherent communication schemes has been conducted.
In \cite{8692830}, by exploiting quantum error correction techniques, a space-time block code for non-coherent communications has been proposed. The proposed method is observed to achieve comparable or better performance when compared to the coherent and differential non-coherent communication approaches.

As the role of quantum technologies for both the BsC and non-coherent communications has been identified and studied in the recent literature, there is a strong potential in jointly investigating the scope of quantum technologies with BsC and non-coherent communication methods.

\begin{table*}[t]
  \centering
  \caption{Brief summary of Challenges and open research topics related to backscatter and non-coherent communications}
    \begin{tabular}{|p{26em}||p{30em}|}
\hline
    \textbf{Non-Coherent Communications} &
    \textbf{Backscatter Communications (BsCs)}
\bigstrut\\
\hline \hline
    Reducing training overhead (e.g., for pilot decontamination in mMIMO systems, for high-mobility vehicular communications where channel coherence-time is very limited, etc, in co-existence with coherent communications) &
    Multiple access management (accurate resolution of concurrent backscatter signals received from multiple illuminated BsC devices,
    trade-off optimization for energy harvesting, backscattering, original active transmission, and channel utilization).
\bigstrut\\
\hline
    Non-coherent channel capacity characterization and enhancement for mMIMO systems (Spectrally efficient non-coherent communications) &
    Full-duplex for BsCs (with simultaneous active transmission and energy  harvesting)
\bigstrut\\
\hline
    Accurate modeling and characterization of non-coherent channels to understand their fundamental limits (e.g., for different spectrum ranges, propagation environments, mobility conditions, etc) &
    MIMO for BsCs (utilization of multiple antennas for diversity, multiplexing, and/or beamforming in BsCs)
\bigstrut\\
\hline
    Scope of employing learning methods for non-coherent communications (e.g., deep learning for non-coherent receivers)  &
    System design enhancement for improving energy sensitivity level and decreasing system loses
\bigstrut\\
\hline
    Adaptive/intelligent/smart switching between coherent and non-coherent schemes to achieve an optimum performance trade-off  &
    Interference management and suppression (decreasing interference to licensed spectrum, accurate backscattered information detection, etc)
\bigstrut\\
\hline
    Ability for integration with dynamic spectrum sharing schemes (e.g., spectrum sensing with non-coherent transmissions) &
    Communication range and channel capacity characterization and enhancement (e.g., improving multi-hop links, efficient routing in B2B networks etc)
      \bigstrut\\
    \hline
    Characterization of non-coherent, non-orthogonal, and cell-free (distributed mMIMO) communications &
      Addressing security, privacy, and jamming vulnerability in BsCs
\bigstrut\\
\hline
    Devising advanced non-coherent modulation and detection methods (e.g, joint or in dependant exploration of spatial, permutation, differential, index modulation etc)  &
      Localization precision improvement (for logistics, surveillance, and other IoT applications)
\bigstrut\\
\hline
    Performance enhancement for high SNR conditions &
    Provisions of learning ability to backscatter devices
\bigstrut\\
\cline{2-2}
   (for low SNR conditions, performance of non-coherent is usually regarded as better than coherent communications)  &
    BsC for spectrum beyond microwave and RF spectrum (e.g., mmWave, teraHz, visible light, and acoustic BsCs)
\bigstrut\\
\cline{2-2}
     &
    Joint scope of BsCs with intelligent reflective surfaces  (reconfigurable) for energy efficiency improvement
\bigstrut\\
\hline
    \multicolumn{2}{|p{50em}|}{Quantum technologies assisted backscatter and non-coherent communications}
\bigstrut\\
\hline
    \multicolumn{2}{|p{50em}|}{Integration with blockchain (jointly investigating blockchains, non-coherent transceivers, and BsCs for devising massive-IoT framework)}
\bigstrut\\
\hline
    \multicolumn{2}{|p{50em}|}{Backscatter and non-coherent communications for cognitive radio}
\bigstrut\\
\hline
    \end{tabular}%
  \label{tab_challenges_summary}%
\end{table*}%



\subsection{Multiple Access Challenges in BsC}

For enabling multiple users to access the wireless channel, the conventional Orthogonal Multiple Access (OMA) techniques such as TDMA, Frequency Division Multiple Access (FDMA), space division multiple access (SDMA), or Code Division Multiple Access (CDMA) rely on the concept of slicing and diving the time, frequency, space, or code resources, respectively, among the multiple users.

In BsCs, the accurate resolution of concurrent backscatter signals received from multiple illuminated BsC devices necessitates the deployment of appropriate multiple access mechanisms. In SDMA based BsC systems, beam steering with directional antennas can help in illuminating multiple BsC nodes. Directional antennas being expensive and large in size increase the overall system complexity and cost. In FDMA-based BsC systems, the BsC nodes are required to alter the frequency of the backscatter signals (modulated or unmodulated) in order to achieve distinct frequency for each BsC node. This imposes high signal processing burden for frequency mixing and Fourier transform.
In TDMA-based BsC systems, each node transmits its information at a pre-assigned time-slot. However, achieving tight time synchronization in such systems require coordination between the nodes or extra synchronization equipment. In CDMA-based BsC systems, each node utilizes its preassigned unique orthogonal code to transmit their information. However, the near-far problem necessitates an appropriate power control which imposes an high system complexity \cite{liu2019next}.


In the recent years, NOMA has been unfolded as a potential technology to meet the massive connectivity requirement of 5G and B5G/6G wireless networks. NOMA offers significant benefits in terms of increased spectral efficiency, increased number of simultaneously served users, and better QoS due to non-orthogonal sharing of resources (time/frequency ) among users/devices in the power and/or code domain \cite{6692307,6868214}. In contrast to OMA, the fundamental working principle of NOMA is to void the orthogonality constraint for serving multiple users in the same time/frequency/code resource \cite{6868214}. In the downlink, the information of users being served is superimposed by the serving station (e.g., BS), while each user decodes its own data by incorporating SIC.
In NOMA-based BsC systems, the distance-based  power control and SIC enables multiple concurrent communications \cite{8636518}. However, high signal processing and computational power requirements for SIC and distance-based power control increases the system complexity. Nevertheless, there is a vital need to develop efficient multiple access schemes for BsC systems, which exploit the inherent nature of BsC and IoT networks.


\subsection{Other Challenges and Research directions}

In this subsection, some other challenges and potential topics are highlighted.

\subsubsection{Challenges and Research Directions for Non-Coherent Communications}

As highlighted in Table II of Section III, existing works have studied non-coherent communications in different settings such as the characterization of non-coherent channels, non-coherent modulation and detection, non-coherent SIMO/MIMO systems, non-coherent signal processing and Quantum-assisted non-coherent communications. Regarding the  non-coherent channels, it is crucial to understand the fundamental limits of various non-coherent channels and to characterize them as the existing system design and protocols mostly consider the coherent channels. As illustrated in \cite{Zheng2000info}, the capacity gain of non-coherent channel in multi-antenna settings is given by $M'(1-M'/T)$ bps/Hz, with $M'=\min\{M,N,LT/2\}$ as compared to the capacity gain of coherent multi-antenna channels given by $\min\{M,N\}$. In this direction, future research should investigate the characterization of non-coherent channels in various practical settings while considering channel uncertainties and the effect of transceiver impairments.

In the context of non-coherent modulation and detection, non-coherent SM, the combination of SM with the differential modulation, the IM and detection techniques are being investigated in the recent literature. Besides these techniques, another promising approach could be to employ an adaptive method which can switch between non-coherent and coherent SM schemes \cite{xu2019adaptive}. Similarly, in the domain of Non-coherent SIMO/MIMO systems, various non-coherent channels such as non-coherent SIMO, non-coherent MISO, non-coherent MIMO, temporally-correlated non-coherent MIMO Channels and space-time codebook design have received significant attentions in the literature. In this direction, future research may focus on investigating and characterizing the non-coherent channels in mMIMO settings, mmWave systems, TeraHertz systems and dynamic spectrum sharing systems, which are considered important for upcoming 5G and beyond systems.

\subsubsection{TeraHz Communications}
TeraHz bands hold strong potential in meeting the crucial channel capacity requirements of 6G wireless networks. Despite that, the propagation characteristics of such radically high-frequency bands limit their applications to short-range LoS applications, teraHz communications have made various interesting use-cases such as inter-chip communications, THID, etc. Research work to extend the usable radio spectrum beyond the conventional microwave bands has recently resulted in the launch of a few mmWave radio frequency bands in 5G. A huge amount of mmWave, sub-teraHz, and teraHz radio spectrum is still unutilized which, if made usable, holds strong potential in resolving the challenges of B5G/6G wireless networks. Moreover, low-cost substrate and all-passive structure-based tags design for identification applications using BsC concepts is a potential application of teraHz bands. Such THID concept (similar to RFID) is studied in \cite{7750893,8454696,8823734,6548163}. An efficient transceiver design for B5G networks that can operate over a wide range of frequencies (including teraHz) is a vital research challenge. For coherent transceiver design, the imbalance between the in-phase and quadrature-phase branches leads to an imperfect image-frequency filtration, which emerges as more prominent for high carrier frequencies, e.g., mmWave, sub-teraHz, and teraHz bands. In this context, non-coherent communications over teraHz bands is a more suitable proposal in various applications context. There exist various other interesting futuristic teraHz communication applications, e.g., inter-chip communications, healthcare, etc.

\subsubsection{Deployment and Operational Cost of Wireless Networks}

The deployment cost of 5G networks is significantly higher than the capital expenditure of mobile network operators, which is anticipated as a primary delaying factor in the development of 5G network infrastructure (especially in rural areas) \cite{8951153}. The revolutionary technologies being introduced in 5G networks can only be fully benefited from if there are opportunities for revenue generation available in proportion to the required deployment cost.
In addition, the affordability of 5G network services is regarded as a vital issue by the International Telecommunication Union (ITU) broadband commission \cite{8300495}.
There is a vital need to re-engineer the technologies and explore multiple horizons to reduce the overall network deployment cost and affordability from customers' point of view. For example, design of the optimized spectrum sharing techniques can significantly assist in reducing the spectrum licensing cost for the operators, sharing of infrastructure (active and passive) between multiple operators can reduce the deployment cost per operator, reducing the power utilization through efficient device design and communication technique can decrease the power consumption cost, data (network and users) sharing can help in laying the foundations of viable business models, etc.
Zhang et al. in \cite{zhang20206g} has argued that the key to success of 6G lies in delivering a 1000x reduction in the cost from the customer's point of view compared to that of 5G.

Non-coherent communication devices being free of the requirement of synchronization-equipment (for phase) can be regarded as a cost-efficient solution for the device design (especially for massive-IoT devices). Moreover, BsC for connectivity at massive scale (e.g., massive-IoT devices) is an ultra-low power consumption and spectrum sharing communication primitive, which can help in reducing the power and spectrum resources consumption costs. Together, the benefits offered by BsC and non-coherent communications can be exploited in devising a comprehensive framework for green and cost-effective air-interfaces for e-mMTC services in B5G and 6G wireless networks.
\subsubsection{Blockchain for IoT}
Blockchain is a kind of distributed ledger technology that spreads across the entire distributed system. It has an enormous potential in various dynamic areas beyond its popular application in the area of cryptocurrency.
Along with various other notable applications of blockchain technology in finance, security, logistics, etc, IoT is also regarded as one of its most synergistic application.  In IoT, the challenges associated with the nature of decentralized architecture, heterogeneous nodes, heterogeneous data, and complex network topology makes blockchain a natural application.
Blockchain technology can extend various promising provisions for ensuring interoperability in IoT nodes; security, traceability, and reliability of data; automation in node-to-node operations, etc.
In \cite{8731639}, the integration of blockchain in IoT networks is strongly motivated along with a through a literature survey.
There exist various comprehensive review articles in the literature which strongly motivate the adoption of blockchain technology to foster different features in IoT, see e.g., \cite{8370027,7467408,8012302}.

In Table \ref{tab_challenges_summary}, a summary of the discussed open research challenges and research directions related to backscatter and non-coherent communications is provided. Considering the reviewed enormous independent potential of blockchain, BsC, and non-coherent technologies for addressing different problems of different nature encountered in massively connected networks (e.g., mMTC/IoT), it is strongly recommended to thoroughly investigate their potential in a joint application framework.
\section{Conclusions}\label{Sec_Conclusion}
Starting with the comprehensive review of innovative 5G technologies and services, various attention-seeking research challenges that may emerge in the B5G era have been identified in this paper. The identified notable emerging challenges include the scarcity of network capacity to support the anticipated ultra-massive connectivity that will arise from the advent of mMTC services, scarcity of usable frequency spectrum and the need for spectral-efficient communication techniques, and high network deployment and operational cost. Considering the high diversity in the requirements of 6G wireless networks, it is undeniable that no single physical layer technology can stand as the enabler of all services and use-cases. The solution lies in the appropriate amalgamation of multiple promising modern technologies besides the development of new innovative technologies.

In the above context, a framework for enabling ultra-massive connectivity at the physical layer for delivering e-mMTC services in B5G wireless networks has been proposed in this paper. The proposed framework is mainly based on the joint exploitation of non-coherent and BsC. Considering the promising features of these promising technologies, their state-of-the-art has been comprehensively reviewed. Furthermore, the potential applications and use-cases of these technologies in 6G wireless networks have been identified. Subsequently, the potential of the proposed framework in applying together with other promising technologies for mMTC services have also been thoroughly discussed; which include, quantum technologies, ML, VLBsC, AcBsC, UAVs, and multi-antenna systems -- to name a few. Moreover, various potential future research directions have been identified and discussed.

\bibliographystyle{IEEEtran}
\bibliography{References}

\begin{IEEEbiography}[{\includegraphics[width=1in,height=1.25in,clip,keepaspectratio]{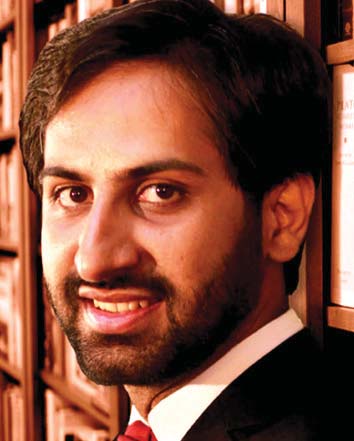}}]{Syed Junaid Nawaz} (S'08--M'12--SM'16) received the Ph.D. degree in electronic engineering from Mohammad Ali Jinnah University, Islamabad, in February 2012. Since September 2005, he has worked on several research and teaching positions with COMSATS University Islamabad (CUI), Pakistan; Staffordshire University, UK; Federal Urdu University, Pakistan; The University of York,
UK; and Aristotle University of Thessaloniki, Greece. He is currently working as an Assistant Professor with the Department of Electrical and computer Engineering, COMSATS University Islamabad (CUI), Islamabad, Pakistan.

His current research interests include physical channel modeling, channel estimation and characterization, mMIMO systems, adaptive signal processing, machine learning, compressed sensing, mmWave channels, airborne internet, internet of things, and vehicle-to-vehicle communications.
\end{IEEEbiography}

\begin{IEEEbiography}[{\includegraphics[width=1in,height=1.25in,clip,keepaspectratio]{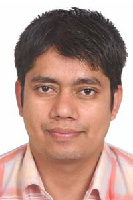}}] {Shree Krishna Sharma} (S'12-M'15-SM'18) is currently Research Scientist at the Interdisciplinary Center for Security, Reliability and Trust (SnT), University of Luxembourg. Prior to this, he held research positions at the University of Western Ontario, Canada, and Ryerson University, Canada; and also worked as a Research Associate at the SnT after receiving his PhD degree in Wireless Communications from the University of Luxembourg in 2014. He has published more than 100 technical papers in scholarly journals, international conferences, and book chapters, and has over 2300 google scholar citations with an h-index of 24. His current research interests include 5G and beyond wireless, Internet of Things, machine-type communications, machine learning, edge computing and optimization of distributed communications, computing and caching resources.

He is a Senior Member of IEEE and is the recipient of several prestigious awards including ``FNR Award for Outstanding PhD Thesis 2015'' from FNR, Luxembourg, ``Best Paper Award" in CROWNCOM 2015 conference, ``2018 EURASIP JWCN Best Paper Award" and the co-recipient of ``FNR Award for Outstanding Scientific Publication 2019". He has been serving as a Reviewer for several international journals and conferences; as a TPC member for a number of international conferences including IEEE ICC, IEEE GLOBECOM, IEEE PIMRC, IEEE VTC and IEEE ISWCS; and an Associate Editor for IEEE Access journal. He co-organized a special session in IEEE PIMRC 2017, a workshop in IEEE SECON 2019, worked as a Track co-chair for IEEE VTC-fall 2018 conference, and published an IET book on ``Satellite Communications in the 5G Era" as a lead editor.
\end{IEEEbiography}

\begin{IEEEbiography}[{\includegraphics[width=1in,height=1.25in,clip,keepaspectratio]{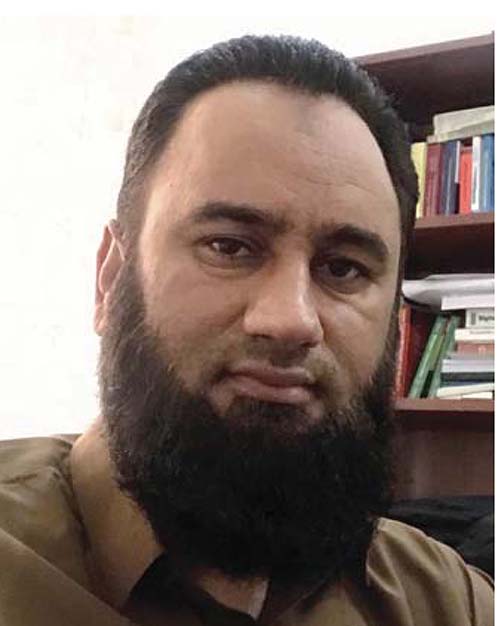}}] {Babar Mansoor} received the B.Sc. degree in Computer Engineering from COMSATS Institute of Information Technology (CIIT), Abbottabad, Pakistan, in 2005, M.Sc. degree in Microelectronic Systems in 2007 from The University of Liverpool, UK and PhD in Electrical Engineering in 2018 from COMSATS University Islamabad (CUI), Islamabad, Pakistan. Presently, he is working as an Assistant Professor with the Department of Electrical and Computer Engineering, COMSATS University Islamabad (CUI), Islamabad, Pakistan.
His research areas include wireless communications, signal processing, and MIMO communication systems.
\end{IEEEbiography}

\begin{IEEEbiography}[{\includegraphics[width=1in,height=1.25in,clip,keepaspectratio]{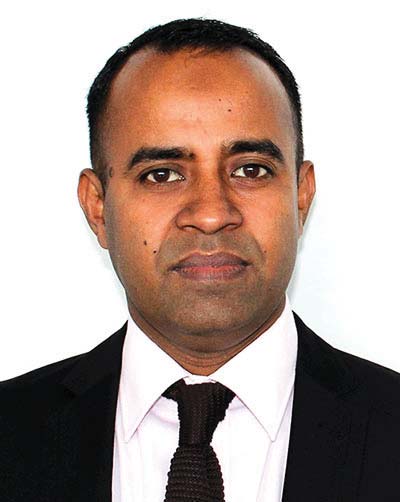}}]{Mohammad N. Patwary}
(SM'11) received the B.Eng. degree (Hons). in electrical and electronic engineering from the Chittagong University of Engineering and Technology, Bangladesh, in 1998, and the Ph.D. degree in telecommunication engineering from The University of New South Wales, Sydney, Australia, in 2005. He was with General Electric Company of Bangladesh from 1998 to 2000 and with Southern Poro Communications, Sydney, from 2001 to 2002, as Research and Development Engineer. He was a Lecturer with The University of New South Wales from 2005 to 2006, and then a Senior Lecturer with Staffordshire University, U.K., from 2006 to 2010. He was then a Full Professor of Wireless Systems and Digital Productivity and the Chair of the Centre of Excellence on Digital Productivity with Connected Services, Staffordshire University, until 2016. From 2016 to 2020, he was a Full Professor and Head of the Intelligent Systems and Networks (ISN) research group with the School of Computing and Digital Technology, Birmingham City University, UK. He is currently a Professor with the Faculty of Science \& Engineering and Director of Centre for Future Networks \& Autonomous Systems, University of Wolverhampton, Wolverhampton, UK.
He is Principal Data Architect for a large scale 5G testbed in the UK to accelerate digital productivity \& develop urban connected community. His current research interests include - sensing and processing for intelligent systems, wireless communication systems design and optimization, signal processing and energy-efficient systems, future generation of cellular network architecture and business modelling for Data-economy.
\end{IEEEbiography}

\begin{IEEEbiography}[{\includegraphics[width=1in,height=1.25in,clip,keepaspectratio]{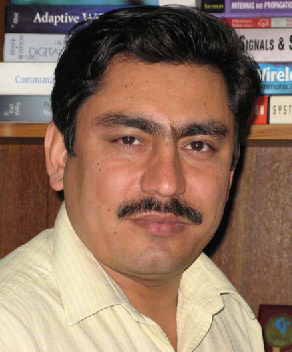}}]{Noor M. Khan} (M'06--SM'12) received his B.Sc. degree in electrical engineering from the University of Engineering and Technology (UET), Lahore, Pakistan, in 1998 and Ph.D. degree in electrical engineering from the University of New South Wales (UNSW), Sydney, Australia in 2006. He held several positions in WorldCall, NISTE, PTCL, UNSW, GIK Institute of Engineering Sciences and Technology, and Mohammad Ali Jinnah University, Pakistan from 1998 to 2015. Currently, he is working as professor and head of the electrical engineering department with the Capital University of Science \& Technology (CUST), Islamabad, Pakistan. He has served as Chair of the Technical Program Committees of IEEE International Conference on Emerging Technologies (ICET) in 2012 and 2017, respectively. He has been awarded Research Productivity Award (RPA) by the Pakistan Council for Science and Technology (PCST) for the years 2011 and 2012. His research interests include channel modeling and characterization, wireless sensor networks, cellular mobile communication networks and ground-to-air communication systems.
\end{IEEEbiography}

\end{document}